ELSEVIER

# The IceCube Data Acquisition System: Signal Capture, Digitization, and Timestamping

R. Abbasi [t], M. Ackermann [af], J. Adams [k], M. Ahlers [y], J. Ahrens [u], K. Andeen [t], J. Auffenberg [ae], X. Bai [w], M. Baker [t], S. W. Barwick [p], R. Bay [e], J. L. Bazo Alba [af], K. Beattie [f], T. Becka [u], J. K. Becker [m], K.-H. Becker [ae], P. Berghaus [t], D. Berley [l], E. Bernardini [af], D. Bertrand [h], D. Z. Besson [r], B. Bingham [f], E. Blaufuss [l], D. J. Boersma [t], C. Bohm [z], J. Bolmont [af], S. Böser [af], O. Botner [ac], J. Braun [t], D. Breeder [ae], T. Burgess [z], W. Carithers [f], T. Castermans [v], H. Chen [f], D. Chirkin [t], B. Christy [l], J. Clem [w], D. F. Cowen [ab,aa], M. V. D'Agostino [e], M. Danninger [k], A. Davour [ac], C. T. Day [f], O. Depaepe [i], C. De Clercq [i], L. Demirörs [q], F. Descamps [n], P. Desiati [t], G. de Vries-Uiterweerd [n], T. DeYoung [ab], J. C. Diaz-Velez [t], J. Dreyer [m], J. P. Dumm [t], M. R. Duvoort [ad], W. R. Edwards [f], R. Ehrlich [l], J. Eisch [t], R. W. Ellsworth [l], O. Engdegård [ac], S. Euler [a], P. A. Evenson [w], O. Fadiran [c], A. R. Fazely [d], T. Feusels [n], K. Filimonov [e], C. Finley [t], M. M. Foerster [ab], B. D. Fox [ab], A. Franckowiak [g], R. Franke [af], T. K. Gaisser [w], J. Gallagher [s], R. Ganugapati [t], L. Gerhardt [f], L. Gladstone [t], D. Glowacki [t], A. Goldschmidt [f], J. A. Goodman [l], R. Gozzini [u], D. Grant [ab], T. Griesel [u], A. Groß [o,k], S. Grullon [t], R. M. Gunasingha [d], M. Gurtner [ae], C. Ha [ab], A. Hallgren [ac], F. Halzen [t], K. Han [k], K. Hanson [t], R. Hardtke [y], Y. Hasegawa [j], J. Haugen [t], D. Hays [f], J. Heise [ad], K. Helbing [ae], M. Hellwig [u], P. Herquet [v], S. Hickford [k], G. C. Hill [t], J. Hodges [t], K. D. Hoffman [l], K. Hoshina [t], D. Hubert [i], W. Huelsnitz [l], B. Hughey [t], J.-P. Hülß [a], P. O. Hulth [z], K. Hultqvist [z], S. Hussain [w], R. L. Imlay [d], M. Inaba [j], A. Ishihara [j], J. Jacobsen [t], G. S. Japaridze [c], H. Johansson [z], A. Jones [f], J. M. Joseph [f], K.-H. Kampert [ae], A. Kappes [t,1], T. Karg [ae], A. Karle [t], H. Kawai [j], J. L. Kelley [t], J. Kiryluk [f,e], F. Kislat [af], S. R. Klein [f,e], S. Kleinfelder [f], S. Klepser [af], G. Kohnen [v], H. Kolanoski [g], L. Köpke [u], M. Kowalski [g], T. Kowarik [u], M. Krasberg [t], K. Kuehn [p], E. Kujawski [f], T. Kuwabara [w], M. Labare [h], K. Laihem [a], H. Landsman [t], R. Lauer [af], A. Laundrie [t], H. Leich [af], D. Leier [m], C. Lewis [t], A. Lucke [g], J. Ludvig [f], J. Lundberg [ac], J. Lünemann [m], J. Madsen [y], R. Maruyama [t], K. Mase [j], H. S. Matis [f,2], C. P. McParland [f], K. Meagher [l], A. Meli [m], M. Merck [t], T. Messarius [m], P. Mészáros [ab,aa], R. H. Minor [f], H. Miyamoto [j], A. Mohr [g], A. Mokhtarani [f], T. Montaruli [t,3], R. Morse [t], S. M.

---

[1] On leave of absence from Universität Erlangen-Nürnberg, Physikalisches Institut, D-91058, Erlangen, Germany.

[2] Corresponding author. Tel.: +1-510-486-5031; fax: +1-510-486-4818; e-mail: hsmatis@lbl.gov.

[3] On leave of absence from Università di Bari and INFN, Dipartimento di Fisica, I-70126, Bari, Italy.

Movit [aa], K. Münich [m], A. Muratas [f], R. Nahnhauer [af], J. W. Nam [p], P. Nießen [w], D. R. Nygren [f], S. Odrowski [o], A. Olivas [l], M. Olivo [ac], M. Ono [j], S. Panknin [g], S. Patton [f], C. Pérez de los Heros [ac], J. Petrovic [h], A. Piegsa [u], D. Pieloth [af], A. C. Pohl [ac,4], R. Porrata [e], N. Potthoff [ae], J. Pretz [l], P. B. Price [e], G. T. Przybylski [f], K. Rawlins [b], S. Razzaque [ab,aa], P. Redl [l], E. Resconi [o], W. Rhode [m], M. Ribordy [q], A. Rizzo [i], W. J. Robbins [ab], J. P. Rodrigues [t], P. Roth [l], F. Rothmaier [u], C. Rott [ab,5], C. Roucelle [f,e], D. Rutledge [ab], D. Ryckbosch [n], H.-G. Sander [u], S. Sarkar [x], K. Satalecka [af], P. Sandstrom [t], S. Schlenstedt [af], T. Schmidt [l], D. Schneider [t], O. Schulz [o], D. Seckel [w], B. Semburg [ae], S. H. Seo [z], Y. Sestayo [o], S. Seunarine [k], A. Silvestri [p], A. J. Smith [l], C. Song [t], J. E. Sopher [f], G. M. Spiczak [y], C. Spiering [af], T. Stanev [w], T. Stezelberger [f], R. G. Stokstad [f], M. C. Stoufer [f], S. Stoyanov [w], E. A. Strahler [t], T. Straszheim [l], K.-H. Sulanke [af], G. W. Sullivan [l], Q. Swillens [h], I. Taboada [e,6], O. Tarasova [af], A. Tepe [ae], S. Ter-Antonyan [d], S. Tilav [w], M. Tluczykont [af], P. A. Toale [ab], D. Tosi [af], D. Turčan [l], N. van Eijndhoven [ad], J. Vandenbroucke [e], A. Van Overloop [n], V. Viscomi [ab], C. Vogt [a], B. Voigt [af], C. Q. Vu [f], D. Wahl [t], C. Walck [z], T. Waldenmaier [w], H. Waldmann [af], M. Walter [af], C. Wendt [t], S. Westerhof [t], N. Whitehorn [t], D. Wharton [t], C. H. Wiebusch [a], C. Wiedemann [z], G. Wikström [z], D. R. Williams [ab,7], R. Wischnewski [af], H. Wissing [a], K. Woschnagg [e], X. W. Xu [d], G. Yodh [p], S. Yoshida [j]

(The IceCube Collaboration)

[a]*III Physikalisches Institut, RWTH Aachen University, D-52056 Aachen, Germany*

[b]*Dept. of Physics and Astronomy, University of Alaska Anchorage, 3211 Providence Dr., Anchorage, AK 99508, USA*

[c]*CTSPS, Clark-Atlanta University, Atlanta, GA 30314, USA*

[d]*Dept. of Physics, Southern University, Baton Rouge, LA 70813, USA*

[e]*Dept. of Physics, University of California, Berkeley, CA 94720, USA*

[f]*Lawrence Berkeley National Laboratory, Berkeley, CA 94720, USA*

[g] *Institut für Physik, Humboldt-Universität zu Berlin, D-12489 Berlin, Germany*

[h]*Université Libre de Bruxelles, Science Faculty CP230, B-1050 Brussels, Belgium*

[i]*Vrije Universiteit Brussel, Dienst ELEM, B-1050 Brussels, Belgium*

[j]*Dept. of Physics, Chiba University, Chiba 263-8522, Japan*

[k]*Dept. of Physics and Astronomy, University of Canterbury, Private Bag 4800, Christchurch, New Zealand*

[l]*Dept. of Physics, University of Maryland, College Park, MD 20742, USA*

[m]*Dept. of Physics, Universität Dortmund, D-44221 Dortmund, Germany*

[4] Affiliated with School of Pure and Applied Natural Sciences, Kalmar University, S-39182 Kalmar, Sweden.
[5] Currently at Dept. of Physics and Center for Cosmology and Astro-Particle Physics, The Ohio State University, Columbus, OH 43210, USA.
[6] Currently at the School of Physics and Center for Relativistic Astrophysics, Georgia Institute of Technology, Atlanta, GA 30332, USA.
[7] Currently at the Dept. of Physics and Astronomy, University of Alabama, Tuscaloosa, AL 35487, USA.

[n] Dept. of Subatomic and Radiation Physics, University of Gent, B-9000 Gent, Belgium
[o] Max-Planck-Institut für Kernphysik, D-69177 Heidelberg, Germany
[p] Dept. of Physics and Astronomy, University of California, Irvine, CA 92697, USA
[q] Laboratory for High Energy Physics, École Polytechnique Fédérale, CH-1015 Lausanne, Switzerland
[r] Dept. of Physics and Astronomy, University of Kansas, Lawrence, KS 66045, USA
[s] Dept. of Astronomy, University of Wisconsin, Madison, WI 53706, USA
[t] Dept. of Physics, University of Wisconsin, Madison, WI 53706, USA
[u] Institute of Physics, University of Mainz, Staudinger Weg 7, D-55099 Mainz, Germany
[v] University of Mons-Hainaut, 7000 Mons, Belgium
[w] Bartol Research Institute, University of Delaware, Newark, DE 19716, USA
[x] Dept. of Physics, University of Oxford, 1 Keble Road, Oxford OX1 3NP, UK
[y] Dept. of Physics, University of Wisconsin, River Falls, WI 54022, USA
[z] Dept. of Physics, Stockholm University, SE-10691 Stockholm, Sweden
[aa] Dept. of Astronomy and Astrophysics, Pennsylvania State University, University Park, PA 16802, USA
[ab] Dept. of Physics, Pennsylvania State University, University Park, PA 16802, USA
[ac] Division of High Energy Physics, Uppsala University, S-75121 Uppsala, Sweden
[ad] Dept. of Physics and Astronomy, Utrecht University/SRON, NL-3584 CC Utrecht, The Netherlands
[ae] Dept. of Physics, University of Wuppertal, D-42119 Wuppertal, Germany
[af] DESY, D-15735 Zeuthen, Germany

---

**Abstract**

IceCube is a km-scale neutrino observatory under construction at the South Pole with sensors both in the deep ice (InIce) and on the surface (IceTop). The sensors, called Digital Optical Modules (DOMs), detect, digitize and timestamp the signals from optical Cherenkov-radiation photons. The DOM Main Board (MB) data acquisition subsystem is connected to the central DAQ in the IceCube Laboratory (ICL) by a single twisted copper wire-pair and transmits packetized data on demand. Time calibration is maintained throughout the array by regular transmission to the DOMs of precisely timed analog signals, synchronized to a central GPS-disciplined clock. The design goals and consequent features, functional capabilities, and initial performance of the DOM MB, and the operation of a combined array of DOMs as a system, are described here. Experience with the first InIce strings and the IceTop stations indicates that the system design and performance goals have been achieved.

## 1. IceCube Overview

IceCube [1],[2] is a kilometer-scale high-energy neutrino observatory now under construction at the Amundsen-Scott South Pole station. Its main scientific goal is to map the high-energy neutrino sky, which is expected to include both a diffuse neutrino flux and point sources [3]. The size of IceCube is set by a sensitivity requirement inferred from the spectrum of Ultra High Energy (UHE) cosmic rays [4]. IceCube is designed to observe and study neutrinos with energies from as low as 100 GeV to perhaps well into the EeV range with useful sensitivity. It can also measure supernova bursts.

Figure 1 is a schematic representation of IceCube. A deep "InIce" array and a surface array "IceTop" are the main components of IceCube. The combination of InIce and IceTop enables the study of cosmic ray composition over a wide range of energies. The AMANDA array [5] is contained within IceCube, as shown in the figure.

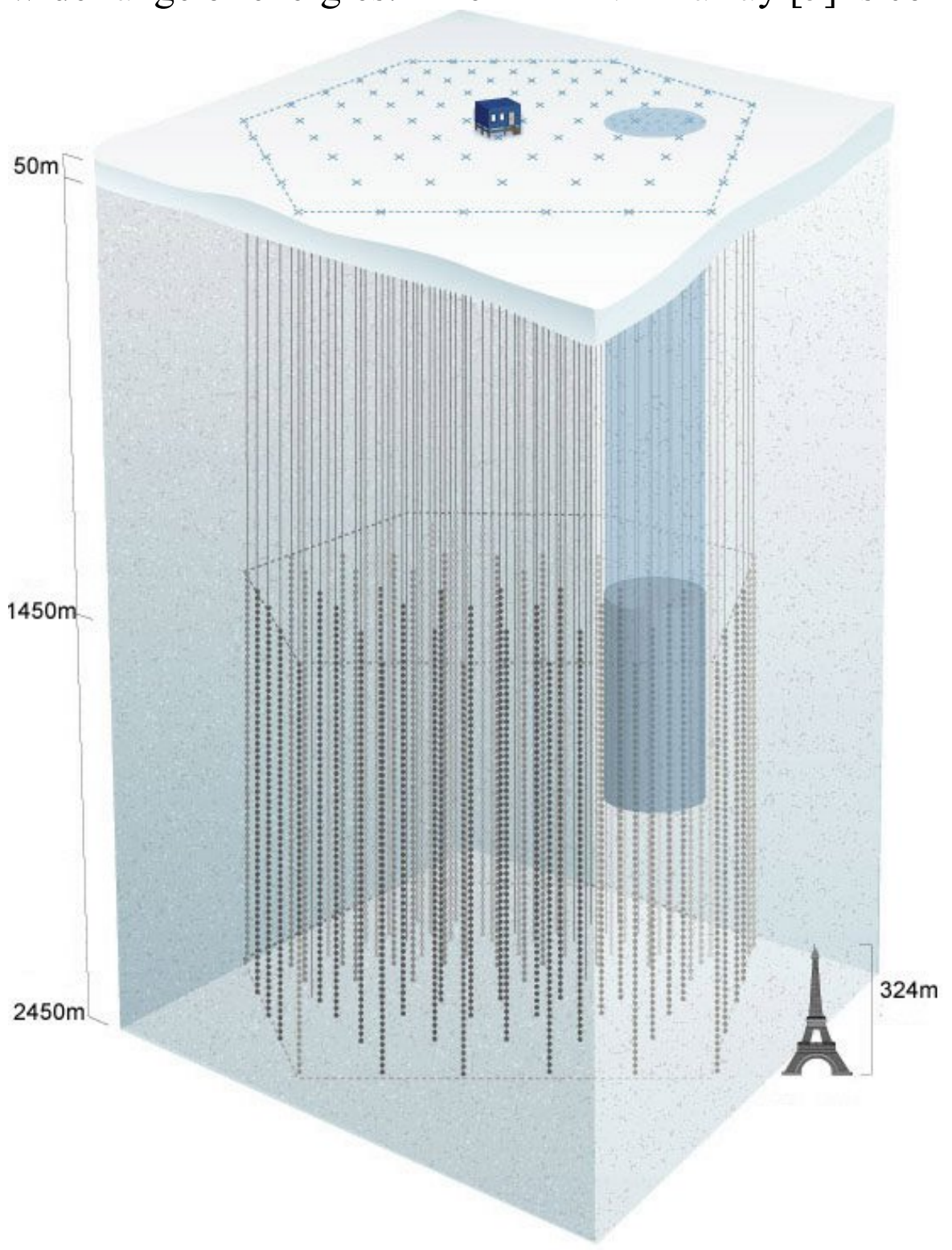

Figure 1. A perspective view of a fully instrumented IceCube detector. The 80 strings for InIce are shown. Each dot represents a DOM. The darker shaded area shows the location of AMANDA, the precursor detector. The IceCube Laboratory is also shown on the surface of the ice.

The detector uses the ~2800 m-thick polar ice sheet to provide a target, an optically clear radiator, and a stable instrument deployment platform. InIce consists of an array of optical sensors called Digital Optical Modules (DOMs), organized in a lattice of ultimately 80 vertical "strings", frozen into the polar ice sheet. Each string includes 60 DOMs, spaced uniformly from a depth of 1450 to 2450 m. There are plans to create a densely instrumented deep core in the center of IceCube by adding six special strings with more closely spaced optical modules concentrated in the clearest ice toward the bottom [6].

A string is deployed into a water-filled hole, which has been bored with a hot-water jet. Once the water refreezes, the DOMs become permanently inaccessible. Heat flow from within the earth introduces a vertical thermal gradient in the ice, leading to a variation in the internal operating temperature of the DOMs from -9°C at the lowest elevation DOM to -32°C at the uppermost DOM.

The IceTop surface air shower array consists of pairs of tanks placed about 25 meters from the top of each down-hole cable and separated from each other by 10 meters. Each tank is instrumented with two DOMs frozen into the top of the ice in the tanks. The DOMs capture Cherenkov light generated by charged particles passing through the tanks. Typical signals are much bigger than signals in the deep ice. For example, a single muon typically generates a signal with a total charge equivalent to 130 photoelectrons, and a large air shower often produces a signal equivalent to tens or hundreds of muons. Operating temperatures for IceTop stations vary seasonally from -40°C to -20°C. The ice temperature is about 10°C lower than the board temperature.

A DOM contains a photomultiplier tube (PMT), which detects the blue and near-UV Cherenkov light produced by relativistic charged particles passing through the ice. Photons travel long distances due to the large absorption length of well above 100 m on average. Scattering in the ice disperses photon arrival times and directions for distances that are large compared to the effective scattering length of ~24 m. The signal shape depends on both the distance from the source and its linear extent. The width in time generally increases with the track's distance from the DOM. In addition, the reconstruction of InIce cascades and IceTop air showers places further requirements on the DAQ architecture. The very wide range of possible energy depositions leads to a demanding requirement on dynamic range in the detectors. Digital information from the ensemble of DOMs allows reconstruction of event topology and energy, from which the nature of the event may be determined.

In this paper, we concentrate on system aspects of hardware and software elements of IceCube that capture and process the primary signal information. We discuss hardware design and implementation, and demonstrate functionality. The clock distribution scheme used by IceCube is novel, and so we early on focused on our ability to do time calibration. Results from that effort are included. Our ability to calibrate the trigger efficiency and the waveform digitizers for charge and feature extraction will be presented in a later paper.

Sections 2-4 provide conceptual overviews, performance goals, and basic technical aspects. Section 5 describes manufacturing and testing procedures, while Section 6 summarizes the performance and reliability up to August 2008.

## 2. DAQ - Technical Design

In the broadest sense, the primary goal for the IceCube DAQ is to capture and timestamp with high accuracy, the complex, widely varying optical signals over the maximum dynamic range provided by the PMT. To meet this goal, the IceCube DAQ architecture is decentralized. The digitization is done individually inside each DOM, and then collected in the counting house in the IceCube Laboratory (ICL), which is located on the surface of the ice.

Operationally, the DOMs in IceCube resemble an ensemble of satellites, with interconnects and communication via copper wire-pairs. The data collection process is centrally managed over this digital communications and time signal distribution network. Power is also distributed over this network. This decentralized approach places complex electronic subsystems beyond any possibility of access for maintenance. Hence, reliability and programmability considerations were drivers in the engineering process.

The goal of this architecture is to obtain very high information quality with minimal on-site personnel needed at the South Pole during both detector commissioning and operation. The IceCube DAQ design relies upon the collaboration’s understanding of how to build a digital system after studying from the behavior 41 prototype DOMs [8] deployed in AMANDA.

### *2.1. The “Hit” – The Fundamental IceCube Datum*

The primary function of the DOM is to produce a digital output record, called a “Hit” whenever one or more photons are detected. The basic elements of a Hit are a timestamp generated locally within the DOM and waveform information. A Hit can range in size from a minimum of 12 bytes to several hundred bytes depending on the waveform's complexity and trigger conditions. A Hit always contains at least a timestamp, a coarse measure of charge, and several bits defining Hit origin.

Waveform information is collected for a programmable interval – presently chosen to be 6.4 μs, which is more than the maximum time interval over which the most energetic events are expected to contribute detectable light to any one DOM.

Hardware trigger signals exchanged between neighboring channels may be utilized by the DOM trigger logic to limit data flow by either minimizing the level of waveform detail within a Hit, or by rejecting Hits that are isolated – *i.e.,* with no nearby Hits in time or space, and hence much more likely to be PMT noise than real physics events.

### *2.2. DAQ Elements Involved in Generating Hits*

The real-time IceCube DAQ includes those functional blocks of IceCube that contribute to time-calibrated Hits:

1. The *Digital Optical Module*, deployed in both InIce and IceTop.
2. The *DOMHub*, located in the ICL, and based on an industrial PC.
3. The *Cable Network*, which connects DOMs to the DOMHub and adjacent DOMs to each other.
4. The *Master Clock*, which distributes time calibration (RAPcal) signals derived from a GPS receiver to the DOMHubs.
5. The *Stringhub*, a software element that, among other tasks, maps Hits from DOM clock units to the clock domain of the ICL and time-orders the Hit stream for an entire string.

Together, these elements capture the PMT anode pulses above a configurable threshold with a minimum set value of ~0.25 single photoelectron (SPE) pulse height, and transform the information to an ensemble of timestamped, time-calibrated, and time-ordered digital data blocks.

### *2.3. The Digital Optical Module – Overview*

The DOM’s main elements are a 25 cm diameter PMT (Hamamatsu R7081-02), a modular 2 KV high voltage (HV) power supply for the PMT, a separate passive base for PMT operation, the DOM Main Board (MB), a stripline signal delay board, and a 13 mm thick glass sphere to withstand the pressure of its deep deployment. A flexible gel provides support and optical coupling from the glass sphere to the PMT’s face. Figure 2 is an illustration of a DOM with its components.

The assembled DOM is filled with dry nitrogen to a pressure of approximately ½ atmosphere. This maintains a strong compressive force on the sphere, assuring mechanical integrity of the circumferential seal during handling, storage, and deployment. The DOM provides built-in electronic sensing of the gas pressure within the assembled DOM, enabling the detection of a fault either in the seal or failure of the PMT vacuum.

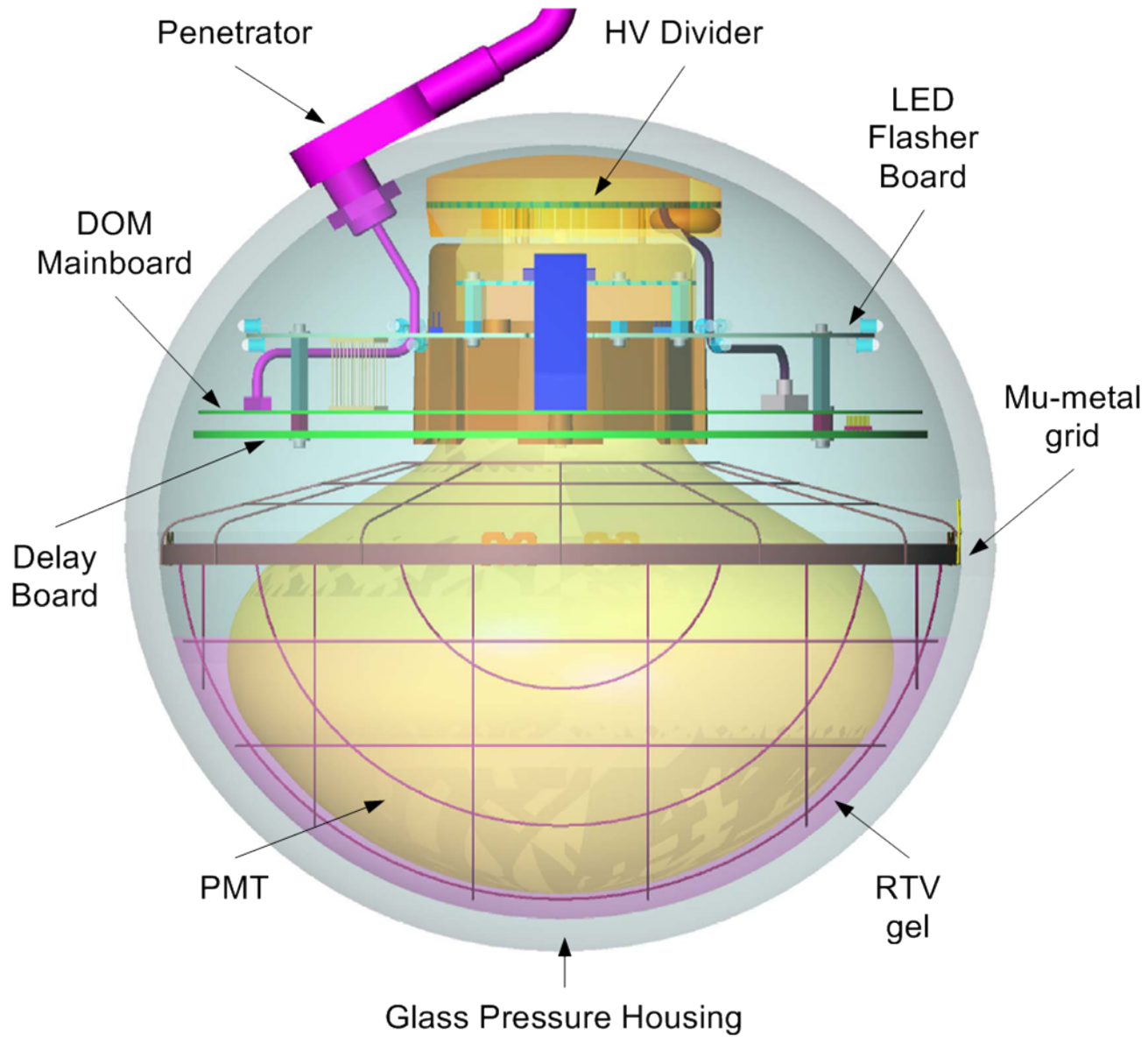

Figure 2. A schematic illustration of a DOM. The DOM contains a HV generator with divides the voltage to the photomultiplier. The DOM Mainboard or DOM MB digitizes the signals from the phototube, actives the LEDs on the LED flasher board, and communicates with the surface. A mu-metal grid shields the phototube against the Earth's magnetic field. The phototube is optically coupled to the exterior Glass Pressure Housing by RTV gel. The penetrator provides a path where the wires from the surface can pass through the Glass Pressure shield.

The PMT is operated with the photocathode grounded. The anode signal formation hence occurs at positive HV. This analog signal is presented to the DOM MB signal path, DC-coupled from the input to a digitizer. At the input, the signal is split to a high-bandwidth PMT discriminator path and to a 75 ns high quality delay line, which provides enough time for the downstream electronics to receive a trigger from the discriminator.

The DOM MB (Figure 3), the "central processor" of the DOM, receives the PMT signals. After digitization, the DOM MB formats the data to create a Hit. High-bandwidth waveform capture is accomplished by an application specific integrated circuit (ASIC), the Analog Transient Waveform Digitizer (ATWD) [9]. Data is buffered until the DOM MB receives a request to transfer data to the ICL.

In addition to the signal capture/digitization scheme, the use of free-running high-stability oscillators in the DOMs is an innovation that permits the precise time calibration of data without actual synchronization, and at the same time creates negligible impact on network bandwidth. Timestamping of data is realized by a Reciprocal Active Pulsing (RAPcal) [10] procedure, which is described in Section 4.7.

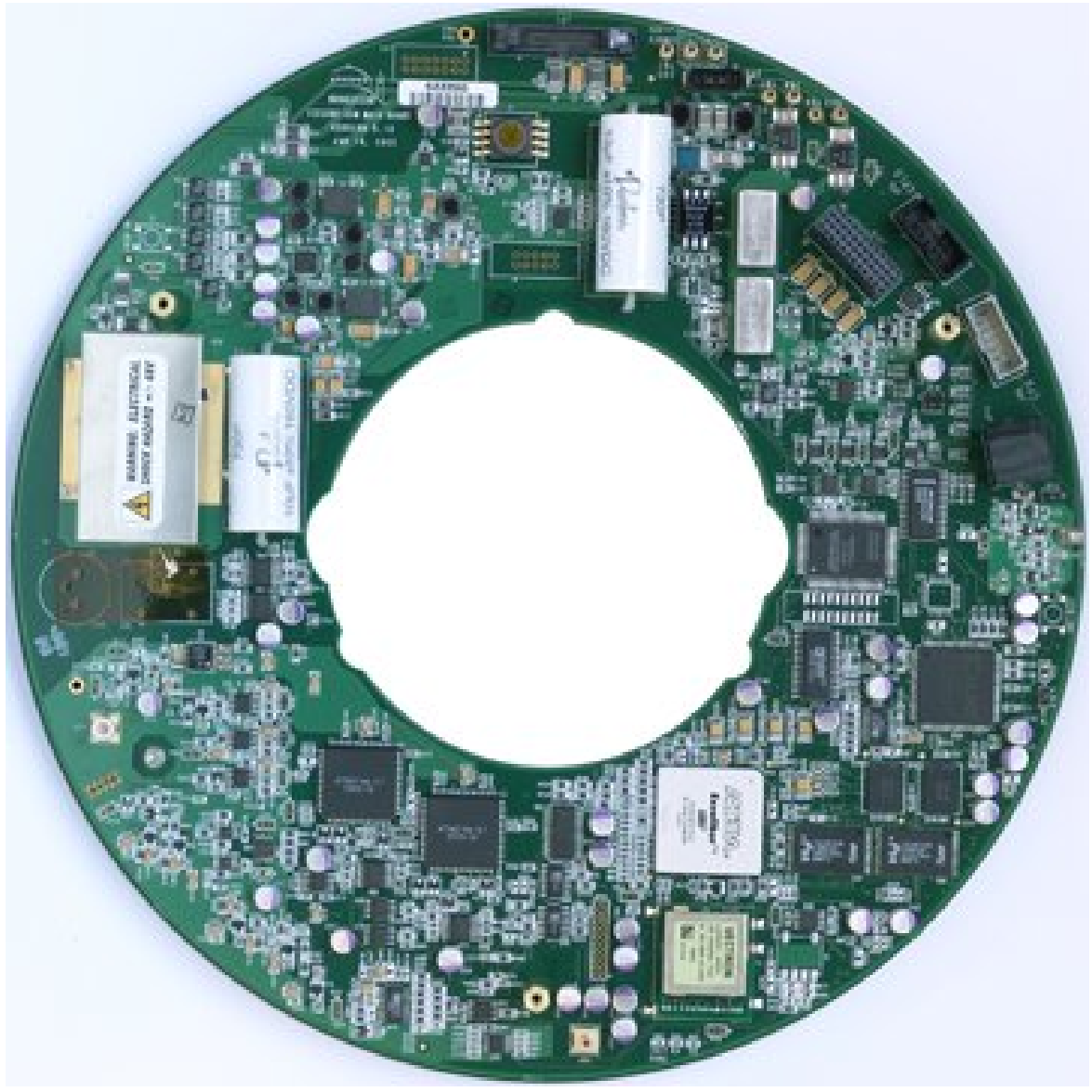

Figure 3. A photograph of the DOM MB. The diameter of the circuit board is 274 mm. This circular circuit board communicates with the surface and provides power and drives the other electronics board inside the DOM. This photograph shows the location of the components, which are described in the text.

The DOM includes a "flasher" board hosting 12 LEDs that can be actuated to produce bright UV optical pulses detectable by other DOMs. Flasher board LEDs can be pulsed either individually or in combinations at programmable output levels and pulse lengths. They are used to stimulate and calibrate distant DOMs, simulate physical events, and to investigate optical properties of the ice. In addition, the DOM MB is equipped with an "on-board LED", which delivers precisely timed, but weak signals for calibration of single photoelectron pulses and PMT transit times. A complete description of the DOM MB, including its other functions, can be found in the next section.

### *2.4. DOM MB Technical Design*

The DOM MB's primary components are identified in Figure 4, while the functional blocks are shown in Figure 5. The top of Figure 5 shows that the analog signal from the PMT is split into three paths at the input to the DOM MB. The top path is for the trigger. Below it is the main signal path which goes through a 75 ns delay line and is then split and presented to three channels of the two ATWDs after different levels of amplification. Finally, a part of the PMT signal is sent to an ADC designed for handling longer signals and a lower sampling speed than the ATWDs. These signal paths, the electronics needed to digitize them and send the Hits to the surface, and some additional circuitry are described in the following subsections.

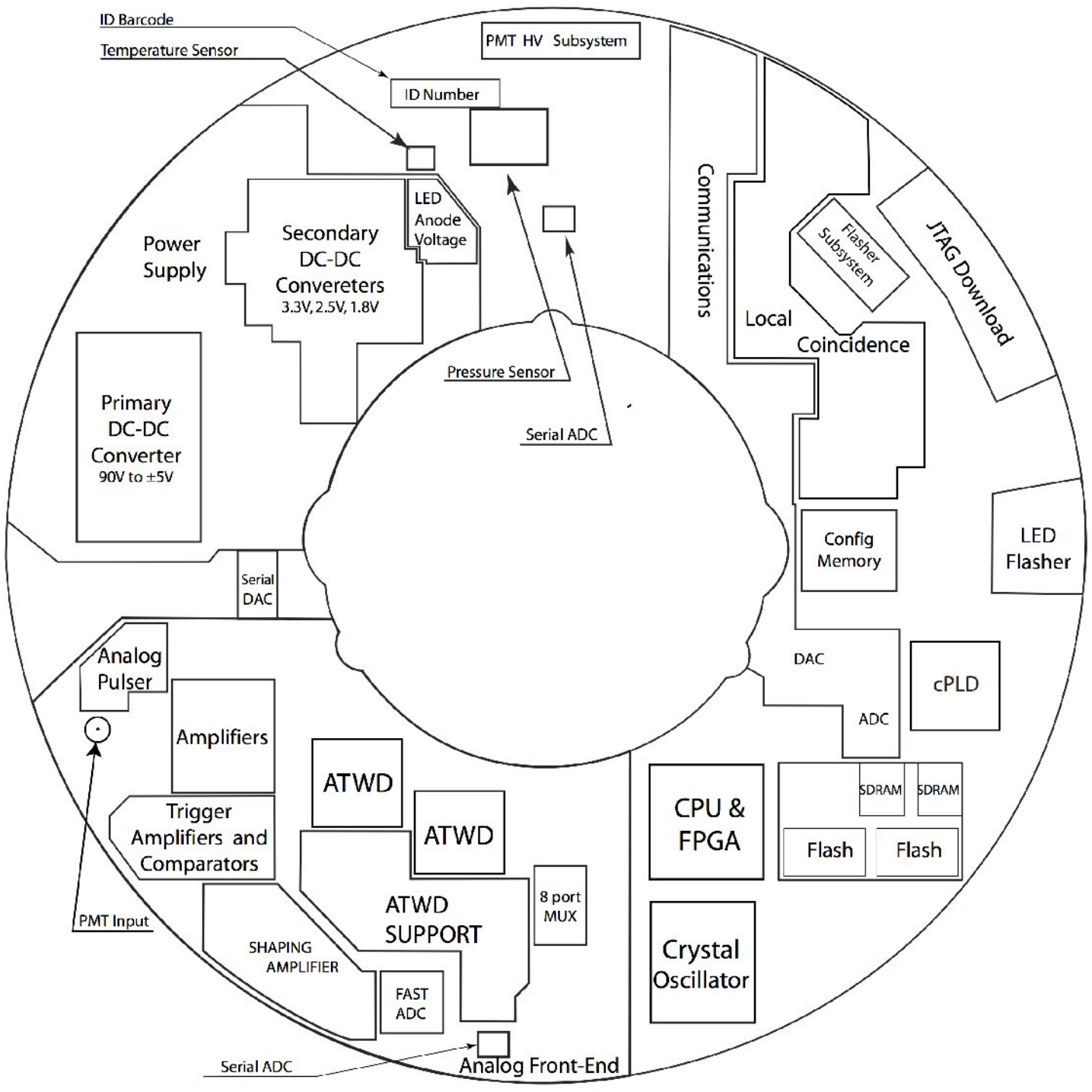

Figure 4. The location of the DOM MB functional blocks. This figure shows the subsystems on the DOM MB. It has the same orientation as Figure 3 so it is possible to locate these items on the photograph.

### *2.4.1. Power Supply*

A single twisted pair carries communications, power, and timing signals to the DOM from the surface electronics. The pair connects to a power filter network that steers the bidirectional, differential signals to the DOM's communication interface, and provides 96 V DC power to the main DC-to-DC converter. Low equivalent series resistance ceramic capacitors and ferrite power filters at the DC-DC converter input and output effectively suppress the switching noise components up to several hundred MHz. The DC-to-DC converter provides +5 V and -5 V to circuits on the DOM main board, to a mezzanine board interface, and to a secondary DC-to-DC converter, which produces 1.8 V, 2.5 V, and 3.3 V for all on-board digital electronics. When power is applied to the DOM, the on-board power management supervisor circuit in conjunction with logic in the Complex Programmable Logic Device (CPLD) initiates the boot sequence of the DOM MB from the serial configuration memory.

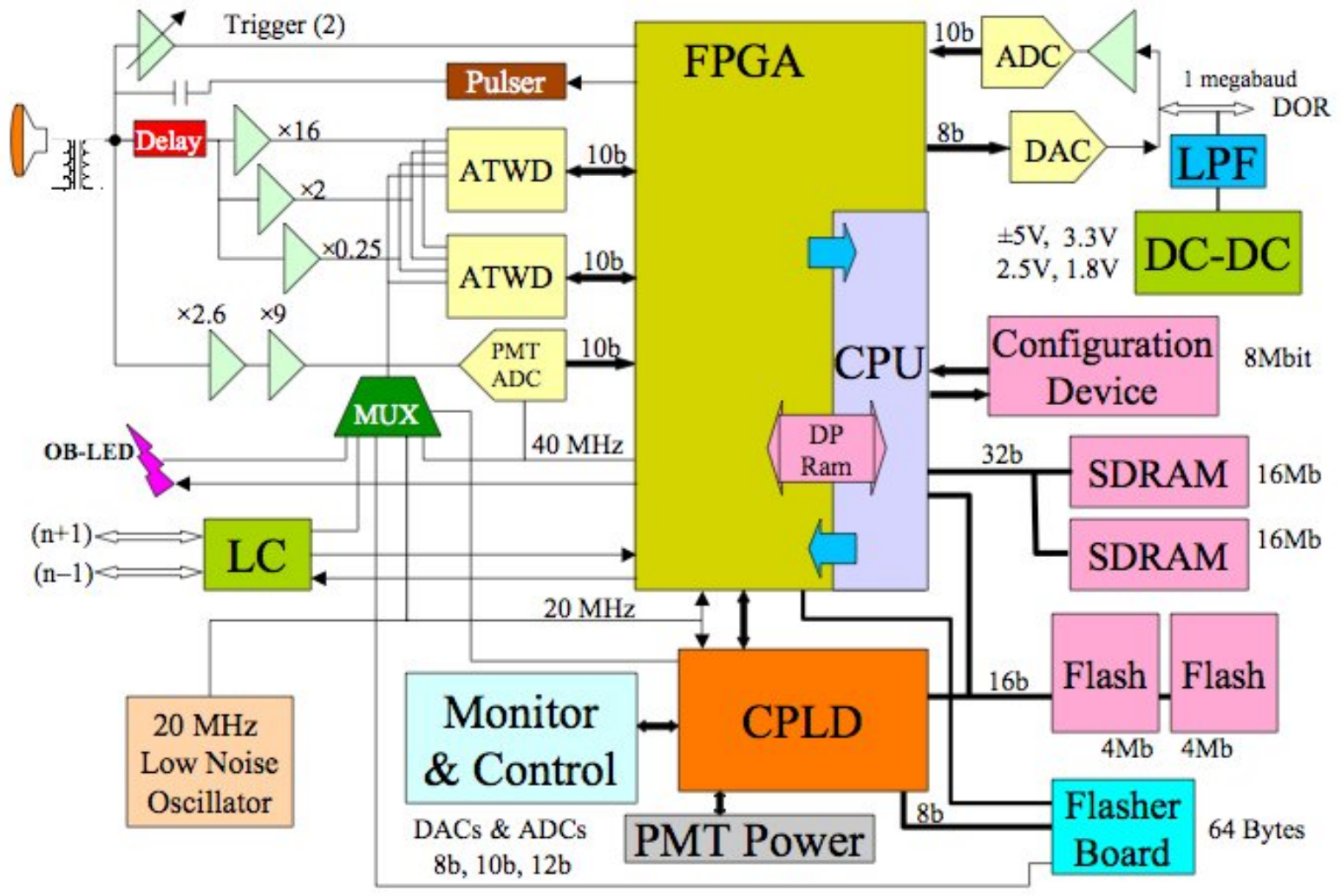

Figure 5. Block diagram of the DOM MB. The triangle with an arrow in the upper left is a comparator with a variable threshold. A photon hits the photomultiplier, which is in the upper left. This signal from the photomultiplier is delayed and split to the ATWD and PMT ADC. The FPGA controls the readout. Full details of the operation of the components are described in the text.

Conservative engineering practices dictate that the PMT photocathode be operated at ground potential with respect to the DOM MB. With capacitive coupling, the signal droop limitation would require an impractically large value (~1 μF for a 50 Ω termination). Furthermore, leakage currents in faulty/degraded high-voltage ceramic capacitors can produce noise resembling PMT pulses. An analysis of the signal and power supply loops reveals that, with transformer coupling, HV power supply noise couples much more weakly into the DOM MB input than with capacitor coupling.

A wide-band high-voltage pulse transformer satisfies the engineering requirements. The 30 pF of anode to front-end capacitance reduces the risk of damage to the DOM MB by discharge in the PMT base because the available energy is small.

The transformer exceeds the pulse rise-time requirements for short pulses (<8 ns FWHM). Good performance depends on shunting the primary winding with a 100 Ω resistor, which also provides back-termination for the DOM MB input circuit and damps ringing in the PMT anode circuit. It is important to note that long time-constants can be employed in the DOM because the average pulse rate is very low; otherwise, field build-up in the core would cause a significant baseline shift.

The time constants of the transformer pass the high-frequency components of the signals with negligible loss, but lead to a droop after large amplitude signals. The DOM MB digitizer pedestals are set at ~10% of the maximum scale, to permit the capture of waveforms with below-baseline excursions.

### *2.4.2. Analog Input Amplifiers*

The amplifiers for the trigger subsystem tap into the decoupled PMT signal right at the DOM MB input coax connector. Also from this input, the signal is passed through a serpentine 75 ns delay line, embedded in a custom printed circuit board made with superior signal propagation materials. The delayed signal is split to three separate wide-band amplifiers (×16, ×2, and ×0.25), which preserve the PMT analog waveform with only minor bandwidth losses. Each amplifier sends its output to separate inputs of the ATWD. The amplifiers have a 100 MHz bandwidth, which is roughly matched to the 300 MSPS ATWD sampling rate

The circuitry confines the ATWD input signal within a 0 to 3 V range. If the input voltage were below –0.5 V, then the ATWD could be driven into latch-up; an input signal above 3.3 V would drive the ATWD into an operating condition from which it would recover slowly. Resistor-diode networks protect the inputs of the amplifiers from spikes, which might be produced by the PMT, or from static discharge.

*2.4.3. ATWD*

The ATWD, which is a custom designed ASIC, is the waveform digitizer for four analog inputs. Its analog memory stores 128 samples for each input until it digitized or discarded. Three amplified PMT signals provide the input to the first three ATWD channels. In addition, two 4-channel analog multiplexer chips, which can be individually selected, are the fourth input channel. The ATWD is normally quiescent, dissipating little power, and awaits a trigger signal before it converts the data to a digital signal

A transition of the PMT discriminator initiates the waveform capture sequence by triggering an ATWD capture. The actual ATWD launch is resynchronized to a clock edge to eliminate ambiguity in timestamps. Capture results in 128 analog samples in each of the four channels. After capture is complete, digital conversion is optional, and may be initiated by the FPGA's (see Section 2.4.8) ATWD readout engine only if other logical conditions are met, as determined by the local coincidence settings and operating mode of the array (see Section 4.4). If the subsequent trigger-to-conversion conditions are not met, firmware in the FPGA resets ATWD sampling circuitry in two counts of the 40 MHz clock.

If trigger conditions lead to ATWD digitization, 128 Wilkinson 10-bit common-ramp analog to digital converters (ADCs) internal to the ATWD digitize the analog signals stored on a selected set of 128 sampling capacitors. The digital data are stored in a 128-word deep internal shift register.

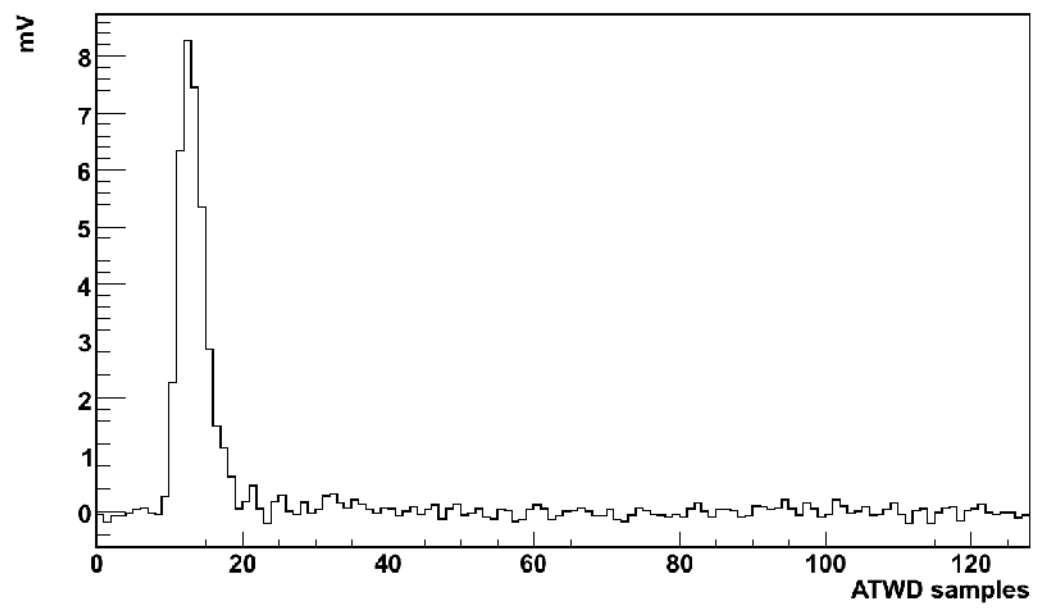

Figure 6. A typical single photoelectron waveform. This graph shows the measurement by the ATWD of a photoelectron produced by a photon in the ice. A few samples are digitized before the signal and many afterwards. These samples can be used to determine the normal operating baseline.

After conversion, another part of the readout engine transfers the data into the FPGA. The ATWD channel driven by the ×16 amplifier is converted first. To provide good overlap between ranges for larger signals, the ×2 gain channel is digitized, if any sample in the most sensitive channel exceeds 768 counts. If this next channel overflows, the ×0.25 channel is digitized. Figure 6 shows a typical waveform.

Including the analog to digital conversion, transfer to the ATWD, and incidental overhead, the ATWD takes 29 μs to digitize a waveform after capture. These parallel signal paths have the dynamic range of a 14-bit 300 MSPS ADC while consuming only ~150 mW of power without any high-speed clock or digital memory requirement. To minimize dead time, the DOM is equipped with two ATWDs such that while one is processing input signals, the other is available for signal capture.

*2.4.4. High-speed Monitoring With the ATWD Multiplexer*

The fourth ATWD channel is driven by either of two 4-channel analog multiplexers, permitting measurement from eight signal sources on the DOM MB. Multiplexer channel 0 carries the 20 MHz signal from the internal clock oscillator, as a sine wave. Channel 1 carries the frequency-doubled output of the internal FPGA phase-locked loop (PLL). These signals allow the ATWD capture sampling rate calibration, the verification of the phase of ATWD capture, and the clock phase for PMT waveform capture by the PMT ADC (see Section 2.4.5).

Signals from the on-board and the off-board LED flasher switch circuits, multiplexer channels 2 and 3, make possible the measurement of PMT transit time, and the timestamping of flasher signals targeted at neighboring DOMs in the array.

Signals from the local coincidence transceivers (see Section 4.4) appear on channels 4 and 5. These signals provide diagnostic waveforms for assessing possible fault conditions with the local coincidence subsystem.

The communications transceiver signal on channel 6 allows calibration of the Hit timestamp with respect to the precisely timed RAPcal calibration pulses transmitted to the DOM from the ICL.

Channel 7 monitors the output of an arbitrary waveform generator, driven by FPGA code.

Bits written to registers in the DOM MB CPLD separately enable each multiplexer, and select the appropriate channel to be digitized. To save power, the multiplexers are shut down when not in use.

*2.4.5. PMT ADC*

Some physics signals last longer than can be captured by the ATWD. To obtain this information, there is a fourth PMT signal path. This path consists of a three-stage waveform-shaping amplifier with a 180 ns shaping time. This path drives a low power, high-speed, 10-bit wide, parallel output, and pipelined "PMT ADC". The PMT ADC continuously samples the bandwidth-limited PMT output signal at 40 MSPS.

DOM MB electronics downstream can record an arbitrarily long PMT ADC record in response to a trigger, but the length of the raw PMT ADC record is chosen to be 6.4 μs. The DOM MB is capable of triggering two clock cycles after digitizing a previous event.

An SPE signal from the PMT produces approximately a 13-count value above the ADC's baseline, sufficient to detect its presence. This relatively low gain allows the PMT ADC to offer reasonable dynamic range.

*2.4.6. PMT Trigger Discriminator*

The DOM MB can trigger on signals from the PMT. To do this, it uses a signal from the PMT, which drives the amplifier stages preceding two low power, high-speed comparators (discriminators). Each comparator has a different sensitivity. The high-resolution comparator, with a nominal 0.0024 PE/DAC count, has a narrow operating range, and is intended to sense SPE pulses. The low-resolution "multi-PE" comparator has a resolution that is coarser by a factor of 10 and therefore has a wider operating range.

At the nominal InIce PMT gain of $1 \times 10^7$, the low system noise and low device noise allow the PMT trigger level to be set as low as 1/6 SPE. At this threshold, there is no significant increase in trigger rate due to electronic noise.

Also, these two comparators enable the implementation of multi-level event recognition. For example, the IceTop high gain DOMs ($5 \times 10^6$) utilize the multi-PE discriminator for triggering, with its threshold set to an amplitude corresponding to a pulse of 10 PEs. In addition, these two comparators are used in self-local coincidence mode, which is described in Section 4.4.3.

*2.4.7. Local Coincidence Trigger Circuit*

When a trigger comparator fires, two state machines are activated to send local-coincidence digital signals to adjacent DOMs (above and below) through bidirectional, fully duplex transceivers connected to a dedicated network of twisted wire pairs.

Local coincidence receivers on each DOM deliver signals into the trigger system of the FPGA and into a relaying state machine. This local coincidence relaying engine, if enabled, forwards a message beyond the nearest neighbor DOM. The scheme makes possible coincidence between nearest neighbors, next-nearest neighbors, and so on. If a DOM originates and transmits a local coincidence (LC) signal, it will not relay the redundant LC signals from its neighbors.

*2.4.8. FPGA and ARM CPU*

The Altera EPXA-4 FPGA handles signal and communications processing. The CPU handles data transport, system testing and monitoring. The CPU initiates FPGA reconfiguration, in real-time, as dictated by the requirements of data acquisition and system testing. This highly integrated system on a programmable chip (SOPC) architecture confines high speed, high data bandwidth signals to a single die on the DOM MB, which reduces noise and saves power.

The Hit processing portion of the FPGA contains trigger logic, an ATWD readout engine for each ATWD, a Hit record building engine, a data compression engine, and a direct memory access (DMA) controller. A DMA engine transfers Hit records into main memory for subsequent transmission to the ICL. It also implements on-board and off-board flasher control logic, PMT ADC data handling, communications protocol state machines, communications ADC, and DAC data handling.

The communications processing portion of the FPGA contains a half-duplex signaling protocol engine and modulation and demodulation function blocks, which drive a communications digital to analog converter (DAC) and monitor a communications ADC respectively.

The Altera chip also provides IceCube with a supernova (SN) search capability. A SN event, if yielding a sufficiently intense flux of MeV neutrinos at earth, will cause a global increase in the ambient light deep in the ice. A rate increase, seen by all DOMs, provides an unambiguous signal for such an event. The chip records the rate of Hits - binning Hit times at the few ms level.

*2.4.9. Memory (SDRAM, Flash, Flash Files)*

The EPXA-4 architecture supports SDRAM (synchronous dynamic random access memory) for main memory, an SRAM (static random access memory) interface for low-performance memory, memory-mapped peripherals, and flash memory. The DOM MB includes two 16 MB SDRAM memory chips, two 4 MB flash (non-volatile) memories, and a 4 Mb configuration memory.

When the DOM is powered up, the serial configuration memory uploads a configuration into the EPXA4's FPGA and program code into the EPXA4's CPU's memory. The configuration memory can only be programmed before the DOM is sealed and cannot be reprogrammed after deployment. Its contents include a base-band communications package for the FPGA, a utility program for the CPU so that the flash memory chips are loaded, and a command interpreter for rebooting to the boot-block in either flash memory. This is analogous to a desktop computer booting to block 0 of its hard disk.

The two 4 MB boot-block flash memories are organized as a log structured flash memory file system. The file system stores CPU programs, FPGA configuration files, and interpreter scripts. Various files support production testing, integration after deployment, and Hit data acquisition. Any and all files may be updated after the DOM is deployed into the ice. Each flash memory also contains a 64-bit serial number, from which we derive a unique 48-bit DOM ID. The DOM ID maps to the geometrical position of each DOM. Furthermore, unique DOM parameters can be loaded by DAQ control from a database indexed by the DOM ID.

After reboot, the CPU executes program code copied into SDRAM. Approximately half of the SDRAM is allocated as a circular buffer for DMA Hit record transfers from the FPGA. The CPU accesses registers in the CPLD and Flasher Board interface through the Expansion Bus Interface (EBI). Flash memory also resides on the EBI.

### *2.4.10. DOM Local Oscillator*

The DOM's 20 MHz temperature-compensated crystal oscillator has a certified stability of roughly $1 \times 10^{-11}$ for a sample interval of 5 seconds.[8] In practice, the crystal frequency and phase drifts become significant typically only after several minutes. The 20 MHz oscillator output drives the clock inputs of the FPGA, the communications ADC, the communications DAC, and the reference clock signal port of the flasher board interface. A phase locked loop (PLL) in the FPGA doubles the frequency to 40 MHz. This 40 MHz signal drives a 48-bit local clock within the FPGA, the DOM local clock; it rolls over every 81.4 days. The DOM local clock is used to timestamp Hits recorded by the DOM and the RAPcal timing calibration packets exchanged between the DOM and the DOMHub. The 40 MHz reference also drives the clock input of the PMT ADC.

### *2.4.11. Communication With Surface*

As communication and power must share the same wire pair, it is necessary to separate them at the DOM MB. Balanced L sections at the power filter input confine the communications signals to the communications front-end receiver. A Π section filters the power and isolates the DOM communications interface from switching noise.

Ceramic capacitors couple communications signals between the twisted pair and the communications transformer. The grounded center tap of the transformer accommodates the topology required by the 8-bit current-mode communications DAC clocked at 20 MHz. The choice of current mode DAC allows two DOMs to share the InIce end of the main cable pair. The end-most DOM on the pair must be terminated in the characteristic impedance of the twisted pair. The unterminated DOM typically bridges onto the twisted pair 17 meters upstream from the terminated DOM; the additional stub introduces a small, systematic time error. The modulator produces differential signals with amplitudes of approximately 2 V (depending on communications parameter settings) onto the twisted pair.

The center-tapped secondary of the transformer also drives a ×5 differential amplifier stage characterized by high common mode rejection ratio (CMRR). The high CMRR reduces the susceptibility of the DOM line receiver to interfering signals such as electromagnetic interference (EMI) or radio frequency interference (RFI). The amplifier drives a 10-bit, 2 V input-span, ADC that samples the communications waveform at 20 MSPS. The parallel output stream of ADC data drives the inputs of the communications receiver firmware in the FPGA.

---

[8] Vectron International manufactures the model C2560A-0009 TCXO module specifically for IceCube. Since the oscillator is crucial to DOM performance, the brand and model were selected based on Allen variance performance and power consumption. The procurement specification required 100% Allen variance testing.

The topology resembles that of the commonly used T-1 communications links. So, the natural choice was Amplitude Shift Keying[9] (ASK) with a transmission rate of 1 Mb/s, which includes encoding bits. Since we have selected ASK modulation, and multiple DOMs share wire pairs, it is necessary to utilize half-duplex. The communications master resides in the communications card within the DOMHub in the ICL. That master alternately sends commands to and requests data from the two DOMs sharing the InIce end of the wire pair.

The distribution rate of RAPcal signals from the ICL to all DOMs must accommodate any oscillators with marginal short-term stability. The stability requirement for the DOM local oscillator is that RAPcal signals be distributed to all DOMs in the array at least as often as every 5 seconds. However, the complete DOMHub-to-DOM-to-DOMHub exchange of RAPcal signals consumes less than 1.5 ms, so the sacrifice of system communications bandwidth to time calibration is negligible.

#### *2.4.12. CPLD*

The DOM depends upon certain higher-level logic functions and state machines that cannot be implemented in the FPGA because the FPGA does not retain its logic configuration through power cycling. Those logic functions are implemented in a CPLD.

The CPLD code contains a state machine that controls the booting of the EPXA4, initiated by the rising edge of the *not-power-on-reset* (nPOR) signal of the power supervisor chip. The logic in the CPLD assures that all power supplies are at voltage and stable, and that internal initialization of other complex components of the DOM has been completed before the configuration memory uploads its contents into the EPXA4's CPU and FPGA configuration.

The CPLD code also contains an interface between the EBI memory bus and the high-speed interface connector used by the external LED Flasher Board. The bus interface prevents possible catastrophic failure of the flasher or a generic daughter card from disrupting the memory bus the CPU relies upon for booting to its normal running–mode configuration which is read from flash memory.

The applications programmer interface (API) of the CPLD appears to the CPU as read-only and write-only memory in EBI address space. Control register bits enable or disable the high voltage power interface for the PMT, the voltage source for the on-board LED flasher, the Flasher Board interface, and the pressure sensor.

The CPLD firmware controls the reading of 24 channels of the slow (serial) ADC used for monitoring and diagnostic purposes. It also sets the 16 slow (serial) DAC outputs used to control ATWD operating parameters, trigger comparator levels, and ADC reference levels. A separate serial interface supports the control and read-out of PMT high voltage module.

The CPU may control whether it reboots from the serial configuration memory (boot to Configboot), or to flash memory (boot to Iceboot), by executing a CPLD function. Another CPLD function actually initiates reboot.

The CPU may boot from either of the two flash memory chips. The Configboot program supports a CPLD function, which virtually exchanges flash-0 for flash-1, allowing the recovery from failure of the flash-0 boot block subsequent to deployment.

Several registers contain read-write scratch-pad bits. Scratch-pad data are retained through reboot, allowing a limited amount of crucial context information that can be retained through reboot.

[9] The communications interface design is compatible with phase modulation schemes, which offer higher data rate, superior noise immunity and timing precision, as well as full duplex communications.

*2.4.13. On-board Electrical Pulser*

Many DOM calibration programs depend upon a built-in signal source that produces waveforms similar to SPE PMT pulses. The pulser injects charge that is stored on a capacitor into the analog input of the DOM MB, at the PMT cable connector. A serial DAC, controls the pulser amplitude up to roughly 40 PE, in 0.04 PE steps.

The shape design was based upon previous measurements. Care was taken to minimize noise into the DOM MB when the pulser was activated. Recent results show that the pulser shape is slightly wider than an actual PMT shape.

*2.4.14. On-board LED pulser*

Each DOM MB includes a pulse forming circuit driving an ultraviolet LED that has a wavelength of 374 nm. The function of the on-board LED pulse is to stimulate the local PMT, whereas the function of the Flasher Board is to stimulate neighboring DOMs.

A state machine in the DOM initiates a trigger, and simultaneously sends a trigger pulse to the LED flasher circuit. The flasher circuit dumps a current pulse into the LED via a "shunt" which produces a voltage pulse across it. The pulse propagates from the pulser circuit to the MUX connected to ATWD channel number 4. The LED flash produces light, which bounces over to the PMT. The ATWD simultaneously captures PMT input and the current shunt input. The time difference between the current pulse and the PMT pulse is the transit time of the PMT, delay line, other circuit components.

The LED flash brightness can be adjusted to produce zero to a few tens of photoelectrons at the photocathode of the PMT. The brightness range of the on-board LED is sufficient to measure the transit time over several orders of magnitude of PMT gain around the typical operating point. Controlled weak flashes (optimally around 1% occupancy) makes possible the assessment of the SPE behavior of the PMT.

*2.4.15. Interface to PMT Power Supply Daughter-card*

The DOM MB controls and powers the PMT HV subsystem, which resides on a mezzanine card atop the Flasher Board. The control signals, the serial bus signals, system ground, and raw power are delivered to the HV module through a ribbon cable.

The CPLD of the DOM MB contains a control register with one bit allocated to enable the high voltage control board, and another bit allocated to enable the high voltage output of the high voltage module on the HV control board. The HV control board contains two synchronous serial (SPI protocol) devices, a serial DAC for high voltage control, and a serial ADC for monitoring the HV module output voltage.

The serial bus clock, control, and data lines of the HV subsystem interface use CPLD pins and firmware that is independent of the on-board monitoring serial bus, which is also supported by the CPLD and firmware. This feature prevents a failure causing disruption of the on-board serial bus from interfering with the operation of the high voltage subsystem, and vice-versa. The CPLD firmware also contains code to read the serial number chip built into the HV interface card.

*2.4.16. Interface to the Flasher Board; a Generalized High-speed Interface*

The DOM MB includes a 48 pin, high-speed, memory mapped interface to the Flasher Board. The connector delivers power, an extension of the EBI memory bus, and control lines from the DOM MB FPGA to the daughter card. The connector also delivers system clock and trigger control signals directly from the FPGAs.

This interface delivers a power-enable line to the daughter card. When this line goes high, the daughter- card is permitted to draw power and connect to the memory bus interface. Likewise, an enable is transferred to the bus extension enable firmware in the CPLD. This feature allows a 30% saving of power, since the calibration capability provided by the high intensity flashers is rarely needed.

When the power-enable line goes low, the daughter card powers off all electronics, and disconnects from the bus extension. If the DOM power supervisor should detect an event that causes nPOR to be asserted, then the power-enable line is set to the default (low state) and the memory bus repeater built into the CPLD breaks connection with the EBI memory bus. This buffering feature protects the primary data-taking capability of the DOM MB from being compromised should there be a catastrophic failure of the flasher card.

The interface has 8 bidirectional data lines, a read line, a write line, and 6 address lines to the daughter card. This is sufficient to control a wide variety of states in the daughter card, as well as permitting the transfer of arbitrarily large data sets in either direction. The data requirements of the Flasher Board are, however, modest.

Control lines from the FPGA of the DOM MB notify the Flasher Board when to initiate a flash. The Flasher Board produces an output pulse whose voltage is proportional to the flash amplitude. That signal is sent through a coax cable to a multiplexer input of the ATWD for timing calibration purposes.

The interface includes a low-voltage differential signal (LVDS) 20 MHz clock pair derived from the DOM MB system clock. This feature is useful for synchronizing it with the DOM MB activity. It also contains a set of control and I/O lines configured as a JTAG[10] programming interface. The interface allows firmware to be updated in the flasher board after the DOM has been deployed in the ice.

#### *2.4.17. Monitoring with the Slow ADCs*

Two serial ADCs (I2C protocol) monitor a total of 24 voltages in the DOM. The ADC readouts are particularly useful in the test stage of newly manufactured DOM MBs. They also provide some diagnostic information should the DOM suffer a partial failure after deployment.

Channels 0, 1, 9, 10, and 11 monitor power supply voltages. Channel 2 monitors the (DOM) pressure sensor on the DOM MB. The voltage measurement on Channel 0 is necessary to calibrate the pressure measurement.

Channels 3, 4, 5, 6, and 7 monitor the current delivered by all major power supplies on the DOM MB. Channels 8 and 12 monitor control voltages of the front-end discriminator thresholds. Channel 13 monitors the reference voltage of the high speed ADC. Channel 14 monitors the voltage delivered to the on-board LED pulser. Channels 15-22 monitor control voltages produced by serial DACs, which control ATWD behavior. Finally, channel 23 monitors the voltage of the front-end test-pulse control.

### *2.5. The Cable Network*

The cable network carries power and signals between the DOMs and the DOMHub. The cable is of sufficient quality that the amplitude-shift modulation scheme reliably yields data rates up to ~900 kb/s for the most remote DOMs in the array. By sharing two DOMs on one pair, only one cable is needed for a string, substantially reducing costs. This cable size, roughly 3 cm in diameter, approaches the practical limits for transportation volume and weight, flexibility, and strength during deployment.

The down-hole cable consists of 20 quads. Fifteen quads are used for signals to the DOMS while one is for LC. Each quad contains two pairs of wires. The DOM Quad services four DOMs. Three twisted pairs from the cable enter the DOM in the location identified as the Cable Penetrator Assembly seen in Figure 2. Two of these pairs are the LC links to adjacent DOMs; the third pair connects the DOM to the DOMHub.

[10] Joint Test Action Group IEEE Standard 1149.1-1990.

A surface cable from ICL connects to a surface junction box located near the top of each hole. The surface cable contains extra wire quads in an inner, shielded core to service the two IceTop tanks associated with each hole. Data rates for IceTop DOMs are higher than for InIce DOMs, so only one IceTop DOM is connected to its corresponding DOR card input. However, to maintain commonality throughout IceCube, the control and communications protocol and baud rate are the same as for cable pairs. IceTop DOMs also exploit an LC communication link between the two tanks at each station.

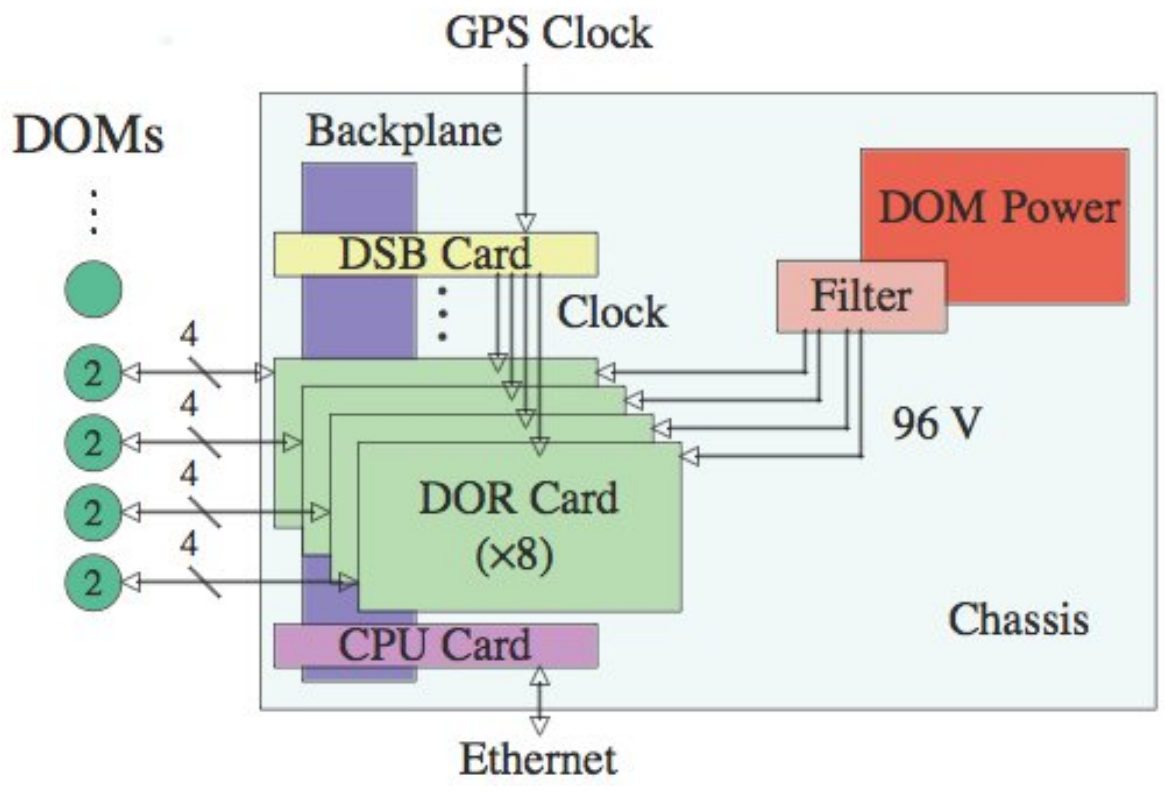

Figure 7. A block diagram of the DOMHub. The left hand side depicts the data from a string of DOMs. The label 2 in the left circles refers to the fact that two DOMS are on each wire pair. The number 4 refers to the fact that each DOR card digitizes four DOM pairs. The DOM Hub contains a power supply, a CPU Card and the DSB card for timing. The DOR cards communicate over the backplane. A DOMHub communicates with the trigger and event builder over Ethernet.

## *2.6. The DOMHub*

The DOMHub is a computer in the ICL that communicates with all of a string's DOM MBs. Its block diagram can be seen in Figure 7. Its components are housed in a standard 24" deep industrial PC chassis. The PCI bus backplane accommodates 8 DOR cards, one DOMHub Service Board (DSB), and one low power single board computer (SBC).

Each DOR card can communicate with eight DOMs, so the DOMHub can host 64 DOMs. In practice, a DOMHub hosts an entire string of 60 DOMs for InIce, or 32 DOMs for IceTop (8 stations).

### *2.6.1. The DOM Readout (DOR) Card*

The DOR card is a full-size full-custom PCI card, shown in Figure 8. Each DOR card handles communications with the DOMs connected to its four wire-pair inputs and the DOR Driver, the lowest level software element of the DAQ chain in the ICL. The block diagram in Figure 9 shows the media access interfaces, and power control functions of the DOR card.

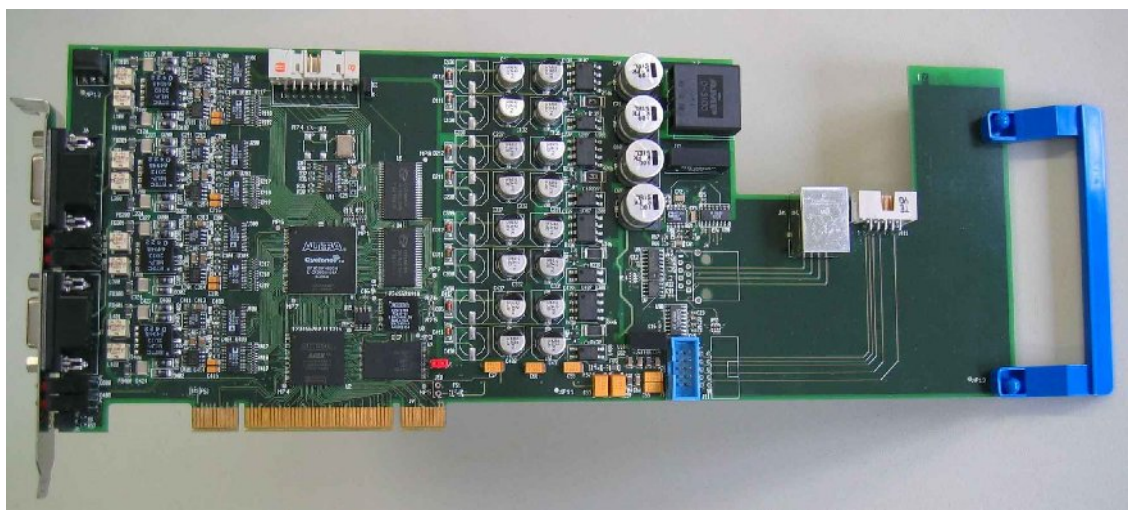

Figure 8. A photograph of a fully functional DOR card.

The DOR card controls power to the DOMs, establishes that boot-up has occurred properly, selects which code the DOM is to run (or downloads new code), establishes a conversation that may include calibration tasks or block data transfer, senses fault conditions, manages time calibration sequences, controls the DOM state, or initiates any of the numerous and diverse DOM actions.

A utility function loads the FPGA configuration files into flash memory. Other functions cause the communications FPGA to be reloaded from flash memory, select the clock source, initiate RAPcal, signal exchanges with DOMs, and other features described in the next section.

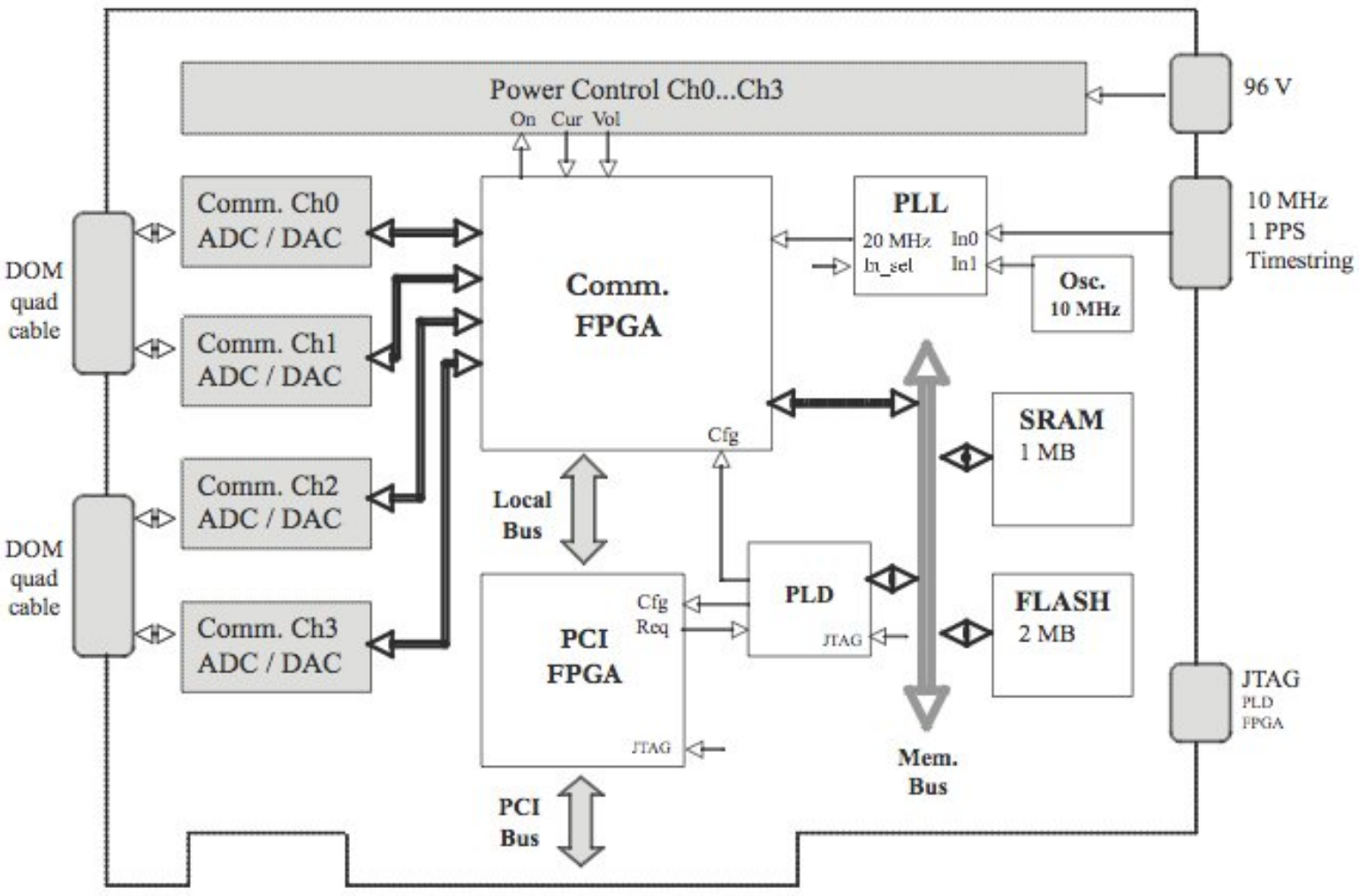

Figure 9. This figure shows the functional blocks of a DOR Card. On the left shows the signals from the DOMs. The COM ADC/DAQ digitizes the signals from the DOMs and sends signals to the DOM over the wire pairs on the quad cable. The Communication FPGA drives these ADCs. Memory is shown as well as the exterior communications and power to the DOR card.

### *2.6.2. PCI Block*

The firmware of the PCI bus interface FPGA includes a commercial VHDL PCI core adapted to suit the requirements of the DOR card hardware and DOR driver. The 32-bit core supports master/slave control logic, and 33 MHz bus speed. The PCI core is compliant with “PCI Local Bus Specification revision 2.2”.

The PCI FPGA executes bi-directional programmed (single bus cycle) 32-bit transfer to thirty-two control and status registers. The firmware also contains code for 32-bit quad-word aligned DMA (Direct memory access) writes of data to main memory on the CPU board in the PC chassis.

*2.6.3. DOM Power Management*

The DOR driver can cause the ±48 V power for any of the 4 wire pairs to be switched on or off. Both wires of a pair are switched for symmetry and safety. Also, for safety, power is applied only when a cable connection is sensed. The switches have built-in slew-rate limiting to reduce component stresses, and suppress power on/off transient noise.

The DOR card has ADC's for monitoring wire pair current and voltage. A "proc file" interface of the DOR driver allows the voltage and current values to be read by user programs, or from the command line. Furthermore, a firmware component detects pair over- current or under-current conditions and then removes power to the pair.

*2.6.4. DSB Card*

The DSB card is a very simple electronics board. Its primary function is to distribute the system timing and reference signals to each of the eight DOR cards in a DOMHub. The inputs to this board include the 10 MHz system clock, 1 Hz, and Global Position System (GPS) reference signal (see Section 2.7). These signals originate in a GPS receiver in the ICL and are distributed isochronously to all DOMHubs.

*2.6.5. Time Calibration*

The DOR card receives the clock signals from the DSB Board. A PLL on the DOR card produces the 20 MHz global clock, which drives the 56-bit clock counter in the communication FPGA of each DOR card. The 1 Hz signal triggers a snapshot of the 56-bit counter value every second. The DOR clock counter rolls over every 114.24 years, and is never explicitly reset. The counter snapshots together with the corresponding encoded time of day provide a cross reference between the DOR card's local clock value and the UTC time.

Software may initiate a time calibration. The DOR firmware completely manages the time calibration process. The cycle produces a RAPcal data packet containing DOR and DOM timestamps and digitized RAPcal received pulse waveforms.

*2.6.6. DOR card Flash Memory and FPGA Configuration*

The communications controller FPGA configuration image resides in page 0, 1, or 2 of a 2 MB Flash memory. The PCI bus interface FPGA image resides in page 3. Loading the communications image into flash depends on the integrity of the PCI image. Consequently, the PCI image is protected.

The two FPGAs and the CPLD share a JTAG chain. At time of DOR card manufacture, first, the CPLD must be programmed via JTAG, and then the PCI FPGA must be loaded via the JTAG. Power must be kept on until flash is loaded. UNIX utility programs write configuration files to flash memory pages through the DOR Driver's Linux file system proc file interface. Validity checking in the DOR Driver prevents invalid images from being loaded into DOR card flash. The JTAG chain provides a means for loading test firmware into either FPGA.

Once flash is loaded, the application of power, or a PCI bus reset, causes a CPLD on the DOR card to read configuration data from flash memory, serialize it, and transfers it to an in-circuit programming interfaces of each FPGA. The communications FPGA may be reloaded at any time by issuing a command to the DOR Driver; the PCI FPGA may be reloaded on the fly, but this is risky in case the new image contains a bug.

Each DOR card, after final assembly and test, has a unique ID, as well as a final test summary, written into the uppermost 64 K byte sector of the flash memory. The code contains the card revision number, the production run number, and the card serial number, which matches the card's label. The production information may be read from the DOR driver proc file interface.

*2.7. The Master Clock and Array Timing*

The two central components of the IceCube timing system are the Master Clock, providing each DOMHub with a high precision internal "clock" synchronized to UTC [11] and the calibration process RAPcal described in Section 4.7. Together, these manage the time calibration as a "background" process, identically for InIce and IceTop.

The Master Clock makes use of the Global Positioning System (GPS) satellite radio-navigation system, which disseminates precision time from the UTC master clock at the US Naval Observatory to our GPS receiver in the ICL. Algorithms in the GPS receiver clock circuit make small, but abrupt, changes to crystal oscillator operating parameters according to a schedule optimized for GPS satellite tracking. The abrupt parameter changes, and phase error accumulation intervals result in deviations from UTC of typically 40 ns RMS over time scales of hours, but not exceeding 150 ns.

The GPS reference time for IceCube is the phase of the 10 MHz local oven-stabilized crystal reference oscillator in the GPS receiver. This oscillator is optimized for excellent Allen variance performance. Altogether, the system delivers an accuracy of about ±10 ns averaged over 24 hours.

The fan-out subsystem distributes the 10 MHz, the 1 Hz (as a phase modulated 10 MHz carrier), and encoded time-of-day data from the GPS receiver to DOMHubs through an active fan-out. The encoded GPS time data, contains second, minute, and hour of the day, day of the year, and a time quality status character.

The principal engineering requirements for the fan-out are low jitter, high noise immunity, simplicity, and robustness. The 10 MHz and 1 Hz modulated 10 MHz signals pass through common mode inductors at the transmitter and receiver end to improve noise immunity and effectively break ground loops between apparatus widely dispersed in the ICL. For the encoded time of day information, RS-485 was chosen for its noise immunity. Each distribution port of the fan-out has its own line drivers to insure that any particular failure has minimal impact on neighboring distribution ports.

The Master Clock Distribution System and interconnecting cables delay the arrival of time reference signals traveling from the GPS receiver to the DOMHub. All signal paths between the GPS receiver's 10 MHz output, via fan-outs to the DSB, and to each DOR card are matched within 0.7 ns RMS. Thus, the DOR cards "mirror" the Master Clock with an accuracy of less than 1 ns.

A static offset must be applied to the experimental data to map IceCube Time (ICT) to UTC. ICT differs from UTC by the master clock distribution cable delay plus the GPS to UTC offset.

The DSB card in each DOMHub distributes the three signals to each of the eight DOR cards. The DOR card doubles the GPS frequency to 20 MHz and drives a clock counter. When the 10 MHz clock signal goes high immediately after the 1 Hz GPS signal goes high, logic in the DOR card's FPGA latches the clock counter value and the time of day data from the GPS receiver from the previous second to form a time sample record. This time sampling engine stores its output in a 10-reading deep first-in first-out (FIFO) memory. Data acquisition software running in a DOMHub's CPU reads these records from a Linux file system proc file and forwards them to a data-logging computer where they become part of the physics data set.

In operation, the clock counter value in each DOR card increments exactly $2 \times 10^7$ counts every second. A module in the DOR card firmware confirms that $2 \times 10^7$ additional counts are registered each second to verify clock integrity, and firmware correctness, and freedom from injected noise.

## 3. Firmware and Software

In the universe of computers, the DOM resembles a hand held device since it does not have a mass storage device like a disk drive. The DOM does not need "processes" in the sense of UNIX, nor "task scheduling." However, interrupt handlers are used for communications and data collection. Therefore, an open source collection of standard UNIX-like single threaded functions called "newlib" [12] was chosen instead of more sophisticated embedded operating systems like Windows CE, embedded LINUX, or NetBSD. Newlib functions streamline common tasks such as memory allocation and string operations. Communications on the DOM side, including a custom communications protocol, are largely implemented in FPGA firmware.

### *3.1. DOM Software*

At power on, firmware embedded in the Altera EPXA4 chip (called the logic master) copies a simple, robust bootstrap program named *Configboot* from a read-only serial configuration memory into internal SRAM, configures the FPGA, and initiates the execution of program code at the start of SRAM. The *Configboot* program interprets a very small set of terse commands that allow the DOM to reboot to the image in the boot block of the primary flash memory, boot from other locations in either flash memory, or boot from a dedicated serial port. A *Configboot* command allows reprogramming flash memories via the communications interface to allow the users to easily upgrade a DOM with new operating software and firmware. Furthermore, if flash chip 0 were to fail, *Configboot* can be configured to boot the DOM's CPU from flash chip 1.

Each DOM uses a reliable, journaling flash file system spanning the pair of flash memories into which a release image is loaded. The release image consists of data files, software programs for test and data acquisition, and FPGA configurations that suit the requirements of the software programs. The release file is built on a server computer from components of a software development archive.

Booting from flash block 0 causes the DOM's CPU to copy a fully featured program, called *Iceboot*, into SDRAM, and start executing it. The program *Iceboot* is built of layers: low-level bootstrap code; Newlib code; a Hardware Access Layer (HAL) used to encapsulate all hardware functions provided by the CPLD; the application program/server; and the FPGA, including the communications interface. The *Iceboot* program presents a Forth language interpreter to the user. From the interpreter's prompt, one can invoke HAL routines directly, write data to, and read data from memory addresses, reconfigure the FPGA, and invoke other applications programs like the DOM Application (*Domapp*) used for data acquisition.

The *Domapp* program implements a simple binary-format messaging layer on top of the DOR to DOM error-correcting communications protocol. Messages are sent to particular "services" within the *Domapp* (e.g. "Slow Control", "Data Access"). Each message targets a particular function (e.g., "Slow Control - set high voltage" or "Data Access - fetch latest Hit data"). Every message to the DOM generates a single response, and no DOM sends data unless queried. Messages are provided for configuring the hardware and for collecting and formatting the physics data from buffers in main memory. Other messages provide for buffering and retrieving of periodic monitoring data describing the DOMs internal state and any exceptional conditions that may occur. The message can be logged or acted upon by the DAQ components in the ICL.

Typically, between runs, the DOM is rebooted into *Iceboot* to guarantee its state at the beginning of each run. Should the DOM become unresponsive, a low-level communications message sent directly to the communications firmware in the FPGA will force the DOM to reboot to *Iceboot*. Power cycling of the DOM is seldom necessary to reinitialize the DOM, and is avoided to minimize electrical stress.

### *3.2. DOM Firmware*

The DOM Firmware suite consists of three different FPGA designs, needed for different actions. The designs are called: the *Configboot* design, the *Simple Test Framework (*STF) design, and the *Domapp* design. Only one of these can run at a given time. DOM firmware is written in VHDL supplemented with several other code generation tools. A communications firmware block is common to all three designs. Only the *Domapp* and *STF* firmware designs manipulate data acquisition hardware. The FPGA firmware design uses about two thirds of the available resources.

In general, the FPGA acts as an interface between the software running on the CPU and the DOM hardware. Beyond this basic logic functionality, the DOM firmware performs time-critical processing of triggering, clock counter, and PMT data (with sub-nanosecond precision); basic data block assembly; DMA of physics data to SDRAM; and communications processing. The FPGA also hosts calibration features, and Flasher Board control functions.

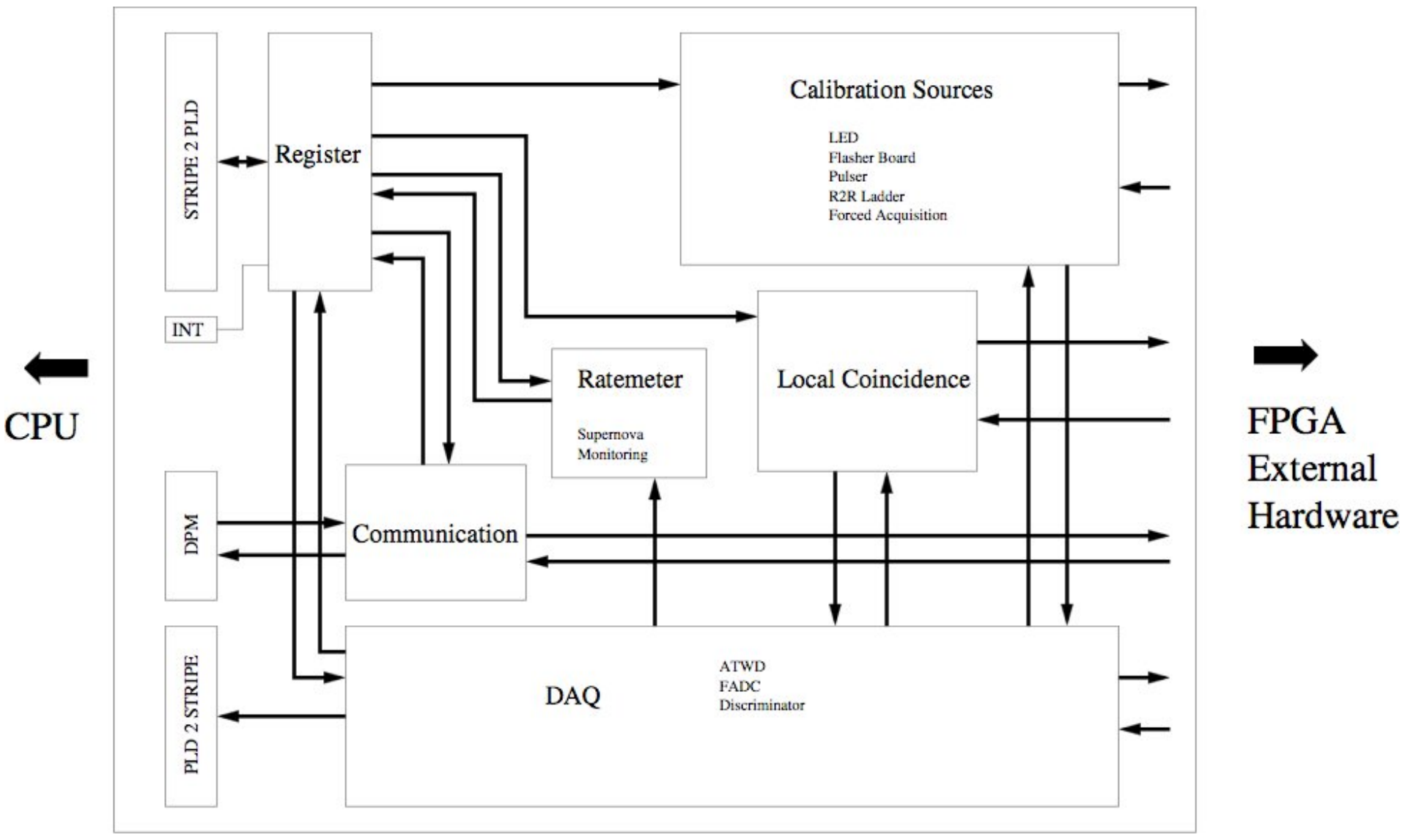

Figure 10. The figure shows a functional block diagram of the firmware modules used in the main data acquisition FPGA, which is used in conjunction with Domapp. The arrows on the right indicate connections to hardware on the DOM MB. The arrows show the direction of data flow. The four boxes on the left are internal interfaces to the CPU. They include communications between the FPGA and CPU (STRIPE2PLD and PLD2STRP), dual port memory (DPM), and interrupts (INT). The other rectangles are code modules for the FPGA. These blocks describe the specific tasks. For instance, the block labeled “Ratemeter” stores information for the supernova trigger, monitoring, and dead time. The module “Register” contains the memory mapped registers for configuration and status information.

The *Configboot* FPGA design contains only minimal required functionality to provide communications to the DOR card. Implementation of a simple, reliable, and robust design was a firm requirement because *Configboot* cannot be upgraded once a DOM has been deployed into the ice.

The *STF* design is used primarily for DOM hardware testing. Software uses the STF program to manipulate, test and verify the functionality of each hardware subsystem.

The *Domapp* design, shown schematically in Figure 10, provides for data acquisition. Based on the settings of bits in memory mapped control registers according to applications programmer interface (API) documentation, the FPGA autonomously collects waveforms from the ATWD and the PMT ADC. It processes the waveforms, builds the hit record according to a hard-coded template, and transfers the data through DMA to a block of the CPU's memory. Parallel to data acquisition, the firmware supplies PMT count-rate meters, which are read from data registers in the API; count-rate metering facilitates the supernovae detection science goal. In addition, the firmware modules control the on-board calibration sources and the Flasher Board.

### *3.3. DOMHub Software*

Software on the DOMHub builds on the rich set of DOR hardware and firmware functionality. DOMHub software consists of a C-language kernel level device driver ("DOR-driver") for the DOR cards and a user-level Java application called *Stringhub* on a Linux server operating system. *Stringhub's* task is facilitating the higher-level configuration control and communications functions to the rest of the IceCube Surface DAQ.

Specific requirements for DOR-driver include the following:

- support for a few dozen control functions specific to the DOR cards,
- a clean interface between the hardware and user applications, and
- concurrent and error-free communications to all attached DOMs at or close to the maximum throughput supported by the hardware.

The first two items are addressed by implementing a tree of control points via the Linux proc file system, a hierarchy of virtual files used for accessing kernel functions without requiring native system calls. This somewhat unusual approach allows the Java-based *Stringhub* to be written without unwieldy native interface modules. This simplifies the DOMHub software architecture and provides the added advantage of making it very easy to operate DOMs interactively or through a wide variety of test software.

The utilization of cyclic redundancy codes (CRCs) in the DOM-to-DOR communications stream ensures that corrupt data is identified. Acknowledge/retransmit functions similar to TCP/IP in both the DOM software and in DOR-driver insure that no communications packets are lost. Packet assembly/disassembly, transmit, acknowledge, and retransmit are carried out on all DOMs in parallel, with periodic time calibration operations seamlessly interspersed. The asynchronous activity of 8 DOR cards, 60 DOMs, multiple user applications, and many control functions create substantial risk of race conditions that are identified and eliminated.

Extensive testing is crucial for a large device driver such as this one operating on custom hardware (in lines of code, DOR-driver is comparable to a Linux Ethernet driver). The first phase of testing addresses long-term stability under normal operating conditions with maximum throughput; the second phase of testing emphasizes "torture tests" designed to expose unexpected races and edge conditions in the driver and firmware. Every driver-release candidate must pass the verification test suite prior to deployment into production DOMHubs.

*3.3.1. Communications with the DOM*

The FPGA firmware on the DOR card contains eight instances of the communications protocol. The PCI-bus-interface FPGA relays data between the communications FPGA and the PCI bus. The half-duplex communications protocol transmits ASCII character encoded data. Transformer and capacitor coupling of the signal dictate a DC-balanced modulation. A 1 μs wide bipolar pulse represents *logic 1*; absence of modulation represents *logic 0*. Simple threshold detection delivers satisfactory bit-error rates

The communications protocol includes the following commands:

- *Data Read Request:* This provides automatic data fetching from the DOMs in a round robin arbitration schema.
- *Buffer Status:* DOM/DOR data buffer synchronization, i.e. the DOM blocks DOR transmission when the receive data buffer is full. Similarly, the DOR defers data fetches from the DOM if its receive buffer is full. (When operating normally, deferred data fetches rarely result in loss of physics data as the DOM's circular data buffer can store many seconds of data.)
- *Communication Reset:* This is used to initialize the communication.
- *DOM Reboot*: This is initiated by a higher-level software command. It causes the reloading of the DOM FPGA, which results in a temporary loss and reestablishment of communication.
- *RAPcal Request:* A higher-level software command initiates a RAPcal sequence, causing the exchange of timing waveforms between DOR and DOM (about 1.5 ms duration), producing a time calibration data packet.
- *Idle:* If no data need to be transmitted to a DOM, a simple read request is transmitted. If the DOM has no data to transmit, it reports an empty queue. Absence of an expected packet constitutes a communications breakdown or an interruption, which may trigger logging or intervention.
- *CRC Error handling:* Data packets with CRC errors are not written into the receive buffer. The transport layer software is responsible for data packets retransmit. Control packets with CRC errors are ignored.
- *Hardware Timeout:* Interruption of either the data transfer or the idle packet stream for more than 4 seconds constitutes a hardware timeout. If this happens, the full system bandwidth will be utilized for communication with the remaining DOM on the pair. This feature is typically exercised when power is cycled, allowing pairs with one connected DOM to utilize the full data carrying capacity.

*3.3.2. Stringhub*

The Linux operating system, the top-level software element of the DOMHub, provides a computing environment for programs such as the *Stringhub*, which converts the flow of DOM Hits into physics-ready Hits that are suitable at both trigger and event-building stages of the surface DAQ. In principle, the *Stringhub* program can reside in a computing platform different from the DOMHub, but the DOMHub CPU is sufficient to accommodate, with adequate margin, the transformed Hit data flow rate for a string of 60 InIce DOMs or 32 IceTop DOMs.

The program *Stringhub* applies a time transformation to the coarse *timestamp* accompanying a Hit. These transformations bring all DOM data into a single, ICT-based time domain. Application of appropriate offsets then converts the Hit from ICT to UTC. The *Stringhub* then time orders DOM hits from multiple DOMs on a string, and can apply string-wide trigger filters.

The RAPcal algorithm uses data in the periodic time calibration event stream to time correct the Hit data stream. The API for time calibration has been designed to allow easy substitution of calibration algorithms as the understanding of systematic errors improves. These algorithms must achieve the correct balance between execution, speed, and accuracy. Since the converted times are only used for triggering and ordering operations, and since re-calibration of Hit times can be performed offline, algorithm performance optimizations are possible at this level.

Time calibration algorithms are designed to produce times with the same format and absolute reference. Each transformed timestamp represents UTC time in tenths of nanoseconds since 00:00 January 1 of the year in which data were acquired.

The *Stringhub* caches the full Hit data for later retrieval. It then creates a minimal version of each hit and sends it onto the multi-string trigger handlers. When the final trigger request is sent back to a *Stringhub*, it responds with a list of all hits matching the criteria and flushes all cached Hits, which occurred before the end of the time window from the most recent request.

## 4. DOM Operations

### *4.1. Hit Creation and Data Compression*

Hits always include a 12-byte header with three distinct 4-byte components: the four-byte coarse timestamp (lowest 32 bits)[11], the four-byte coarse charge stamp, and four bytes of trigger and housekeeping information. These four bytes hold 1 bit to mark a compressed hit, 13 bits of trigger information, 4 bits to indicate the included waveforms, if any, 1 bit to identify which of the two ATWDs is used for the hit, and 11 bits that show the hit size in bytes. In the baseline soft local coincidence (SLC) operating mode, if a Hit has no LC tags, then no other information is included.

When the LC condition is satisfied, the entire waveform information is transmitted. In this case, to reduce the data flow, a “delta compression” algorithm is used, which exploits the fact that waveform changes from one sample to the next are typically small. The delta compressor works by subtracting each sample from the preceding sample, producing mostly small numbers. The differences are then encoded using 2, 3, 6, or 11 bits, with special codes used to change the number of bits. For typical IceCube data, this typically compresses the waveform data by a factor of 3.8 without any loss of information.

[11] The 16 most significant bits of the timestamp change infrequently, and are hence sent once per data block to the surface DAQ, where they are reconnected to Hits. This tactic reduces the data load on the cables significantly.

*4.2. "Slow" Waveform Capture*

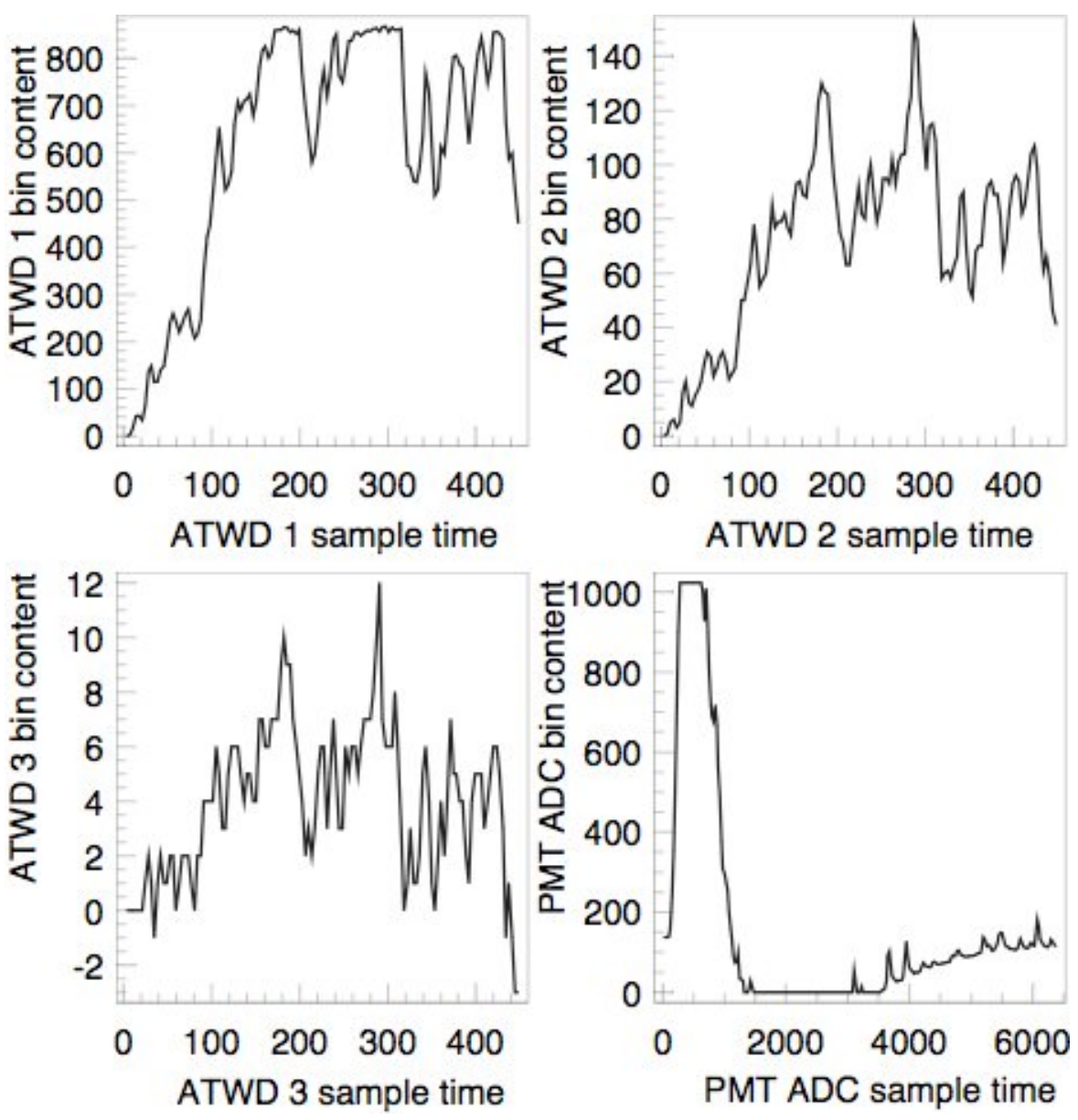

Figure 11. Waveforms from the three ATWD channels and the PMT ADC channel. The data were produced by light from a DOM flasher board in the ice. The horizontal scale is in nanoseconds. Since all four channels sample the same waveform, then structures in one are reflected in the others. The time behavior seen in this figure arises because the light from the flashers has different optical paths due to scattering in the ice. The different amplitudes of these structures can be explained by photon statistics.

Figure 11 shows the signals from all four channels from a sample Hit; the horizontal scales are in nanoseconds. Pulsing the flasher board created this Hit. The ATWD gain is highest in the top-left, medium in the top-right, and lowest in the bottom-left, while the PMT ADC signal (note the different time scale) is in the lower right. The pedestal (the value of the ATWD with no signal) has been removed from each panel. The figure shows that, when one channel saturates, more information can be recovered from a lower gain channel. The PMT ADC channel shows distortion that is the effect of a droop caused by the PMT coupling transformer. The transformer produces both an undershoot and a slow-rising waveform. This effect can be eliminated in software unless the signal drops below 0 ADC counts. The short time-constant transformer used in early DOM production was later replaced with one that produces less distortion and clipping.

*4.3. Synchronous Triggering: "Coarse" and "Fine" Timestamps*

A high-speed comparator detects the threshold crossing of a pulse from an amplified copy of the DOM MB input. The comparator/discriminator transition time is resynchronized in the FPGA to the next well-defined edge of the DOM's 40 MHz clock. The resynchronized, or synchronous, trigger signal launches ATWD capture, and simultaneously latches the DOM's clock counter value ("Hit time") on the next clock edge. Synchronous triggering eliminates the possibility of ±1 count timestamp errors when trigger transitions occur near a clock edge. The latched value of the DOM's 48-bit local clock counter constitutes the coarse timestamp for that Hit. As the DOM local clock runs at 40 MHz, the leading edge of the PMT signal can appear anywhere within a fixed 25 ns window within the ATWD record. The coarse timestamp thus measures Hit time with about 7 ns RMS resolution, which is more than adequate for trigger formation and time ordering of data.

For physics analysis purposes, better time resolution is desired for Hits with waveform data. The determination, *ex post facto,* of the position of the leading edge of the PMT waveform within the ATWD record provides a "fine timestamp", with a resolution better than the 3.3 ns ATWD sample rate, and well within the 4 ns RMS system requirement.

## *4.4. Local Coincidence Modes*

The Local Coincidence (LC) capability is realized by connecting each DOM to its nearest neighbor with a dedicated bidirectional, duplex links (transceivers) over copper-wire twisted-pairs. The LC feature permits DOMs to transmit and receive LC "tag" messages to and from DOMs above or below. A DOM that receives an LC tag can modify and propagate the message further, thereby establishing an LC coincidence length (maximum distance between DOMs contributing to a LC tag).

The local coincidence hardware consists of a pulse generator coupled to a power splitter. The center port of the power splitter connects to the transmission line matching circuit for the off-board twisted pair. The power splitter topology makes possible simultaneous transmission and reception of LC signaling at each DOM. The transceivers support data rates in excess of 10 Mbps, transmitting an LC packet in 350 ns over links ranging from 21 m to 55 m.

When a Hit occurs, a DOM opens a receptive time window, which is typically not more than one µs. If during this window, a tag signal is received from a neighbor DOM, then the local coincidence requirement will be satisfied. Conversely, if a quiescent DOM, *i.e.*, one that is not currently processing a Hit, receives a tag signal from a neighbor, then it will also establish an identical receptive time window to accommodate the possibility that it may also receive a Hit. The processing of LC signals by the DOMs is thus time symmetric, so that LC tag creation does not bias against the time order of Hits. Conversely, FPGA firmware can adjust the window offsets to allow either upward-going or downward-going muons to be favored; neither option is exploited.

Hit information includes the presence or absence of tag signals from neighboring DOMs. Tagged Hits occur at a few percent of the total PMT rate (depending on the coherence length), and are much more likely to have been created by a particle than by PMT noise. An LC tag thus provides an immediate and efficient data selection criterion.

Beyond various testing and commissioning modes of operation, there are only two basic modes of array operation that employ local coincidence signals: "Soft" (SLC) and "Hard" Local Coincidence (HLC). There is a third trigger condition, Self-Local Coincidence (self-LC) that only relies upon a single DOM.

### *4.4.1. Soft Local Coincidence*

In SLC, the baseline-operating mode of IceCube, only those Hits with an LC tag will contain PMT ADC and ATWD waveform data; untagged (isolated) hits contain no ATWD waveform data. In SLC operation, the LC tag rate is typically ~10 Hz, depending strongly on depth of a DOM from the surface, and on the chosen coherence length. The DOM thus digitizes ATWD signals much slower than the ~700 Hz PMT SPE rate.

The justification for SLC is that isolated Hits are about two orders of magnitude more likely to be PMT noise pulses than physics event signals. Thus, the SLC mode significantly reduces both the dead time, and the recorded data flow from noise Hits, while sacrificing only a small fraction of real Hit waveforms.

*4.4.2. Hard Local Coincidence*

HLC requires every Hit to have an LC tag. This allows a high level of background rejection and reduction in data flow by discarding all PMT triggers without LC tags. In HLC, isolated Hits are lost. In addition, DOMs on either end of the string and DOMs with non-functioning neighbors suffer reduced trigger efficiency. This operating mode is used during commissioning and initial science operations.

*4.4.3. Self-Local Coincidence*

Self-LC provides for a DOM to include ATWD waveform information in the Hit when the PMT pulse is significantly larger than a characteristic single pe pulse in that DOM, even in the absence of received LC tags. The two PMT discriminators have thresholds that can be set independently. While one discriminator's threshold is always set to a fraction of a PE, the second discriminator's trigger threshold can be set to trigger on substantially larger pulses, which occur rarely. Triggering of the second discriminator initiates self-LC hit processing, without adding excessively to the data flow. This mode of operation can coexist with SLC or HLC.

Self-LC could also be built into the ATWD readout engine. Resources in the FPGA can be configured to recognize signatures that are more complex.

*4.5. Coarse Charge Stamp*

Monte Carlo simulations show that events do cause isolated hits, and therefore, some extra information about charge (i.e., about the number of photoelectrons collected in a time window) is useful in global trigger formation or in event categorization/reconstruction down stream from the *Stringhub*. To provide this extra information, but with minimum impact on data flow, an FPGA firmware module constructs a coarse charge stamp from a snippet of the PMT ADC record. Every Hit includes a coarse charge stamp, regardless of LC tag.

The 32 bits allocated for this purpose are arranged to include the highest ADC sample within the first 16 samples (400 ns) plus the immediately prior sample and subsequent sample. Only nine bits of the three ADC samples are selected. Another bit specifies the range. Depending on signal amplitude, either the most significant or least significant nine bits are chosen. The four remaining bits specify the index of the highest ADC sample with respect to the beginning of the record.

*4.6. DOM Dead time*

In the baseline SLC mode of operation, dead time is expected only within individual DOMs. No dead time is anticipated to occur due to data transfers from DOMs or due to any messaging activity. Due to the autonomous nature of DOM operation, dead time is distributed throughout the array with negligible inter-DOM correlations.

No dead time is incurred while a DOM is capturing waveform information. However, if ATWD digitization is initiated once the capture phase is over, dead time may occur, depending on the instantaneous circumstances. Because Hits are always created during such occurrences, the dead time intervals are known and can be taken into account during reconstruction of the candidate events. The DOM mitigates dead time in two ways:

1. Use of Local Coincidence: Both SLC and the more restrictive HLC permit ATWD digitization only for those Hits with LC tags. The LC tag rate for Hits with a coincidence length of two is in the range of 2 to 15 Hz, a factor of ~ 100 less than for a mode requiring no tags for digitization. This reduces ATWD dead time substantially. Note that every trigger initiates ATWD waveform acquisition. However, processing by the Hit readout engine in the FPGA is aborted at the end of the LC window when no neighbor LC tag is received.

Afterwards, the DOM is ready for the next trigger in 50 ns. The rate variation tracks the optical properties of the ice, as the rate is highest where the ice is most transparent.

2. Use of two ATWDs: Analog to digital conversion by the ATWD requires 29 μs per channel. As each DOM contains 2 ATWDs, should a DOM retrigger after 6.4 μs while one ATWD is digitizing, the other ATWD is available to start another Hit capture sequence. In SLC mode, a dead time of 50 ns (2 clock cycles) occurs at the end of the local coincidence window during the state transition to the alternate ATWD. Furthermore, a dead time of up to 22.5 μs is accrued if the second ATWD is launched before the first is read out completely. The latter case occurs at roughly 1 Hz. Thus for a random 500 Hz trigger rate, the dead time is $\sim 1 \times 10^{-5}$. However, the dominant sources of PMT noise pulses in the DOMs are scintillations in the glass pressure sphere and the PMT glass, due to $^{40}$K and U-Th decays. These produce a correlated fluorescent emission as much as ~1 ms later. Dead time is hence increased relative to a random flux. We estimate that the total dead time fraction does not exceed ~1 × 10-4. A firmware module in the FPGA counts clock cycles whenever the DOM is neither acquiring data nor ready for a trigger. Thus, the in situ dead time can be precisely measured, but this task has not yet been done.

### *4.7. Reciprocal Active Pulsing (RAPcal)*

The RAPcal method coordinates an ensemble of over 5000 free running clocks with respect to a GPS disciplined reference to establish a common time-base for all Hit data. It has a sequence of six distinct steps that determine the instantaneous frequency and phase (or offset) of the DOM's local clock relative to the Master Clock on the surface. The steps are as follows:

1. The DOR card commands the DOM to enable the RAPcal time-calibration sub-process; the DOM acknowledges receipt of command and enters a quiescent receptive state. However, PMT signals continue to be captured, digitized, and buffered.
2. After the DOM acknowledges readiness, the DOR sends to the DOM a precisely timed bi-polar pulse, the RAPcal signal. At the source, the transition edge rise- and fall-time is 5 ns and is synchronized with the system clock to better than 100 ps. The DOR card firmware latches the value of the 56-bit clock counter exactly when the RAPcal signal begins. The pulse amplitude and width are chosen to produce a robust received signal after attenuation and dispersion in the cable.
3. The DOM's firmware senses the arrival of the dispersed, attenuated pulse as a digital threshold crossing in the communication ADC data stream. The DOM records the entire pulse waveform, plus a few samples of baseline prehistory. The DOM clock-counter value associated with the last pulse waveform sample becomes the coarse time stamp of this portion of the RAPcal record.
4. To insure a quiet condition on the cable, the DOM's RAPcal firmware initiates a short, fixed length idle period "δ" before proceeding.
5. The DOM's firmware then generates its response to the DOR, a pulse identical in shape to the initial DOR pulse. The DOR's firmware senses and timestamps the pulse's arrival as the DOM did in (3) above. (This near-identity of received time-calibration pulse shapes, within natural variations due to components, is termed reciprocal symmetry.)
6. The DOR then requests the pulse waveform and time stamped data from the DOM. The two transmit times, the two received waveforms, and the two received times constitute the complete RAPcal record.

The data from the above steps enable a linear transformation from DOM local time to ICT for all Hits. Identical fiducial points are set for each received waveform, e.g., a leading edge or the crossover point of a bipolar pulse, as in Figure 12. These points define the time a pulse is received - a local time if received by a DOM, an ICT time if received at the DOR. The ratio of time intervals $\Delta T_{DOR}$ between successive pulses transmitted by the DOR and the local time intervals $\Delta T_{DOM}$ at which the DOM receives these pulses determines the ratio of master and local clock frequencies:

$$\nu_{local}/\nu_{master} = \Delta T_{DOR}/\ \Delta T_{DOM}$$

The offset of the local clock with respect to the master clock (the difference in clock values at the same instant of time) can be determined once the one-way propagation time $\tau$ is known for a calibration pulse sent between DOR and DOM. Reciprocal symmetry, or the identity of pulse shapes, results in identical values of $\tau$ for pulses sent in each direction. The value of $\tau$ can therefore be determined from a measurement of the round trip time ($\rho$) minus the known "idle period" ($\delta$) regardless of which waveform feature is taken as the fiducial point:

$$\tau = 1/2\ (\rho - \delta)$$

Reciprocal symmetry is verifiable because the calibration waveforms are digitized in both the DOM and the DOR, and the waveforms, like those in Figure 12, can be compared to determine if any important differences in shape are present. Simple estimators, such as extrapolation of the nearly linear part of leading edge or crossover region to baseline, the midpoint along the leading edge, or a centroid approach, provide precision on the order of 1-2 ns RMS. Repetitive measurements quickly make statistical errors for $\tau$ negligible.

Beyond any possible second-order effects arising from temperature gradients along the 3 km cable, the effects of possible asymmetries resulting from differing electrical component values, different temperatures at the DOR and DOM, and the impedance asymmetry introduced by compression of the quad cable, or the unterminated stub at one DOM were studied on the laboratory bench top by phase locking a DOM's local clock to a DOR card's local clock. Asymmetry effects were, at most, at the level of 0.1 to 0.2 ns, consistent with measurement error. Any remaining asymmetries that are common to all DOMs and DORs would affect only the offset between ICT and UTC.

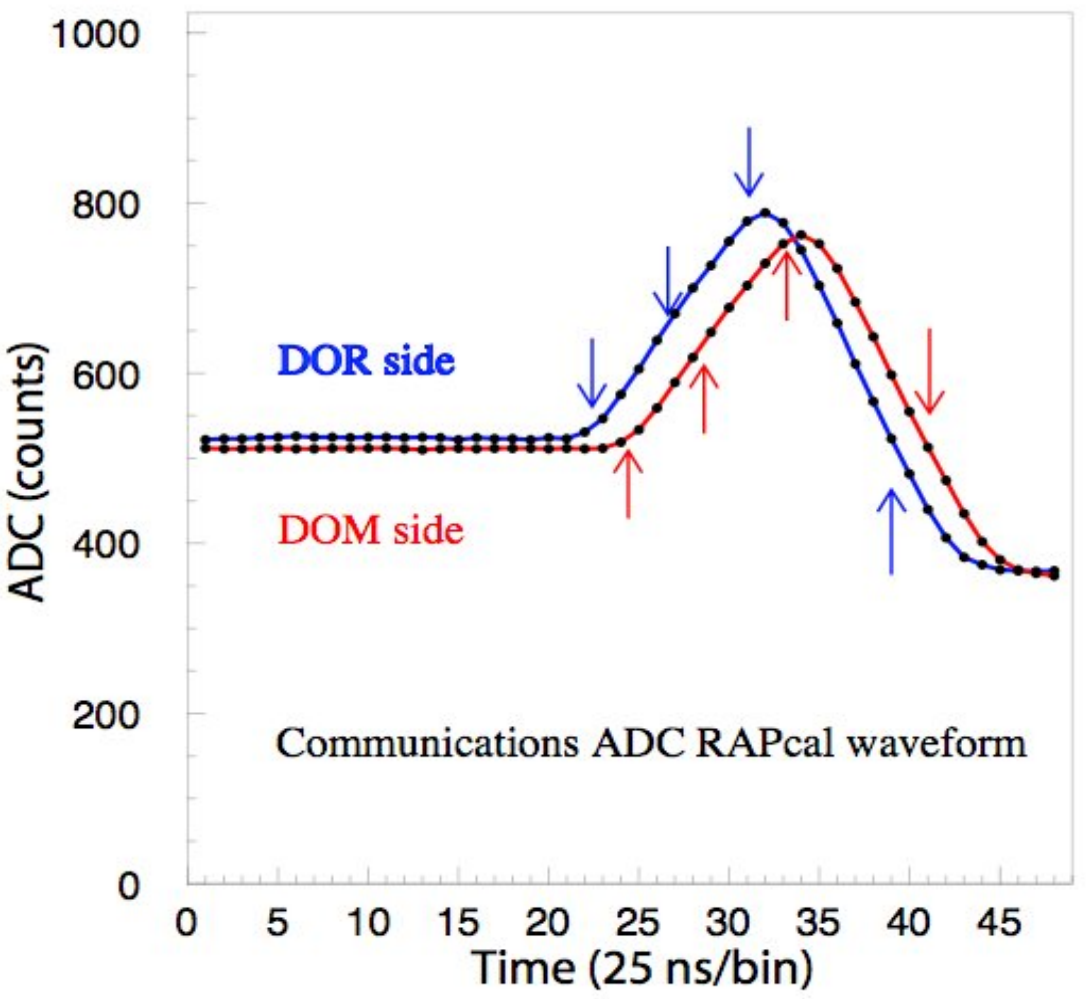

Figure 12. A typical RAPcal waveform, with 4 different time marks (arrows). The top curve (blue or DOR side) on the left is measured at the DOR, while the other (red or DOM side) is measured at the DOM. The time positions of the waveforms were adjusted so they could be easily compared.

The timing precision is limited by electrical noise on the cable and can be estimated by a simple relationship between electrical noise and rise-time:

$$\delta t \sim \delta V/(dV/dt),$$

where δV is the RMS noise/error voltage, and dV/dt is the received pulse rise time, ~ 160 mV/ 0.6 μs. From the observed timing precision and pulse rise-time, the inferred noise voltage is ~ 0.7 mV, which is slightly higher then the communications ADC quantization error of 0.4 mV. Because RAPcal is intrinsically sensitive to high frequency electrical noise, care was taken in the design of circuitry, cables, and operation to minimize induced noise. Since the time calibration procedure is repeated at regular intervals, individual calibrations that have been substantially affected by noise are easily recognized and discarded. In that case, the previous calibration is used for an additional calibration interval.

The RAPcal interrogation rate needed to track the oscillator drift in each DOM depends on the actual stability, which is expected to vary somewhat for each oscillator. Typically, the measured DOM oscillator drift under stable temperature conditions is remarkably good, with $\delta f/f < 3 \times 10^{-11}$. This permits interrogation intervals of a minute, or perhaps more, before drift has accumulated to a magnitude that would affect off-line event reconstruction. However, intrinsic oscillator frequency drifts and phase fluctuations display occasional, minute, abrupt discontinuities. In current practice, the time calibration sequence takes less than ~1.4 ms, and is set to occur once per second. The time spent in time calibration is hence invisible to the main task of data flow.

The coarse timestamp (cf. Section 4.3) is useful for triggering purposes, and is corrected by extrapolation, i.e., by using the two previous RAPcal events. In subsequent data analysis, it is possible to use the events just before and just after the photon's arrival, i.e., by interpolation. At this time, this correction is not needed as the current method provides sufficient precision.

The ice surrounding the DOMs constitutes a massive, stable heat sink for their crystal oscillators. However, on power-up after an extended off period, the oscillators are subject to a ~10°C temperature rise due to heating from the 3.5W dissipated by DOM MB electronics and the PMT HV power supply module within the glass sphere enclosure. Once in steady-state operation, the ensemble of all oscillators in IceCube constitutes a very stable virtual clock, whose stability is expected to exceed by a wide margin the short- and medium-term stability of the GPS receiver, which is affected by algorithmic discontinuities and an evolving mix of satellite signals.

RAPcal requires distribution to the DOMHubs of GPS-derived signals with sub-nanosecond synchronization, and highly coordinated actions in both DOM and DOMHub. The real-time nature of the RAPcal process dictates that code is implemented in firmware, rather than software. The RAPcal data set, acquired repetitively, is sufficient to establish a rolling time transformation of DOM time to ICT.

## 5. DOM MB Manufacturing and Testing Procedures

Because a deployed DOM cannot be repaired, stringent manufacturing and testing procedures were obligatory to minimize failure of a DOM MB during deployment and after it becomes frozen in the ice. The design goal is that not more than 5% of the DOMs shall fail within 10 years of operation, where failure is defined as complete loss of physics-useful data. A quality control program was developed to support the achievement of this goal. The design strategy centered on understanding how, where and why failures occur in boards and associated components when exposed to the operational conditions encountered in IceCube. A flow chart describing the production of the DOM MB can be seen in Figure 13.

The design and fabrication philosophy addressed failure propagation, supplier selection, manufacturing quality level, material restrictions, design control, and configuration control. The power and communications input circuit on the DOM MB is designed to maximize the probability of an open circuit (rather than a short or low-impedance circuit) in the event of a catastrophic board failure. This enables the neighboring DOM on the same wire pair to continue operating. Where possible, the electronic parts were selected from manufacturers that had been vetted by NASA and the Department of Defense as suppliers of high quality components. All components used are required to operate either in the Industrial (down to –40°C) or, preferably, MIL (down to –55°C) temperature range. The component's temperature range depended on availability and cost. Material restrictions minimized inclusion of materials with properties that could potentially shorten the life of the sensors. For example, plastics incompatible with low temperature, cadmium, pure tin, and zinc plating are not appropriate for critical applications, and were only used on the DOM MB if no other option were available.

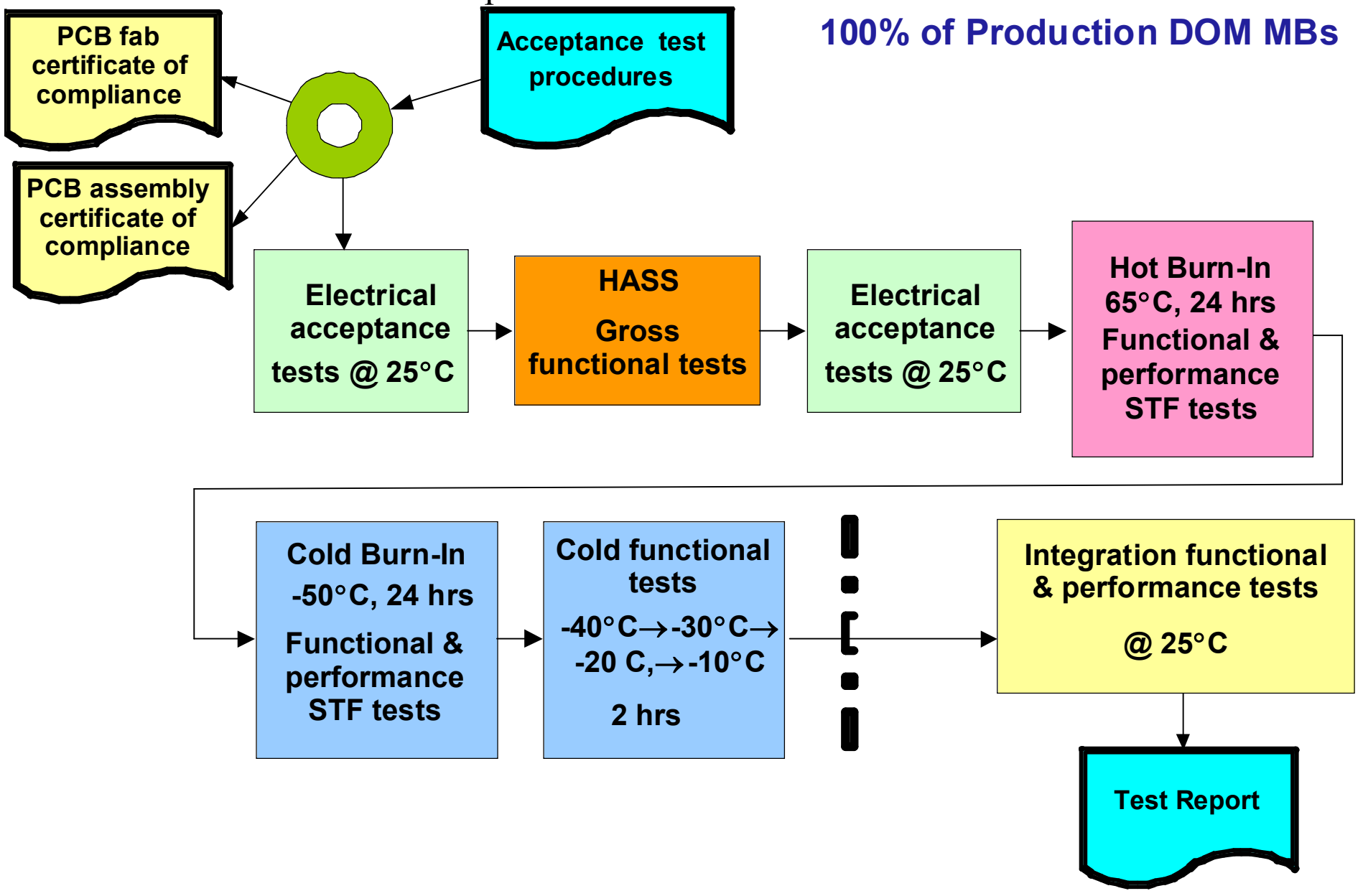

Figure 13. DOM MB production flow. The steps to the left of the dashed-doted line were done at an outside vendor, while the steps to the right were performed at LBNL.

Design, management, and manufacturing controls were put in place to guarantee a consistent product that would meet all system and manufacturability requirements over several separate procurement cycles. DOM design verification was based principally on testing because it provides the highest level of confidence that the actual performance meets the specified requirements. As testing of some requirements such as a life of 10+ years was not practical, verification was done by analysis. Design reliability was addressed by subjecting a sample of pre-production DOM MB assemblies to a stress test method called Highly Accelerated Life Time Test (HALT), which exposed the DOM MB to extremes in vibration and temperature cycling while operational.

The test regimen consisted of a cold and hot temperature stress, rapid thermal transitions, a vibration stress step, and finally simultaneous temperature cycling and vibration. During all portions of the HALT testing, the DOM MB was monitored continuously by running the STF suite. The final result of HALT testing confirmed the suitability of the DOM MB design.

The DOM MB was imaged thermally to determine if there were any excessive hot spots that could cause later failures. The image that is shown in Figure 14 indicates that there is a localized approximate 5°C rise due to heating in the DC-DC converter in an open environment. This figure also shows the power dissipated by the active communications components, the EPXA4, and wide-band amplifiers.

A test stand was used to subject each manufactured DOM MB to an extensive series of tests to be sure that they also met specifications. First, every manufactured DOM MB was tested for function and performance at room temperature, using the STF suite of tests. Following that, the next stage in the DOM MB test process was a less stressful version of HALT, called Highly Accelerated Stress Screening (HASS). HASS was used to test performance from -40°C to +65°C and vibration to 5G. Following HASS, all DOM MBs were tested at +65°C for 24 hours and -50°C for 24 hours. In the next testing stage, the DOM MB was connected to a PMT, a High-Voltage Base board, a Flasher board, and 2500 m of twisted quad cable to test all system interfaces. After a DOM MB assembly passed all these production tests, it qualified to be integrated into a DOM at assembly sites in Wisconsin, Berlin, or Stockholm/Uppsala.

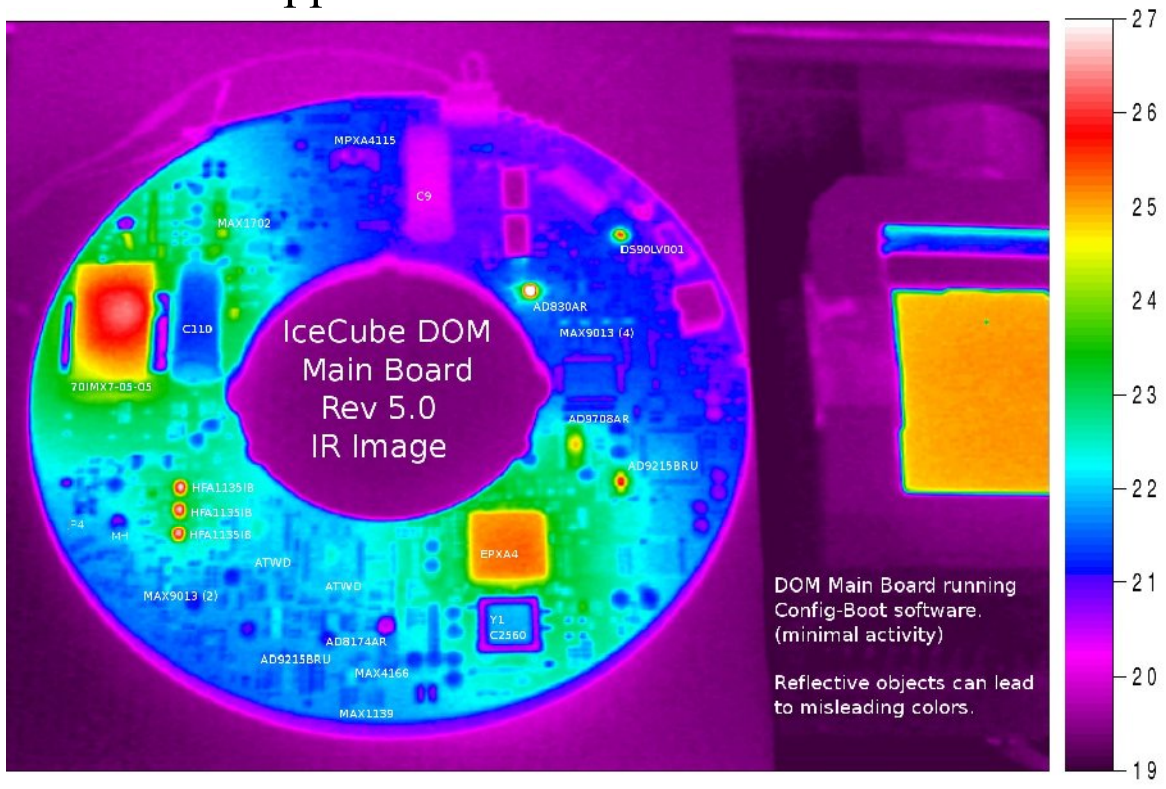

Figure 14. This is an infrared image of a DOM MB that was operating for about 40 minutes. The orange block on the right hand side is a reference temperature source in units of centigrade. Some of the objects in the picture appear colder because they reflect surrounding areas.

The last stage in the production test cycle was to load sealed DOMs into a Deep Freezer Lab (DFL), and run a Final Acceptance Test (FAT) on all units. The FAT lasted 3 weeks, including slow temperature cycles from 25°C to -45°C, periodic STF tests, and a Calibration test suite, all under the control of a test DAQ system. About 85% of the DOMs passed these FAT tests. Failures arose from malfunctions from any of the sub-components.

Once a DOM passes all of these tests, it was ready for shipment to the South Pole. After arrival at the South Pole, it was tested again, to detect any damage during shipment. At this point, the DOM was ready to be deployed into the IceCube array.

## 6. Performance

The operation of IceCube to August 2008 allows a first assessment of IceCube's performance. Tests of the DAQ in situ, as well as normal operation, which involve communicating with deployed DOMs, measure the performance of the communications hardware and protocols. The ability of IceCube to identify point sources of neutrinos, should they exist, depends on the pointing resolution of reconstructed muon trajectories. The pointing resolution depends quadratically on angular reconstruction accuracy, which in turn depends on time resolution for detecting the Cherenkov radiation. Despite the degrading impact of optical scattering by the ice, some photons are nearly “direct”, and their accurate detection is particularly important. Accordingly, time resolution requirements of 4 ns RMS for individual DOMs and 7 ns RMS for the entire IceCube DAQ system were established to ensure that technical aspects would not compromise information quality.

## 6.1. Timing

The accuracy with which the system can determine the time of arrival of a photon at the photocathode has been determined from flasher calibration sources and cosmic ray muons.

### *6.1.1. Timing with Flashers*

A straightforward test of the system in ice is to pulse the LEDs on the flasher board at a known time and measure the arrival time of photons at an adjacent DOM on the same string. Since this measurement depends on the accuracy of the time calibration procedure for both the emitting DOM and the receiving DOM, stochastic errors will combine in quadrature, but some systematic errors may be more difficult to detect.

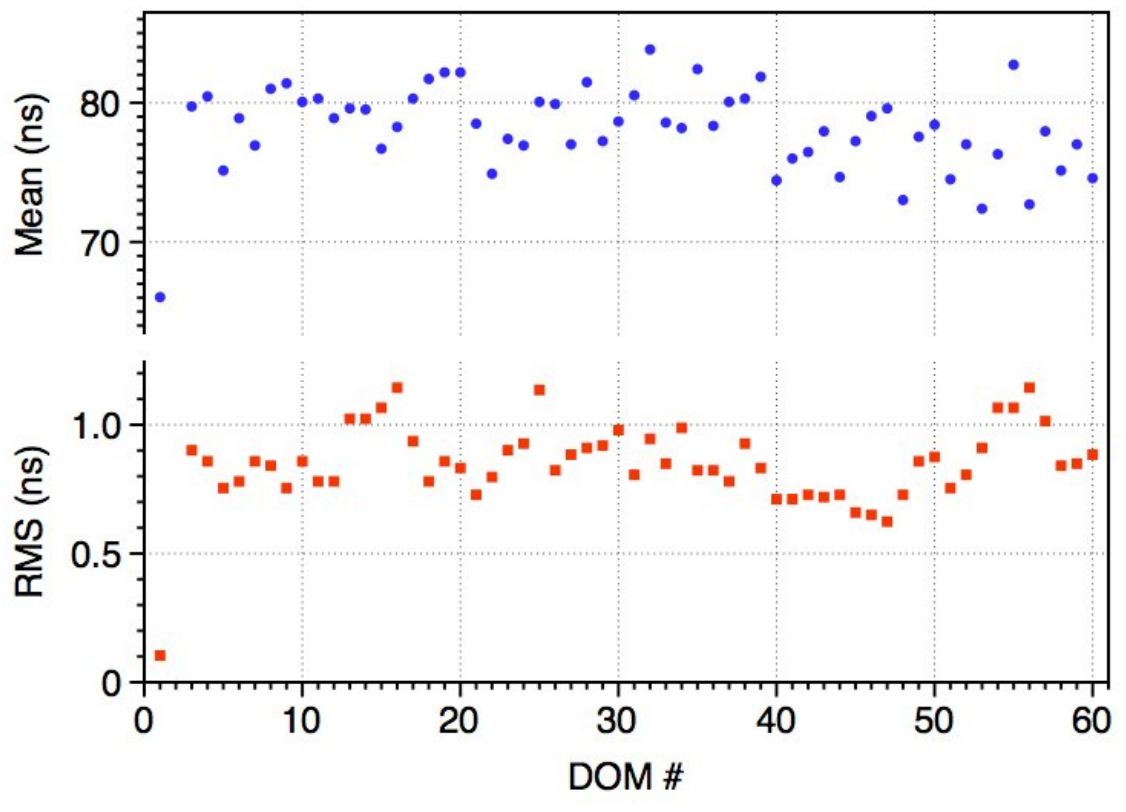

Figure 15. A plot of the mean and the RMS of the difference in time between flashing an LED and arrival of the photons at the receiving DOMs. The flashing DOM is located below the receiving one. The top graph (blue) shows the mean values, while the bottom (red) shows the RMS deviation. DOM 1 is at the top of the string, while DOM 60 is at the bottom. The mean includes the time that the light propagates in the ice.

Figure 15 shows the mean times and RMS values for optical signals received by a DOM when the flashers in the DOM below it are operated. The distance between DOMs (17 m) is small enough that the first photons from high intensity light flashes experience little or no scattering. Thus, time calibration and the response of the DAQ electronics, and not the scattering properties of the ice, should dominate the resolution. Since many photons are detected in a short time, the single-photon timing response of the PMT should not contribute much to the time residuals. As this test measures the difference in time between the LED flash (as determined by calibrating the clock in that DOM) and first photon’s arrival at the second DOM (as determined by calibrating the clock in that DOM), the actual resolution for a single DOM is $1/\sqrt{2}$ of the measured resolution if the dominant contribution to the resolution is error in time calibration.

Most of the RMS values for 59 DOMs in this particular string are less than 1.5 ns (RMS), which indicates that the time calibration error and any other stochastic contributions to the resolution are 1 ns or less. In these tests, the ATWD is used to determine the photon arrival time.

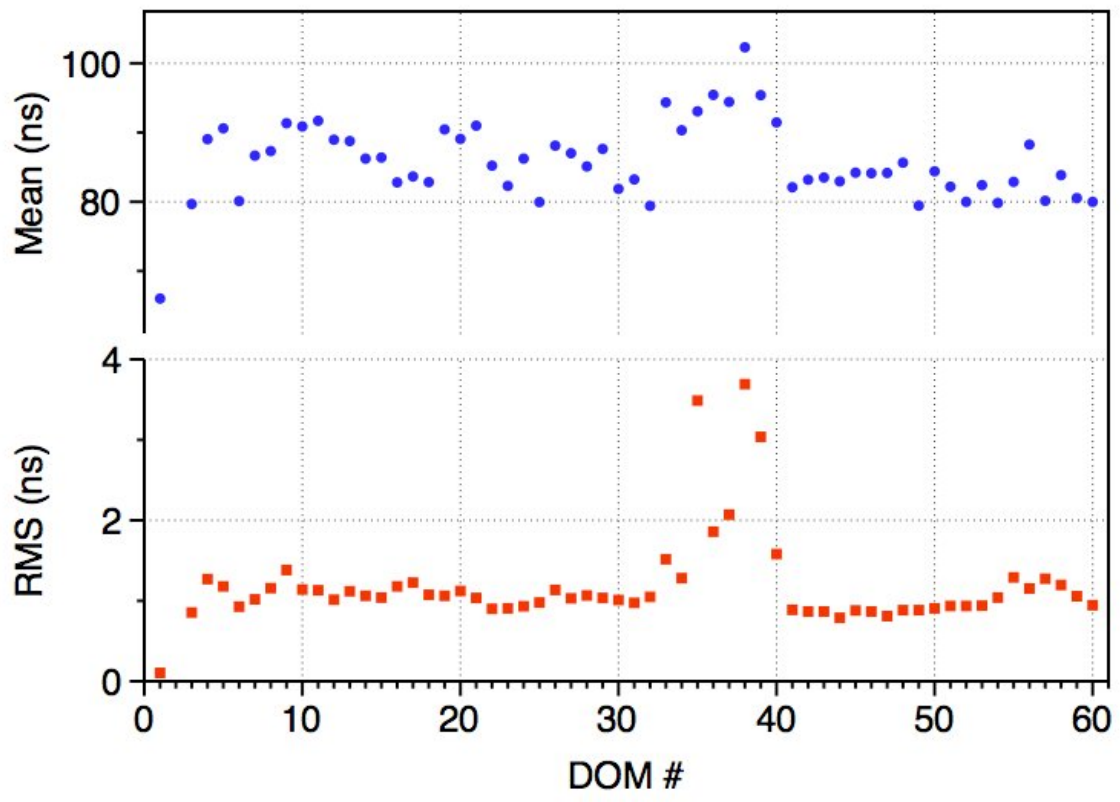

Figure 16. The RMS and mean time difference between photons arriving at two adjacent DOMs located just above a third, flashing DOM. The top graph (blue) shows the mean values, while the bottom (red) shows the RMS deviation.

Systematic errors in timing that involve the properties of the LED flashing system can be eliminated by using two adjacent DOMs to receive light emitted by the flashers on a third DOM located below the two receiving DOMs. The difference in arrival times of photons received in the upper two DOMs emitted in the same light burst is independent of the absolute time at which the burst occurs. The distance traveled by photons to the farthest DOM is now 34 m, which provides additional opportunity for photons to be scattered in the ice with a corresponding delay and jitter in the arrival time. Therefore, the measured time distributions will correspond to upper limits on the time resolution of the system. A $1/\sqrt{2}$ factor applies here as well in estimating the resolution for an individual DOM. The results of such a test are shown in Figure 16. The increase in both the time and the time resolution for DOMs, which are numbered between 33 and 40, arises from a dust layer in the ice[13].

The arrival time of a photon is normally determined from the ATWD waveform. The PMT ADC also records the waveform. Since the ATWD and ADC simultaneously capture the first 420 ns after a trigger, both devices can be cross calibrated.

The ATWD sampling rate is calibrated by making a measurement (using STF and a diagnostic input channel of the ATWD) of the sampling clock used for the PMT ADC. The sampling time offset for the ATWD with respect to the PMT ADC is measured by injecting a short light pulse generated by an on-board LED into the PMT, then measuring the arrival time in both the ATWD and ADC.

Figure 17 shows that the time resolution for the ADC is just under 5 ns, and that the average arrival time for photons determined with these two methods agrees to within 0.6 ns. The observed resolution, 4.7 ns, is about 20% of the bin width for the PMT ADC, which is comparable to the relative resolution seen in the ATWD.

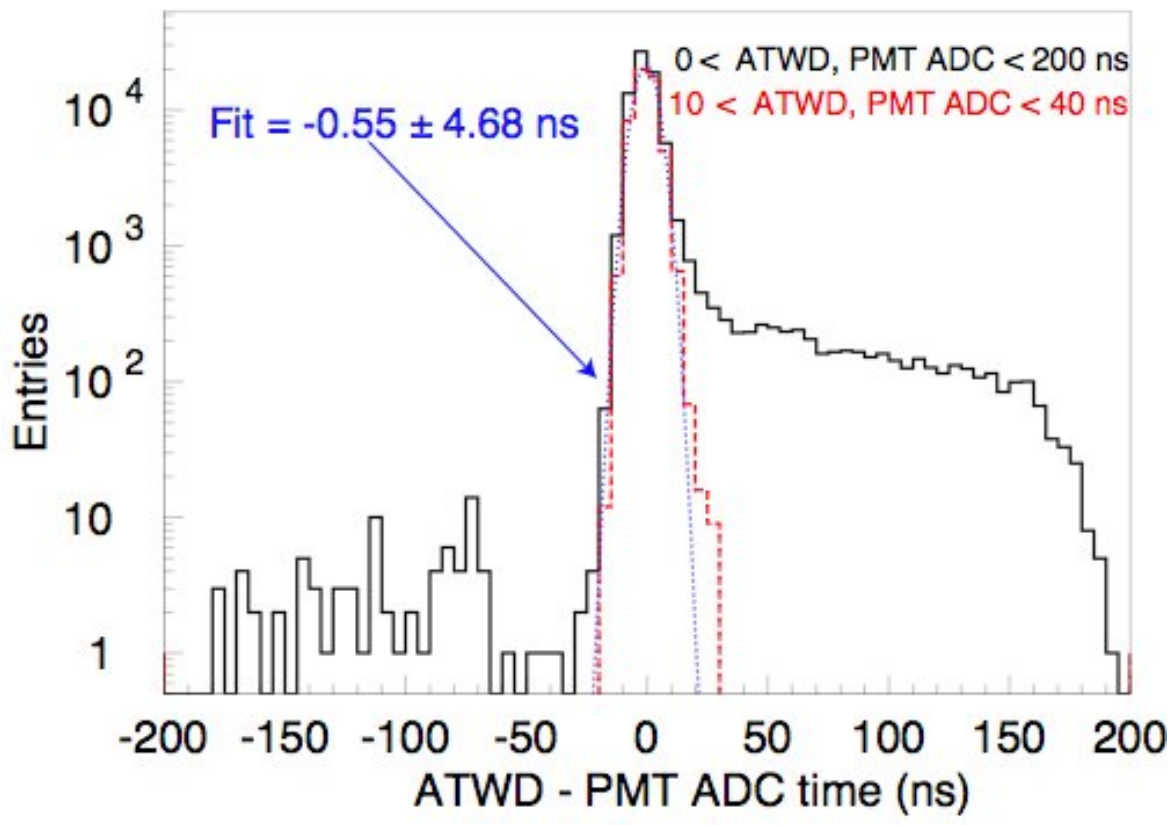

Figure 17. The time difference for pulses reconstructed in both the ATWD and PMT ADC. This difference is -0.6 ± 4.7 ns, dominated by the ADC's resolution. The solid (black) curve is plotted for ATWD and ADC sampling in the range between 0 and 200 ns. The dashed (red) curve has a narrower range of 10 to 40 ns.

*6.1.2. Timing with muons*

The flux of down-going cosmic ray muons penetrating the active volume of the IceCube array enables a test of the timing performance of the DOMs under the same conditions as actual data-taking. Even when imposing cuts, it is possible to perform this timing test with high counting statistics at regular intervals throughout the year.

First a muon track is reconstructed using all hits except for one "test DOM" to avoid any bias in the fit. The predicted arrival time from this track is then compared with the measured arrival time of the photon at the "test DOM". Despite the limited accuracy of the track position measurement, the requirement that the reconstructed track passes within 15 meters of the "test DOM" minimizes the effects of scattering in the ice. This procedure of eliminating one DOM from the reconstruction is repeated for the set of DOMs that are sufficiently close to the muon track.

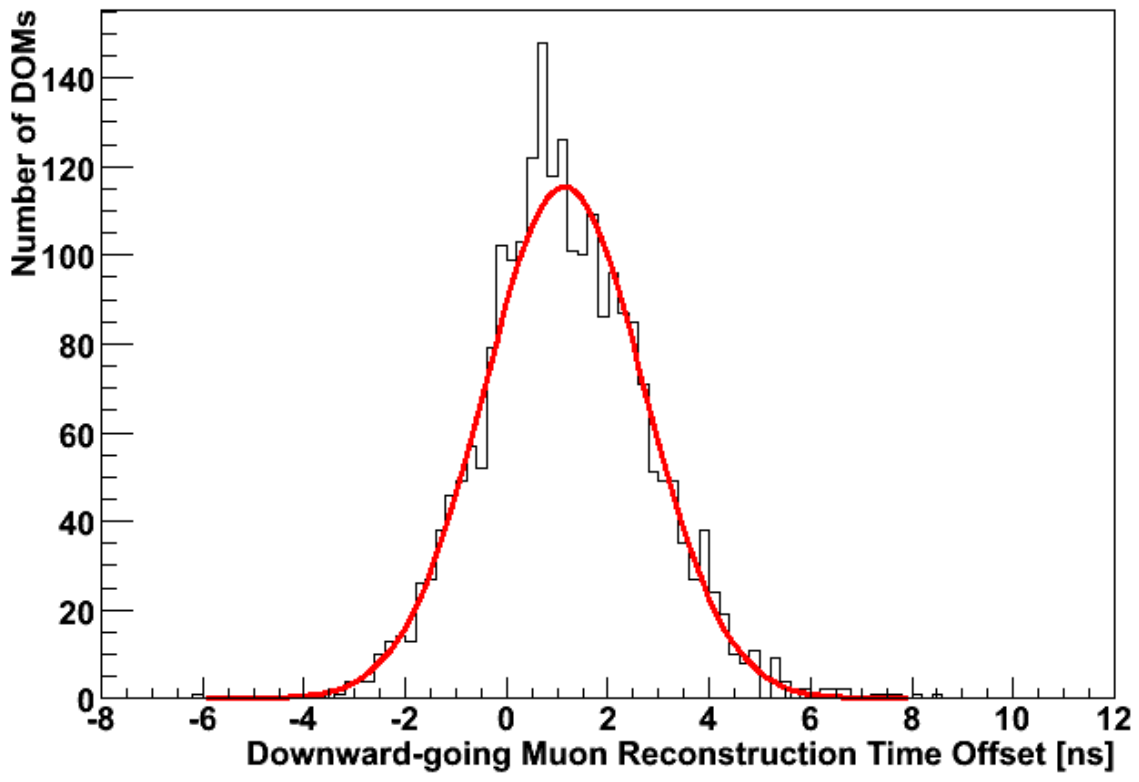

Figure 18. The peak values of the difference between the predicted time (based on fitted muon tracks) and the measured photon arrival time for the active 2341 InIce DOMs out of a total of 2400. This measurement was taken in August 2008. The width, σ, of the Gaussian fit is 1.6 ns.

Figure 18 shows the distribution of the peak values of the difference between the predicted and measured arrival times for the active 2341 out of the 2400 InIce DOMs. This distribution is narrowly peaked at 1.1 ns with a variation of 1.6 ns.

This method also can be used to study the stability of the time calibration. Figure 19 shows the results from data taken in April 2008 and again in August 2008 for a typical string (String 48). The measured time shift for each DOM was found to be less than 1.0 ns, demonstrating good stability over several months of operation.

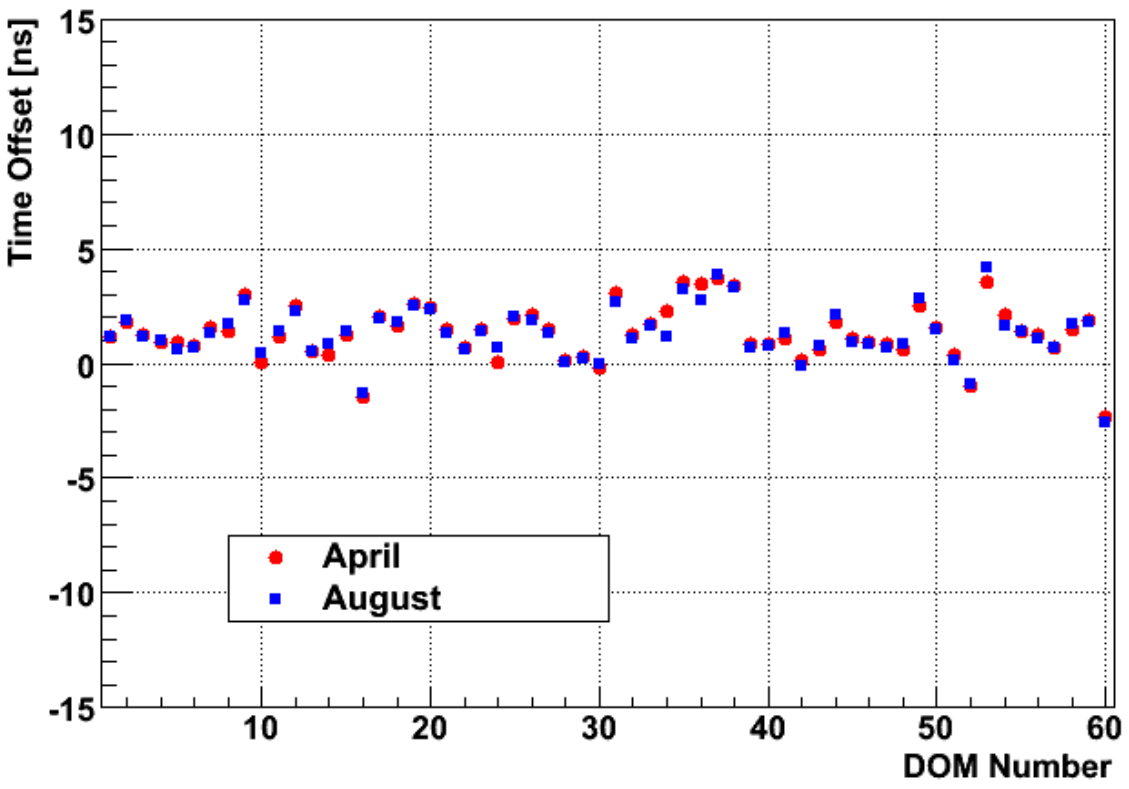

Figure 19. The peak values of the time distribution for the 60 DOMs on String 48 were measured in April 2008 (red circles) and again in August 2008 (blue squares). For all DOMs, the difference between these two sets of date is less than 1.0 ns.

### *6.2. PMT Linearity Measurement*

The PMT SPE signals are well calibrated at a known nominal gain of $1 \times 10^7$. Accurate measurement of the number of photons arriving at a DOM thus translates to the measurement of charge. The charge is given by calculating the area of the waveform peak. The measurement of charge will be affected by the linearity of the electronic signal path for different gains. This can be checked by determining the area of a pulse that falls in a region covered by two ATWD gain stages and comparing the results. This test shows that the calibration of the different ATWD channels is sufficiently accurate to determine pulse height over a wide dynamic range using the LED flashers. Since each DOM contains 12 independently operable LEDs, linearity can be determined by observing the response to individual LEDs, operating one at a time. Once the individual response is known, the response to varying numbers of LEDs pulsed simultaneously can be measured and deviations from linearity determined.

Figure 20 shows the results of such a test for a typical PMT. The deviation from linearity reaches 10% at ~400 photoelectrons/15 ns. Of course, calibrating the response of the PMT in the non-linear region and making the appropriate corrections for larger pulses can extend the dynamic range extended beyond 1000 photoelectrons/15 ns.

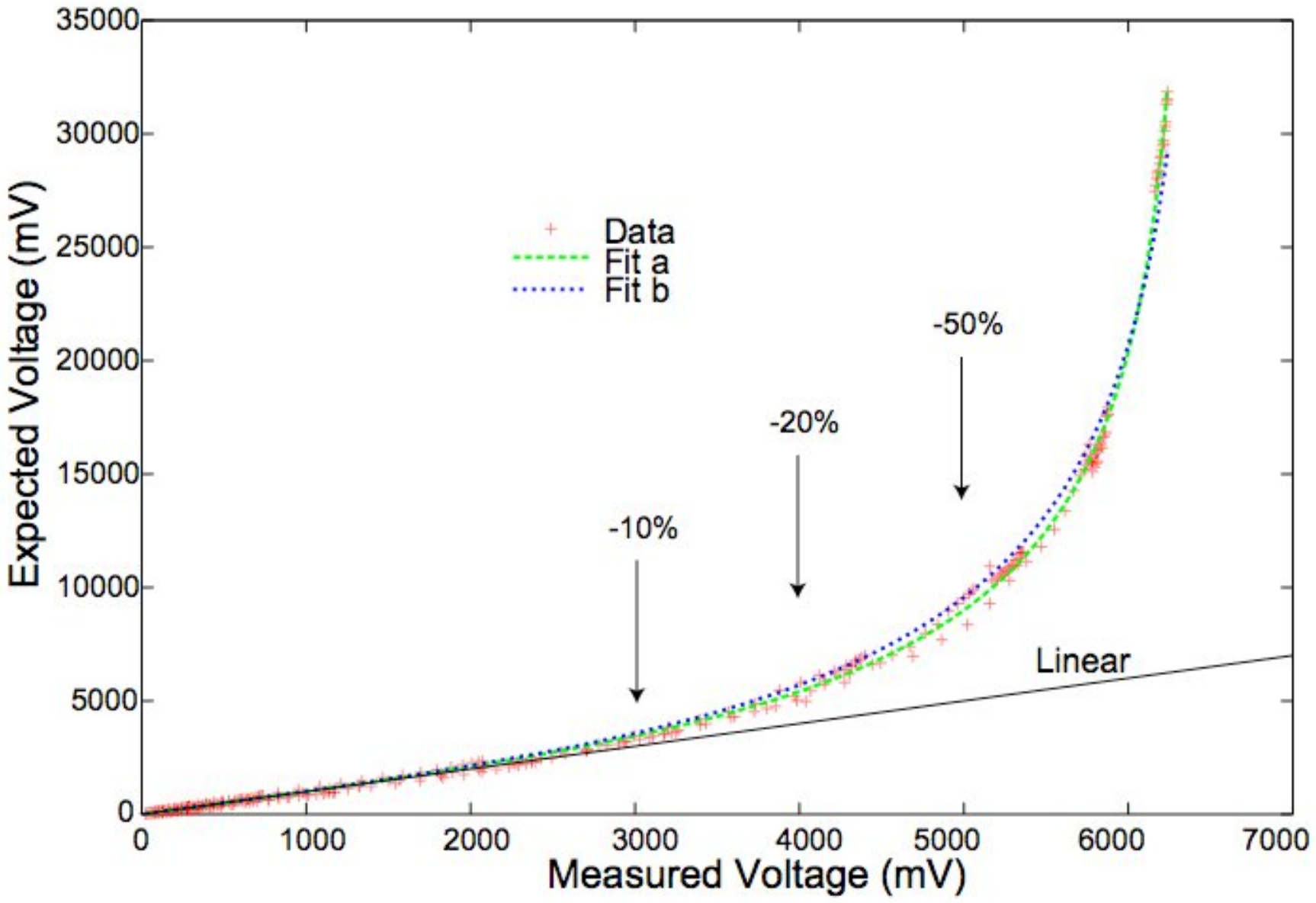

Figure 20. PMT linearity that has been measured for one phototube.

*6.3. Temperature Variation*

IceCube DOMs are deployed in varying thermal environments, which potentially affects instrument response. InIce DOMs could be subject to a systematic position dependant calibration, while IceTop DOMs may experience temporal changes. As described in Section 5, DOMs are tested and calibrated over the full range of IceTop and InIce temperatures. Once deployed, InIce DOMs experience a constant temperature and the calibration needs only to be done at the operating temperature at a fixed location. The timing behavior is constant throughout the year as demonstrated by Figure 19. IceTop presents special challenges. The overall change in IceTop DOM launch rates are of order 20%, of which 15% is associated with day-to-day changes of barometric pressure, which modulates the flux of secondary particles produced in air showers. There remains a 10% seasonal variation that may be due to structural changes of the upper atmosphere, the response of DOM to changing local temperatures, or some other cause. These issues will be discussed in subsequent papers.

*6.4. Reliability*

By the end of the fourth operational year of IceCube, 2560 DOMs had been installed. Twelve of them, 0.5% of the total, are not useable, having failed during deployment or freeze-back of the borehole ice. Evidence suggests that most of these failures were due to stresses on cables and connectors during freeze-back. Three InIce DOMs and one IceTop DOM failed after the ice froze. One of these appears to have failed due to loss of PMT vacuum, as indicated by a pressure decrease in the DOM.

The DOM MB electronics failure rate after deployment – at most 3 in 3260 DOM years – during this notably short operational period suggests that some 97% of the full complement of DOMs may survive in 25 years. This survival rate is much higher than the previously stated design goal of 95% in 10 years. In any case, the relatively small number of failures so far is encouraging and likely attributable to the extensive quality assurance program.

Beyond this relatively small number of failures, there are some 30 DOMs that have minor problems and are temporarily "out of service" but should return to useful operation. At any given time during operation of the IceCube InIce array for data taking, 97% - 98% of the deployed DOMs are operating according to specification. The performance for IceTop DOMs is comparable.

## 7. Summary

The IceCube DOM MB evolved from the circuit board developed for AMANDA’s prototype DOMs. This experience and several newly available components prompted the selection of a more integrated, higher performance CPU-FPGA implementation, a more robust and flexible independent local coincidence transceiver, and a more sophisticated higher performance mezzanine card flasher subsystem interface. The prototype DOM's CPU fetched 8-bit event data from the FPGA's memory, whereas the DMA engine in the IceCube FPGA transfers 32-bit data into the CPU's memory. This change delivers much higher performance with reduced CPU load, resulting in a data rate nearly double that of the prototype DOMs. The DOM's modular real-time software design provides an increased functionality and robustness for a modest increase in resource usage. Since the IceCube DOM's flash file system stores multiple programs and FPGA configuration files, the DOM MB's operating system transitions between them as needed for data acquisition and periodic system testing. To enhance noise suppression in the HV subsystem, a serial DAC and ADC on the mezzanine card replaced the HV analog control. The DOR card and DOMHub were significantly redesigned to improve the performance and packing density and facilitate scaling the system to 80 strings. In addition, there were numerous other improvements that resulted in a very reliable system that can be duplicated many thousands of times.

These improvements produced a DOM MB, which controls all the functionality within the DOM. The DOM MB communicates digitally with the surface by a single twisted pair of copper wires. The main functions of it are PMT signal (waveform) capture and digitization, timestamping of Hits, calibration, coincidence logic, communications, and monitoring. The design and the performance of the DOM, first extensively tested and verified in the laboratory, meet the science-driven design requirements for operation in the deep ice and on the surface in the IceTop tanks. A comprehensive quality assurance and testing program maximizes the probability that deployed DOMs will perform as required. So far, the DOMs are performing very well, with a failure rate (including all sources of failure) of about 1%. On average, 98% of all deployed DOMs participate in experimental data taking for physics purposes. Almost all of the DOM failures are due to cable or deployment issues.

## Acknowledgments

We acknowledge the support from the following agencies: U.S. National Science Foundation-Office of Polar Program, U.S. National Science Foundation-Physics Division, University of Wisconsin Alumni Research Foundation, U.S. Department of Energy, and National Energy Research Scientific Computing Center; Swedish Research Council, Swedish Polar Research Secretariat, and Knut and Alice Wallenberg Foundation, Sweden; German Ministry for Education and Research, Deutsche Forschungsgemeinschaft (DFG), Germany; Fund for Scientific Research (FNRS-FWO), Flanders Institute to encourage scientific and technological research in industry (IWT), Belgian Federal Science Policy Office (Belspo); the Netherlands Organisation for Scientific Research (NWO); M. Ribordy acknowledges the support of the SNF (Switzerland); A. Kappes and A. Groß acknowledge support by the EU Marie Curie OIF Program. Throughout the conception, design, and building of this system, W. Chinowsky has provided invaluable advice and guidance. Michael Phipps and the University Program of Altera Corporation provided us with samples, development tools, advice, and technical support. We also thank the engineering and technical staff at Lawrence Berkeley National Laboratory who were essential to the design, construction and testing of the DOM MB.

Figure 1

Click here to download high resolution image

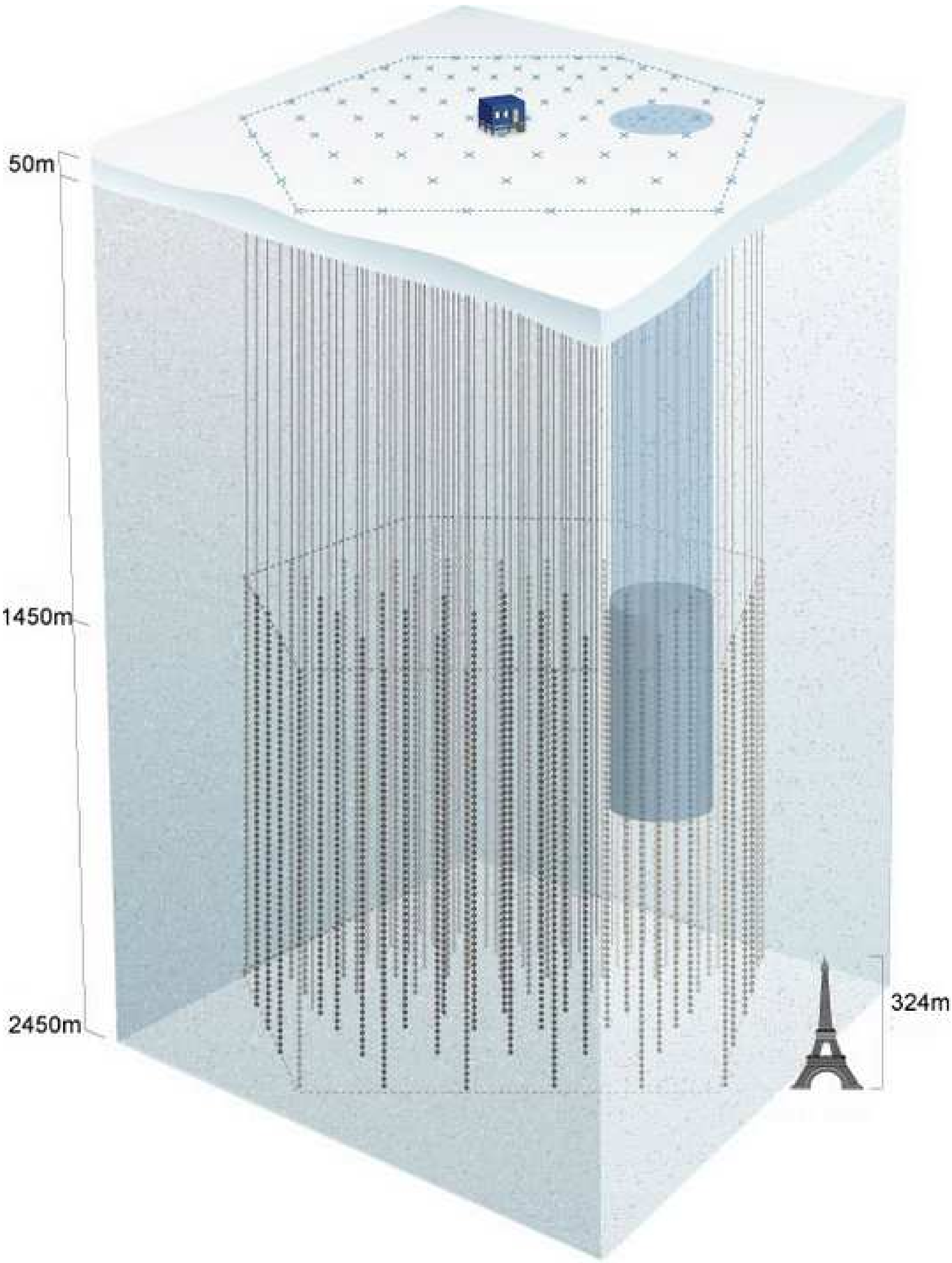

**Figure 2**
Click here to download high resolution image

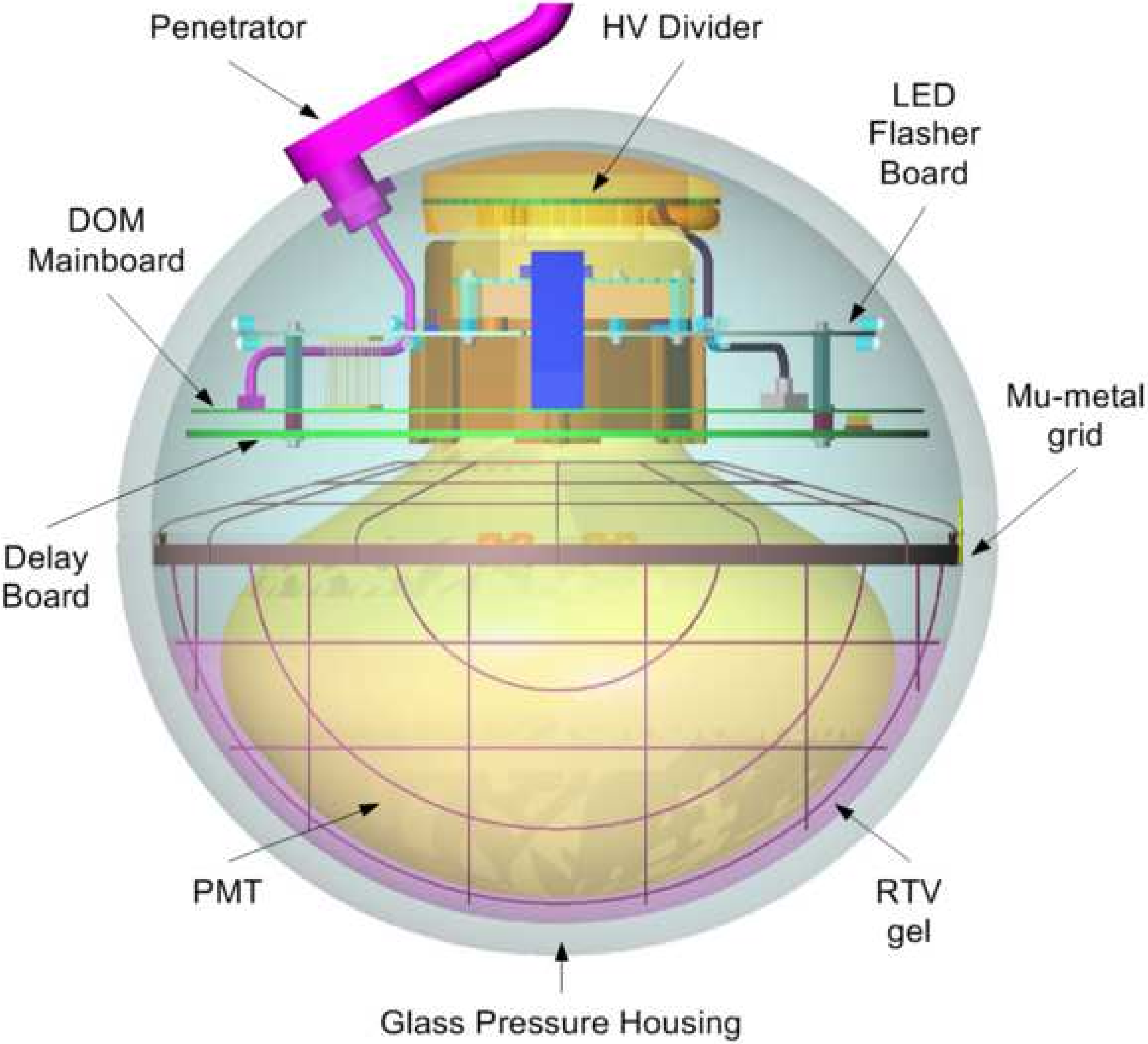

Figure 3
Click here to download high resolution image

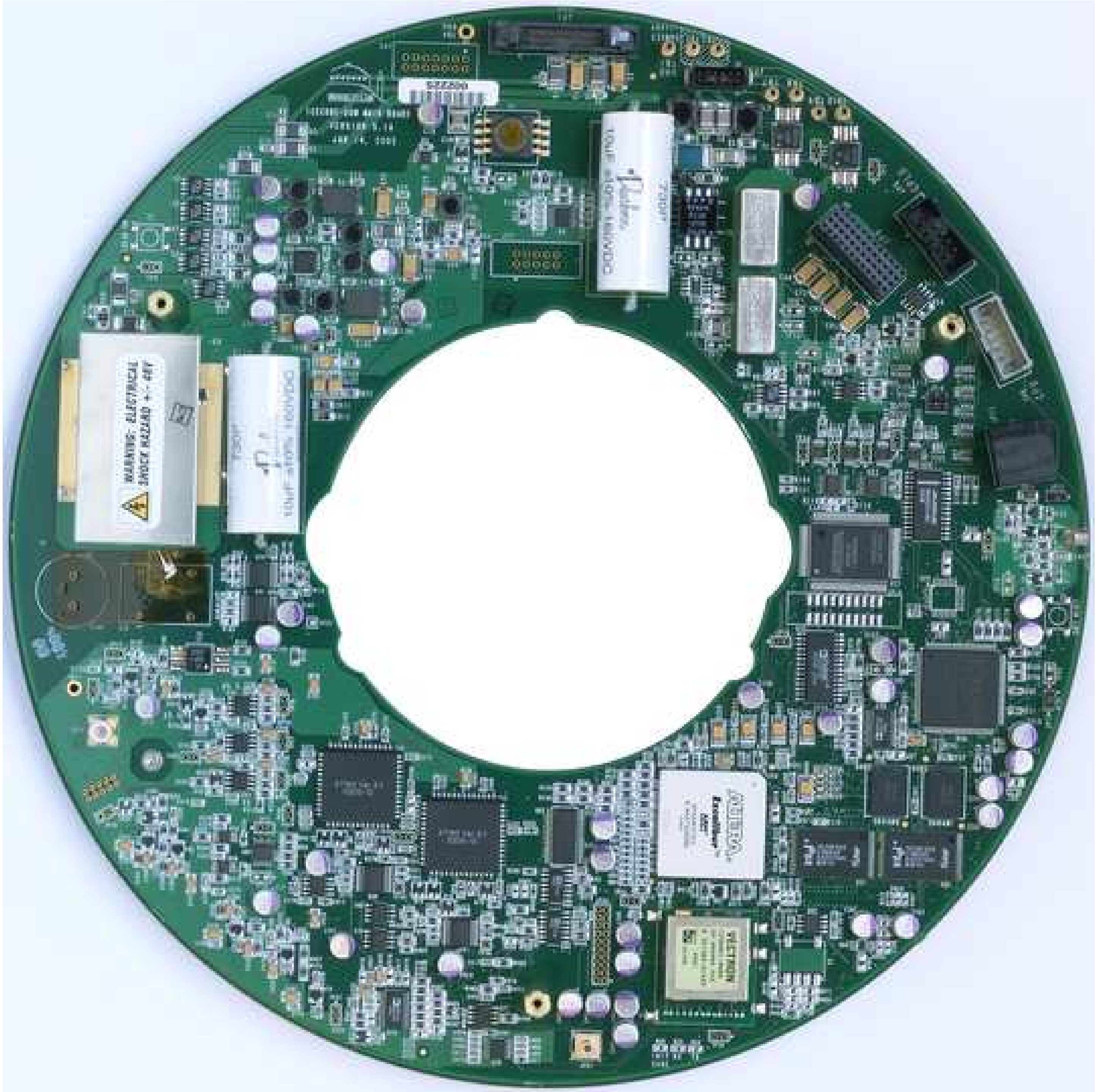

Figure 4

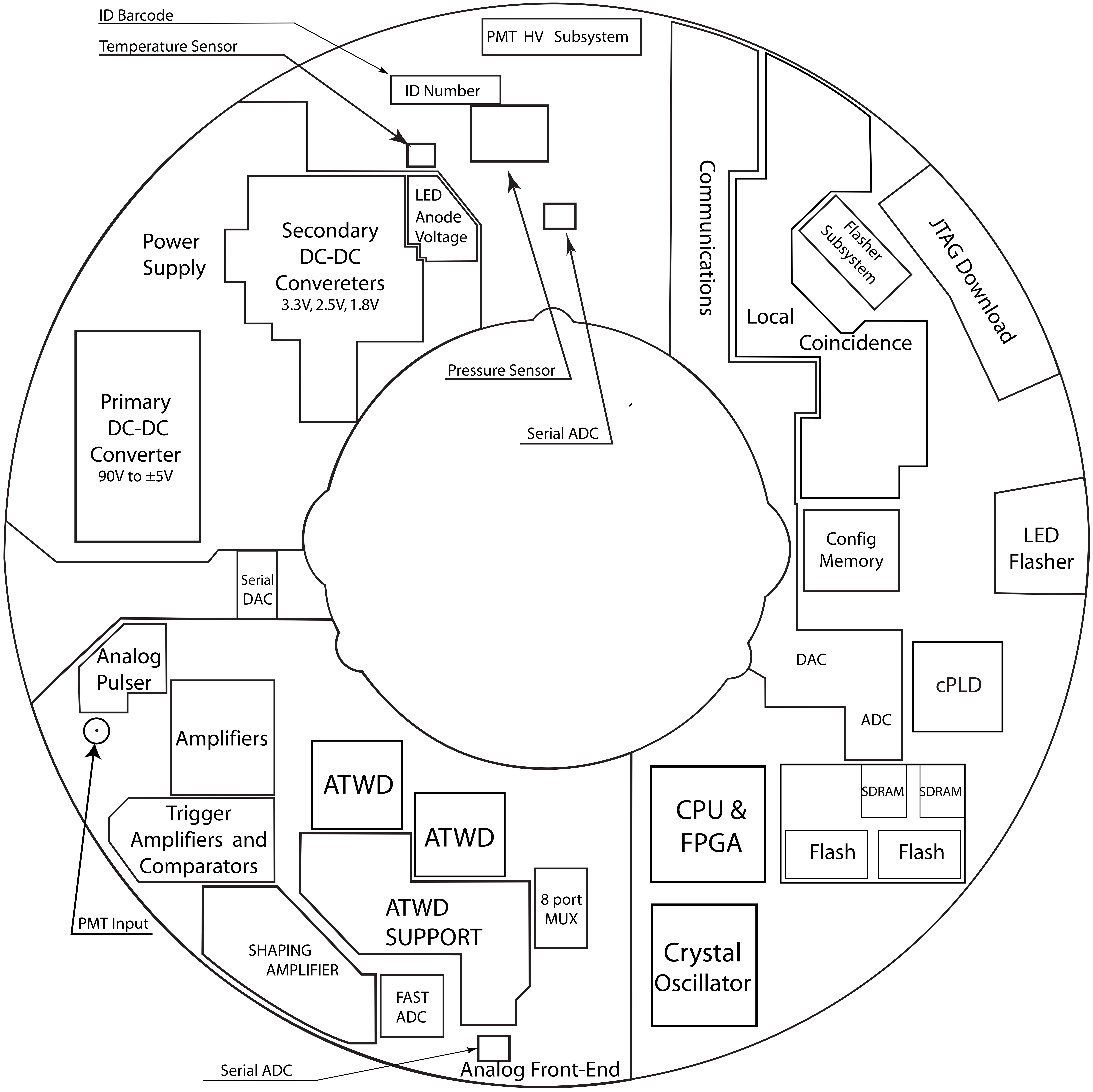

Figure 5

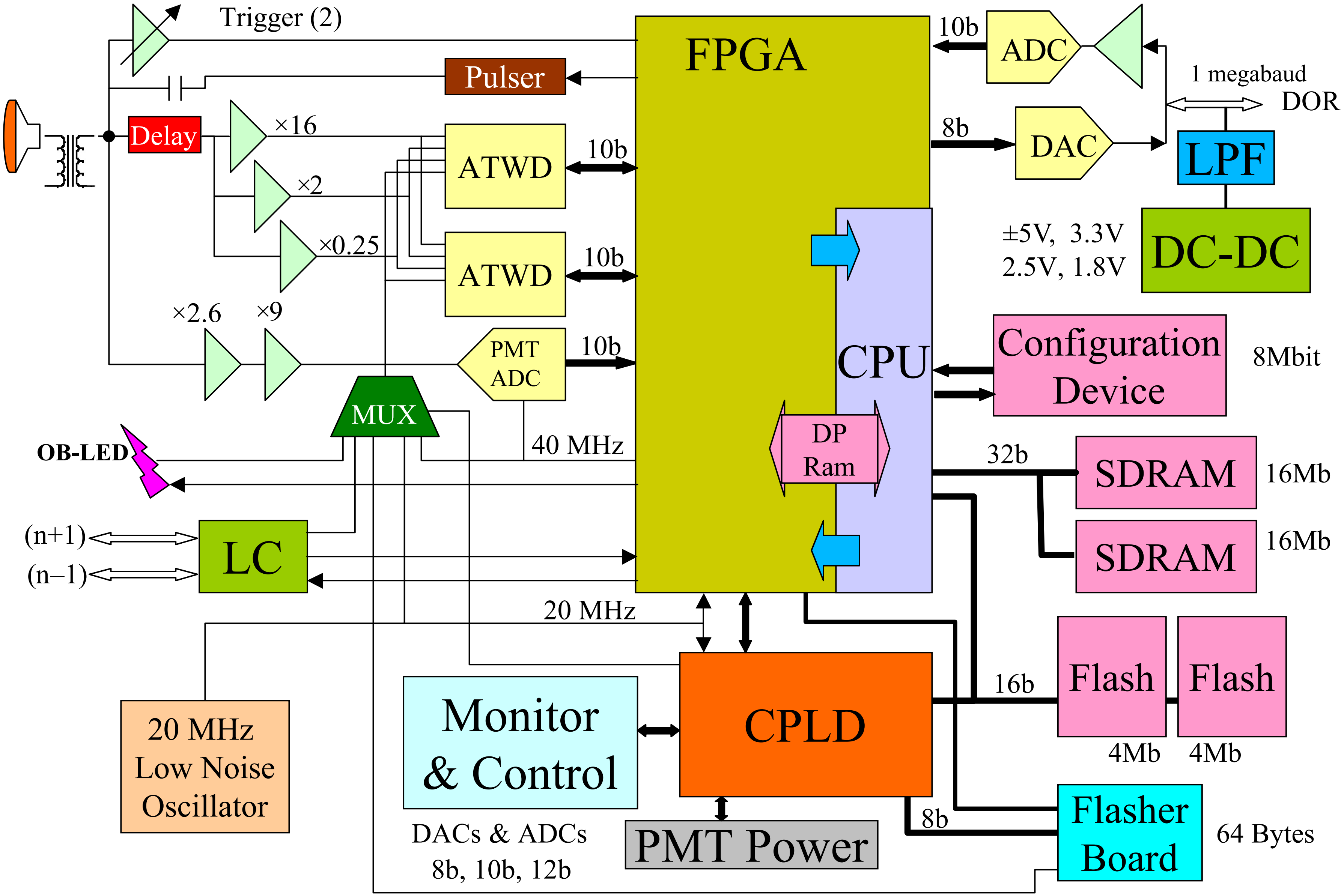

Figure 6

Click here to download high resolution image

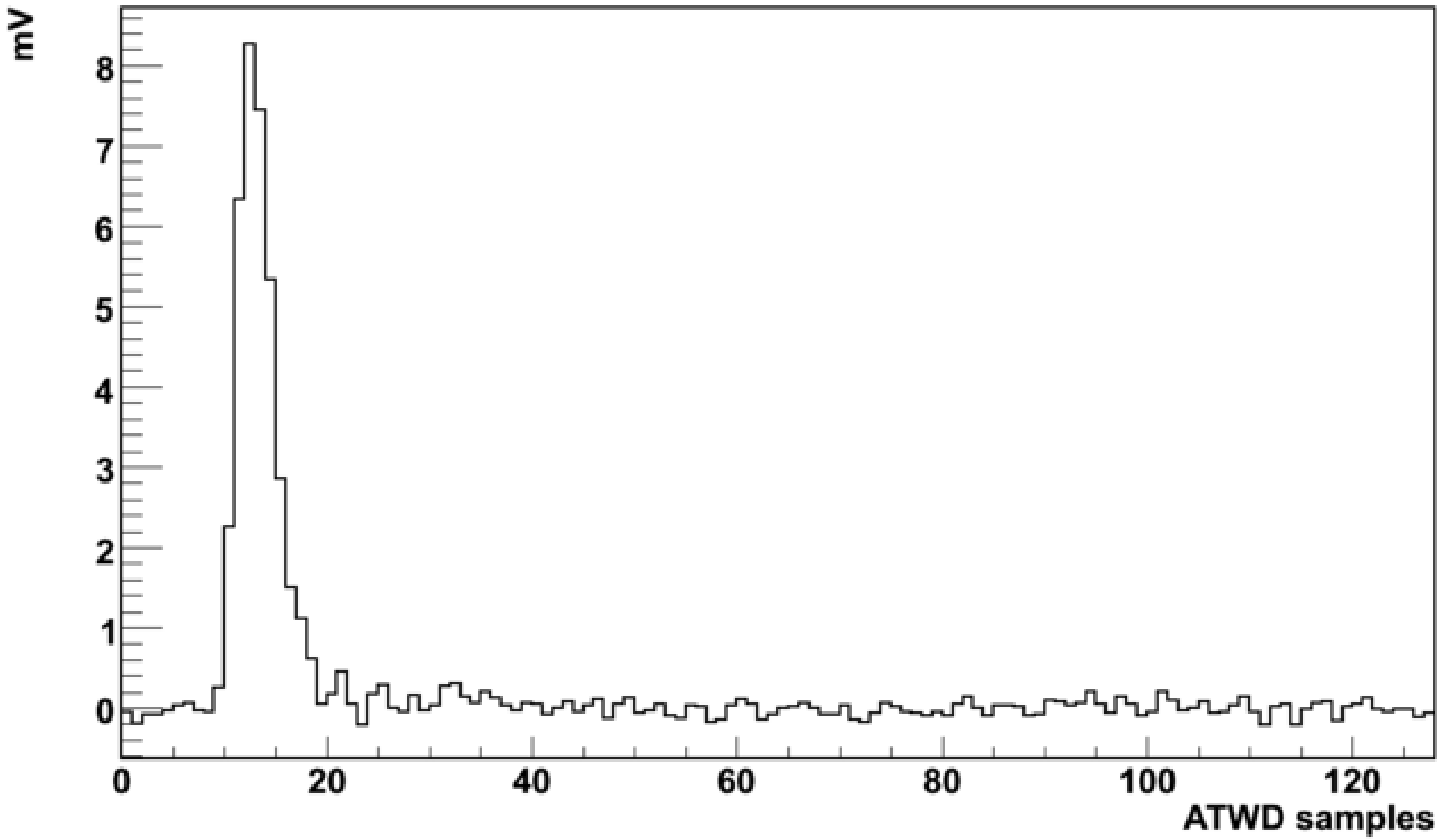

Figure 7

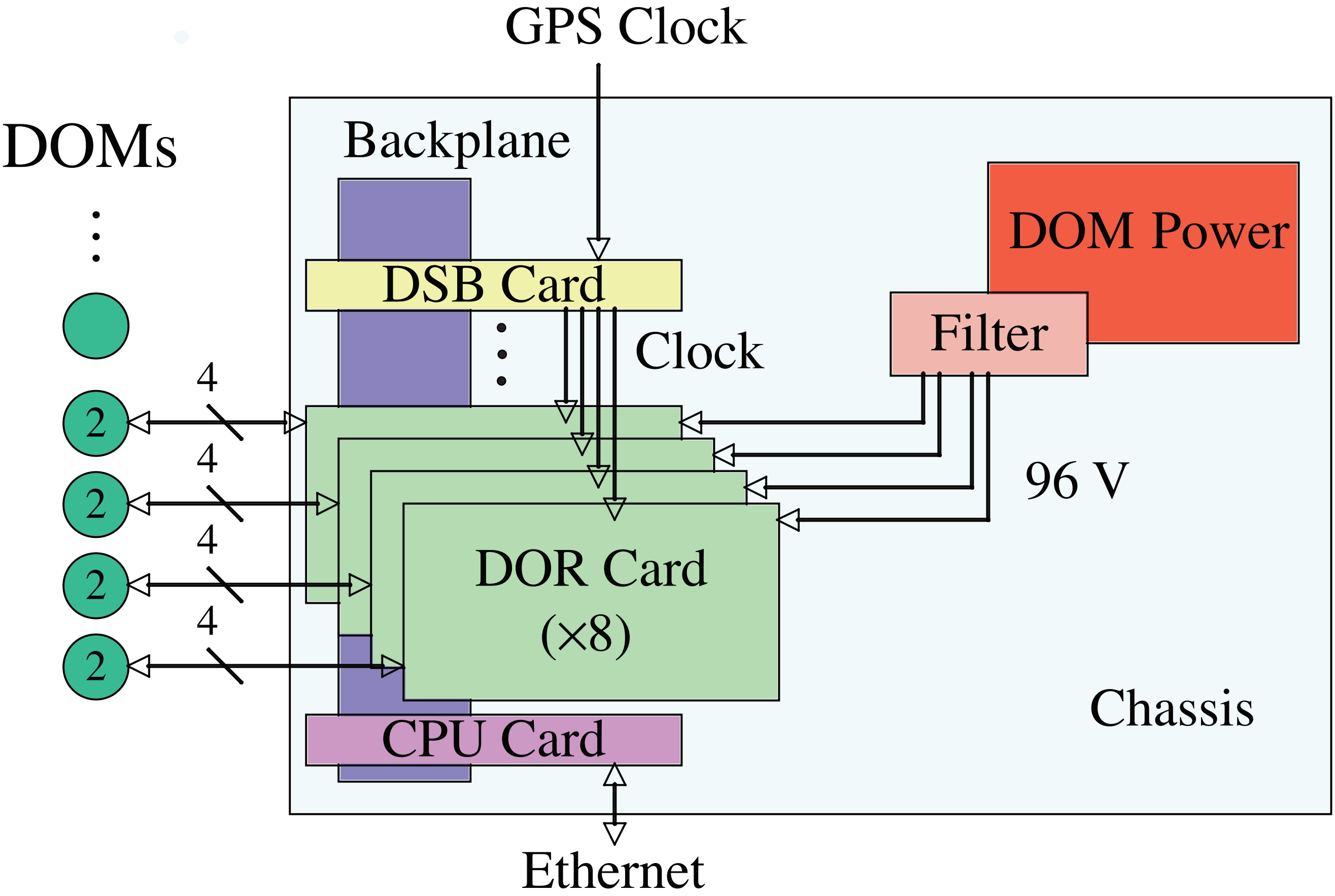

Figure 8
Click here to download high resolution image

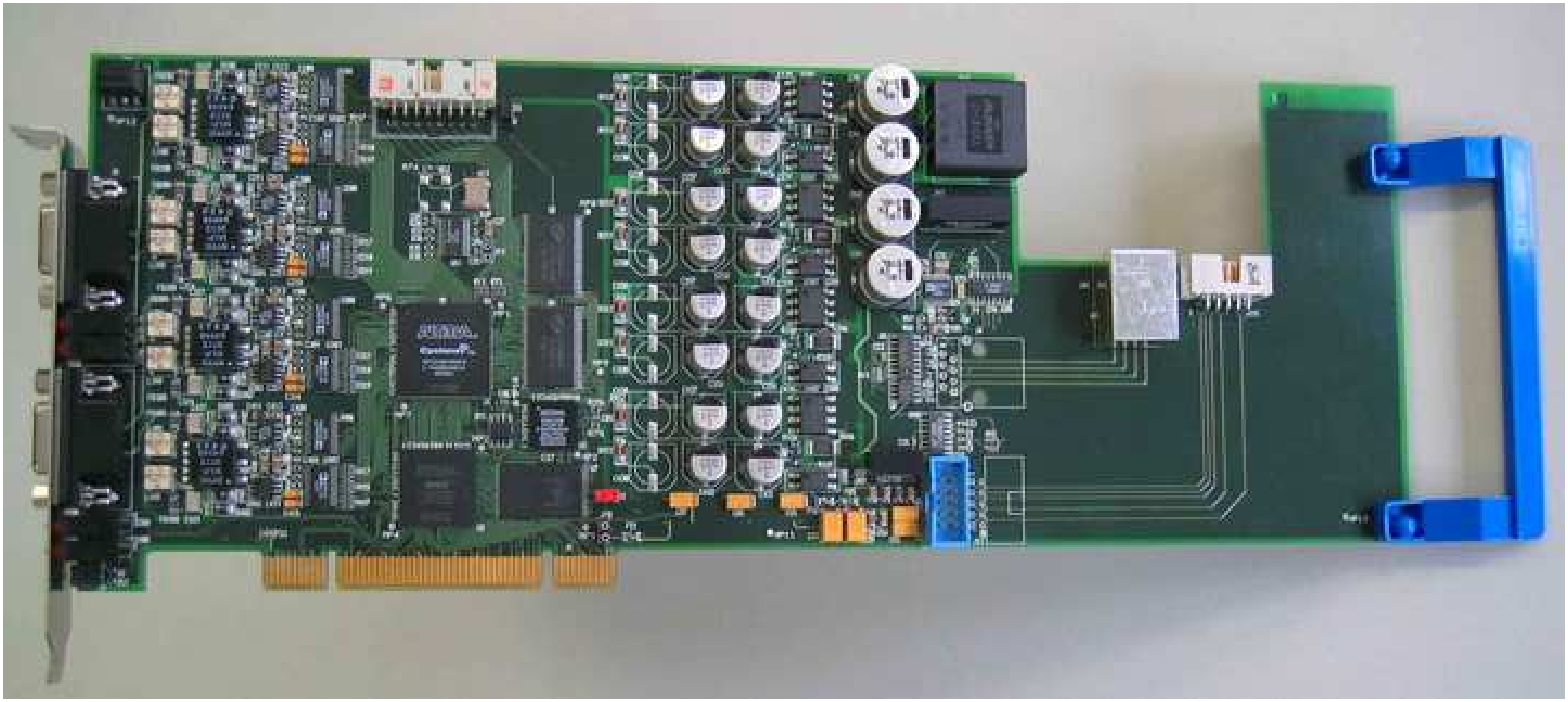

Figure 9

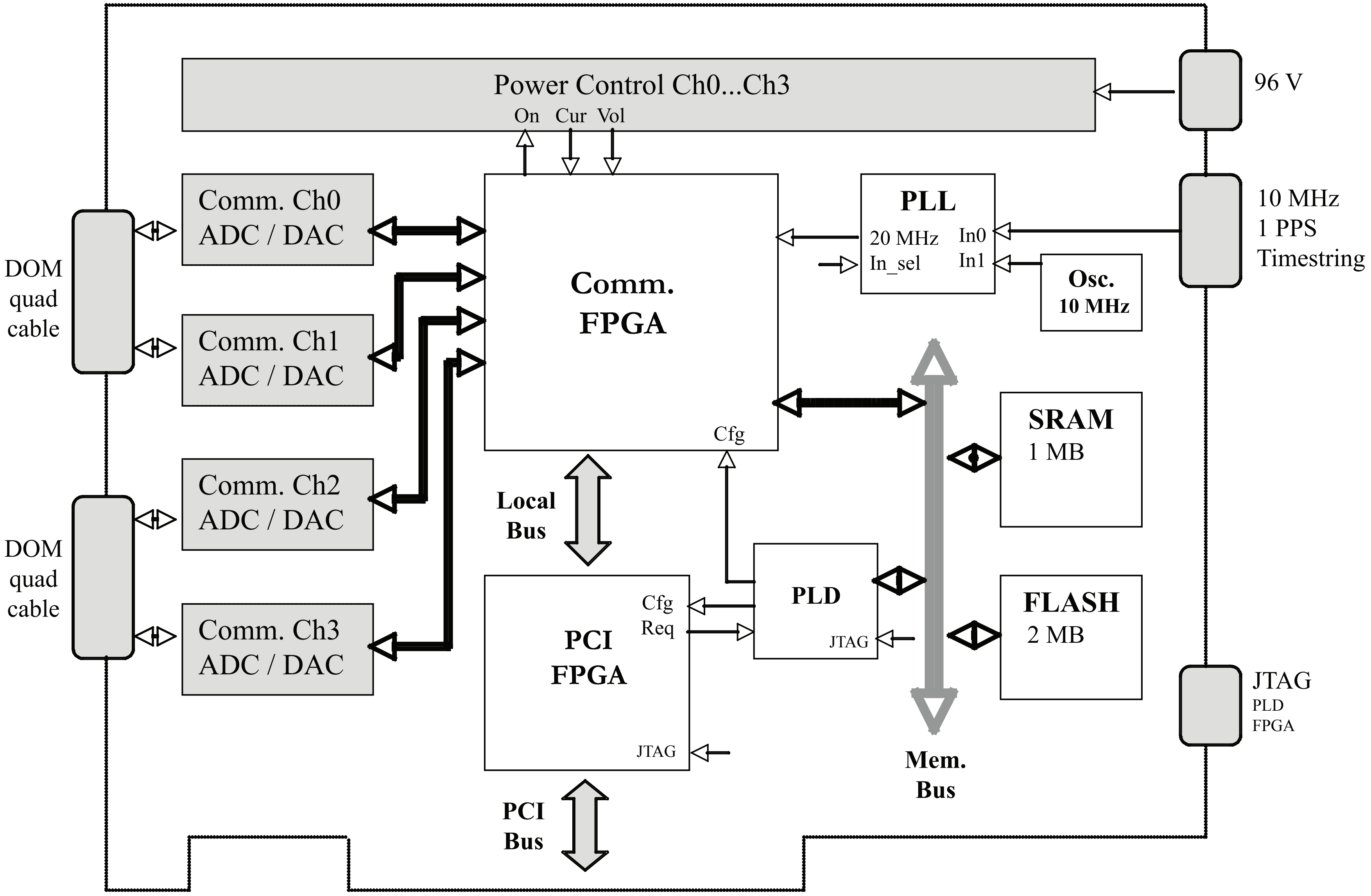

Figure 10

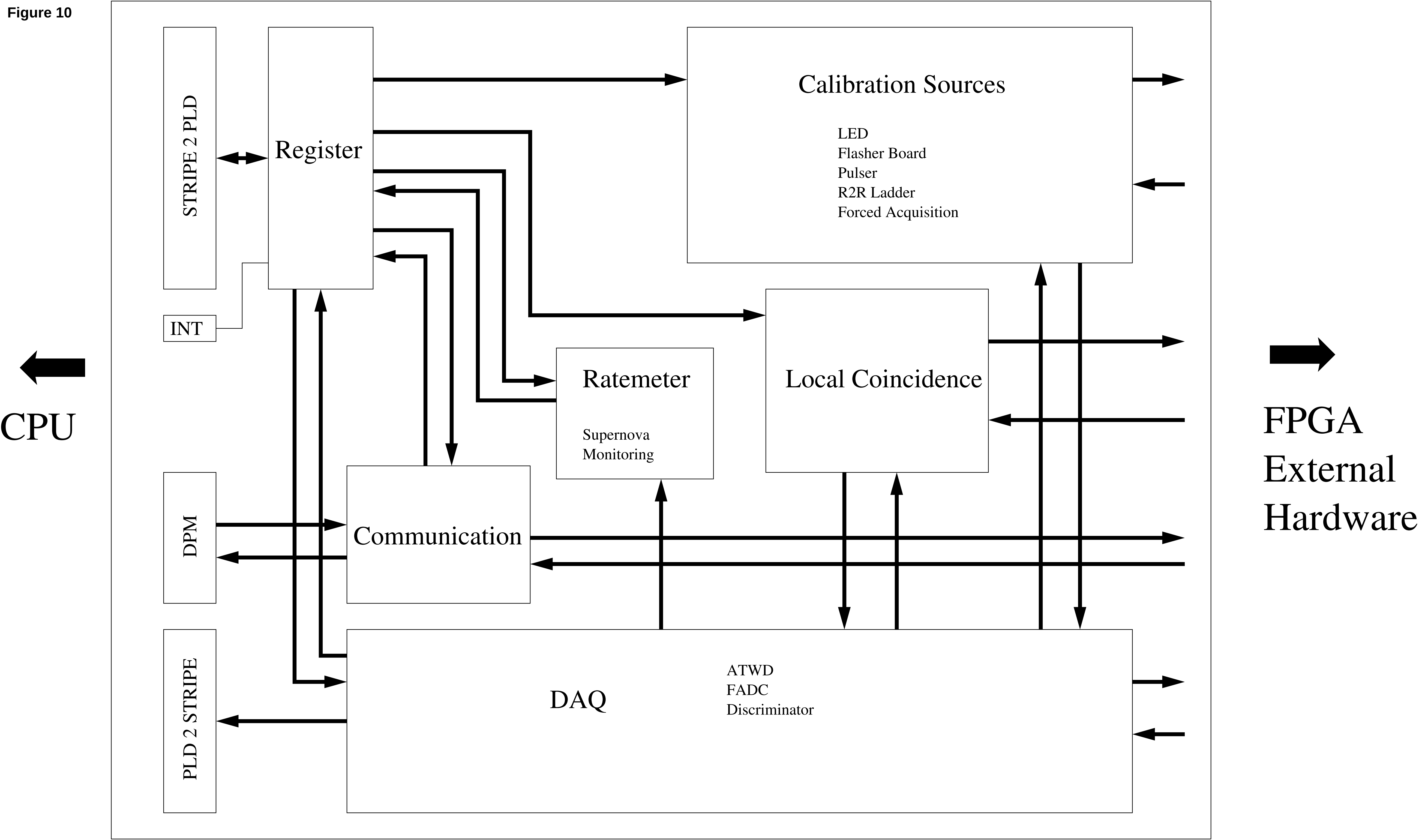

Figure 11

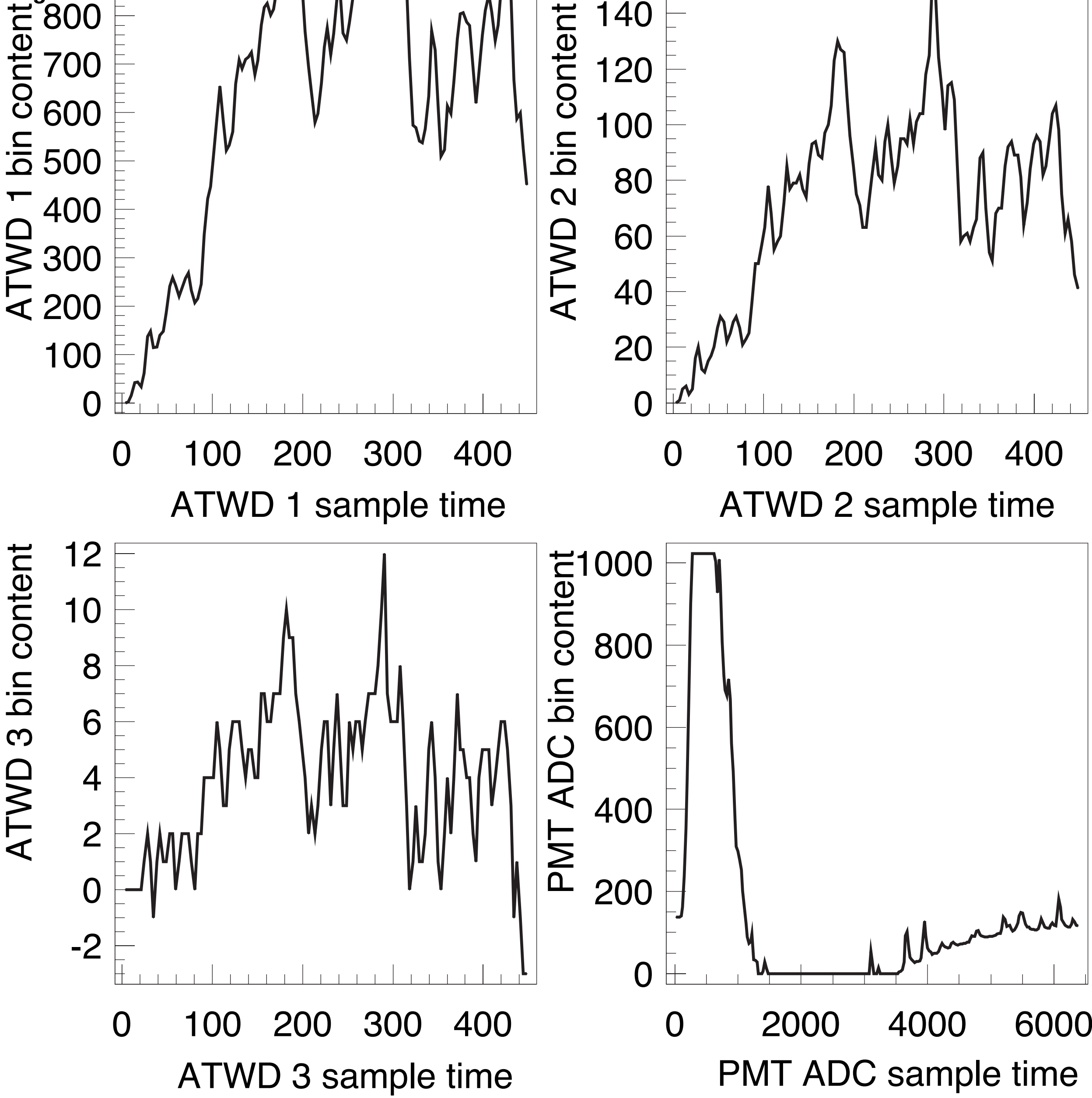

Figure 12

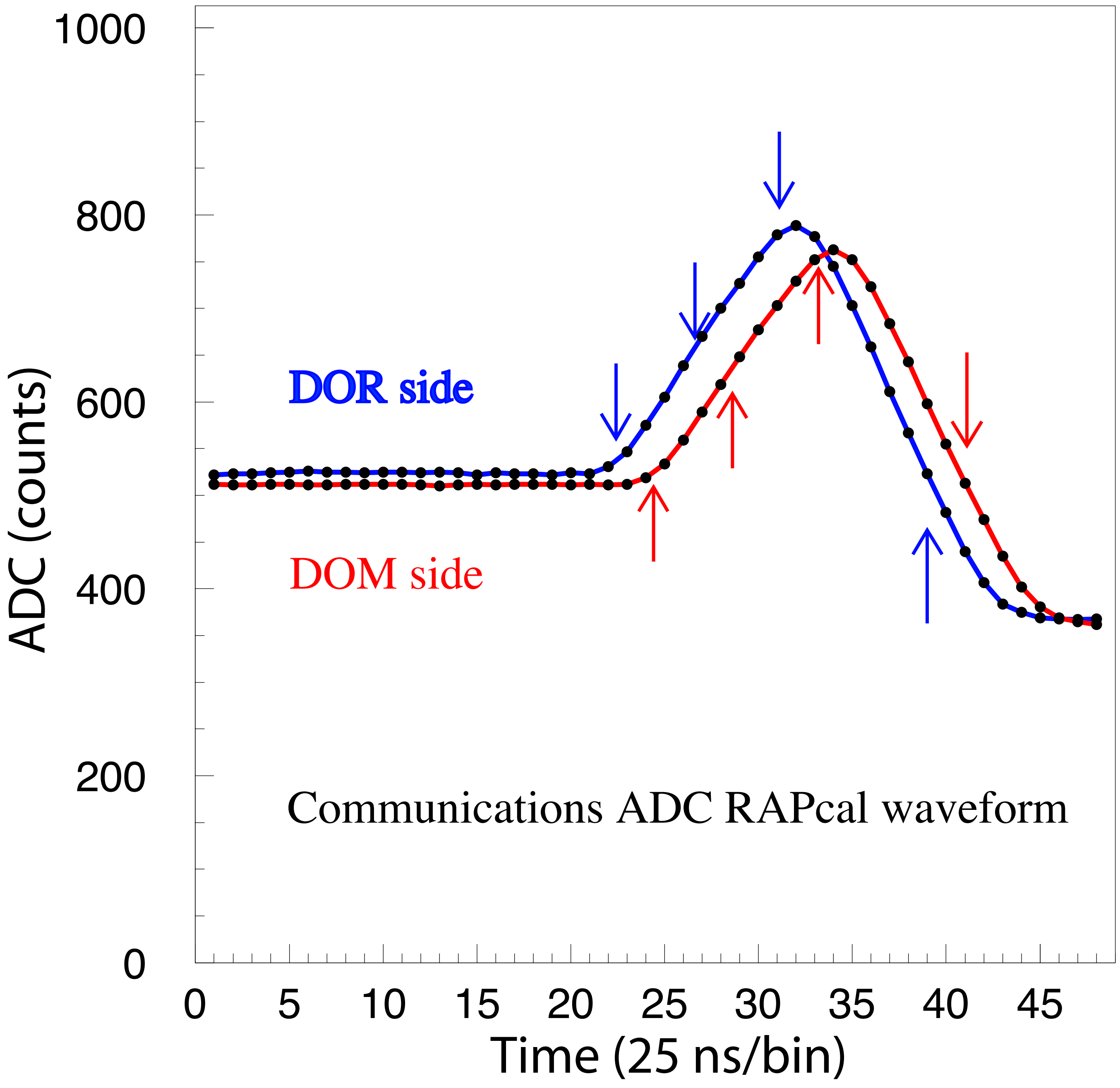

Figure 13

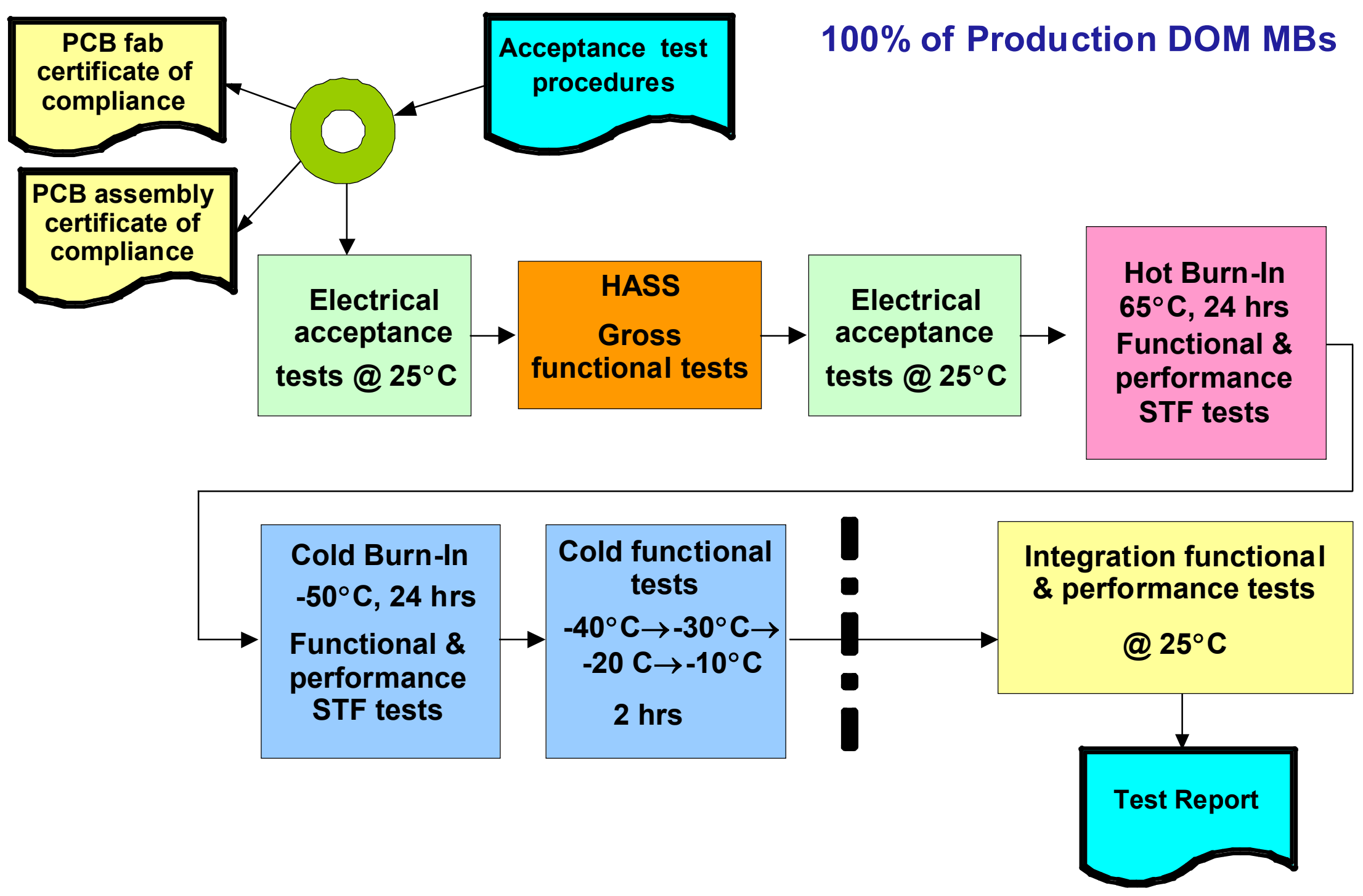

100% of Production DOM MBs
PCB fab certificate of compliance
Acceptance test procedures
PCB assembly certificate of compliance
Electrical acceptance tests @ 25°C
HASS
Gross functional tests
Electrical acceptance tests @ 25°C
Hot Burn-In 65°C, 24 hrs Functional & performance STF tests
Cold Burn-In -50°C, 24 hrs Functional & performance STF tests
Cold functional tests -40°C→-30°C→ -20 C→-10°C 2 hrs
Integration functional & performance tests @ 25°C
Test Report

Figure 14

Click here to download high resolution image

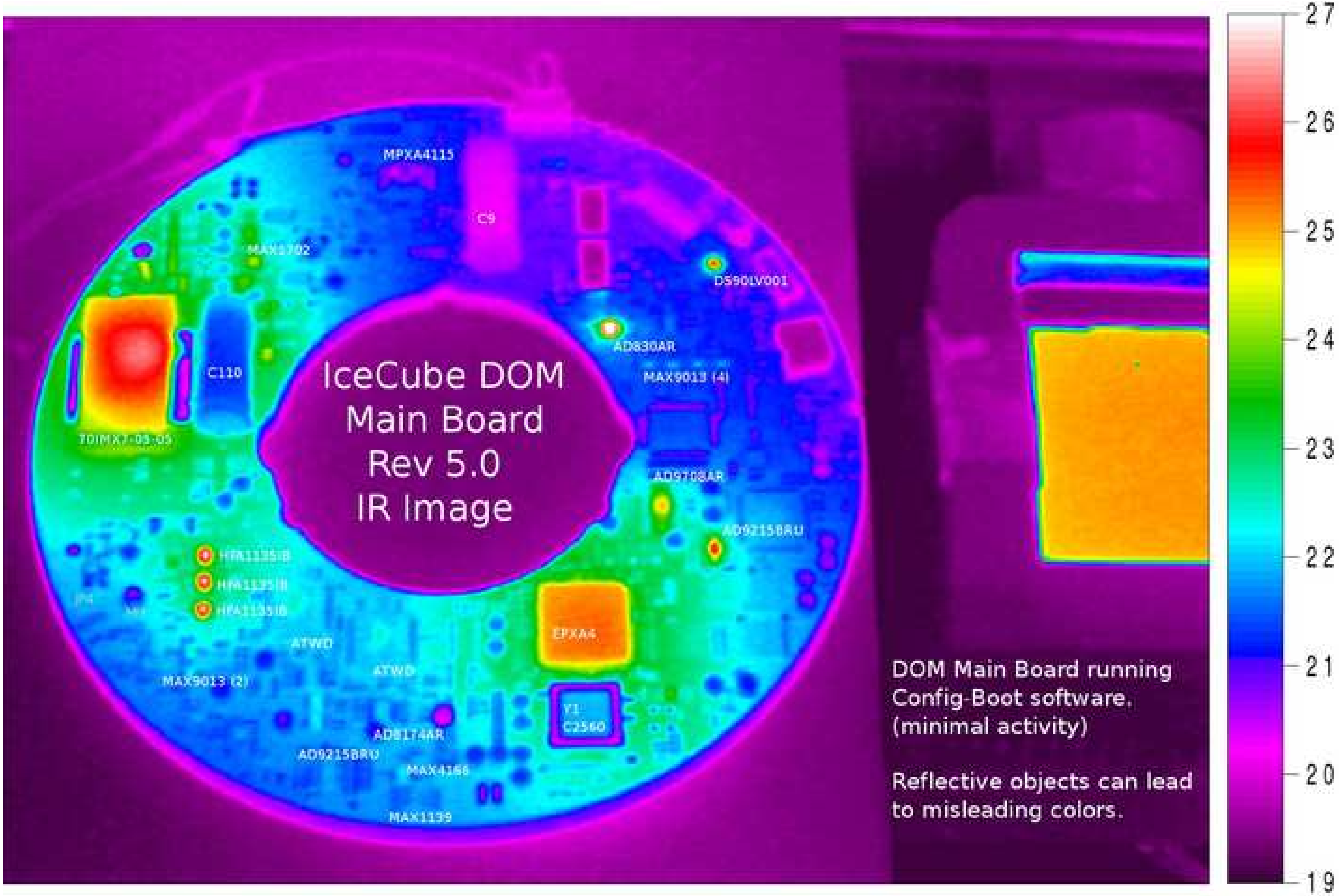

Figure 15

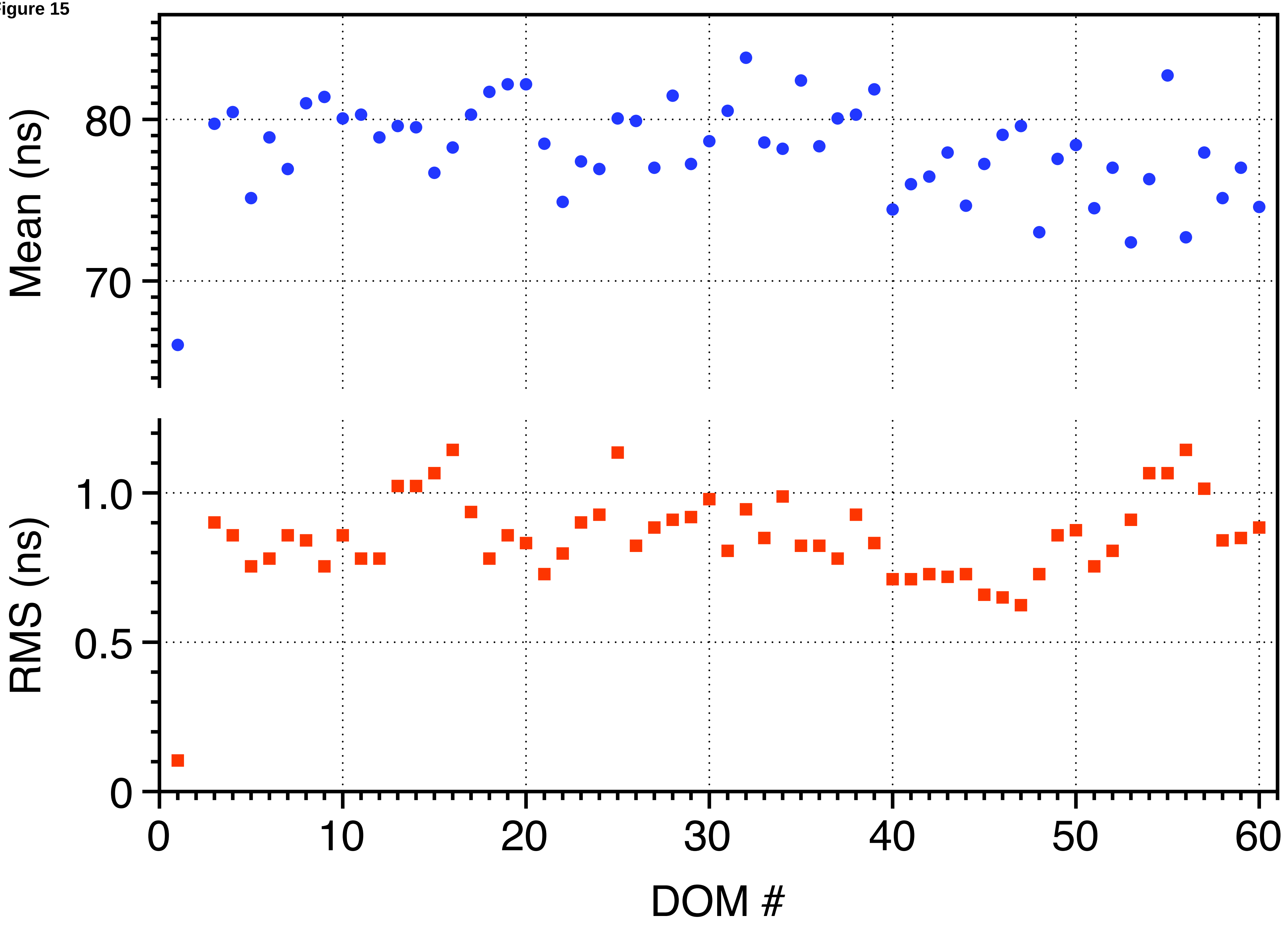

Figure 16

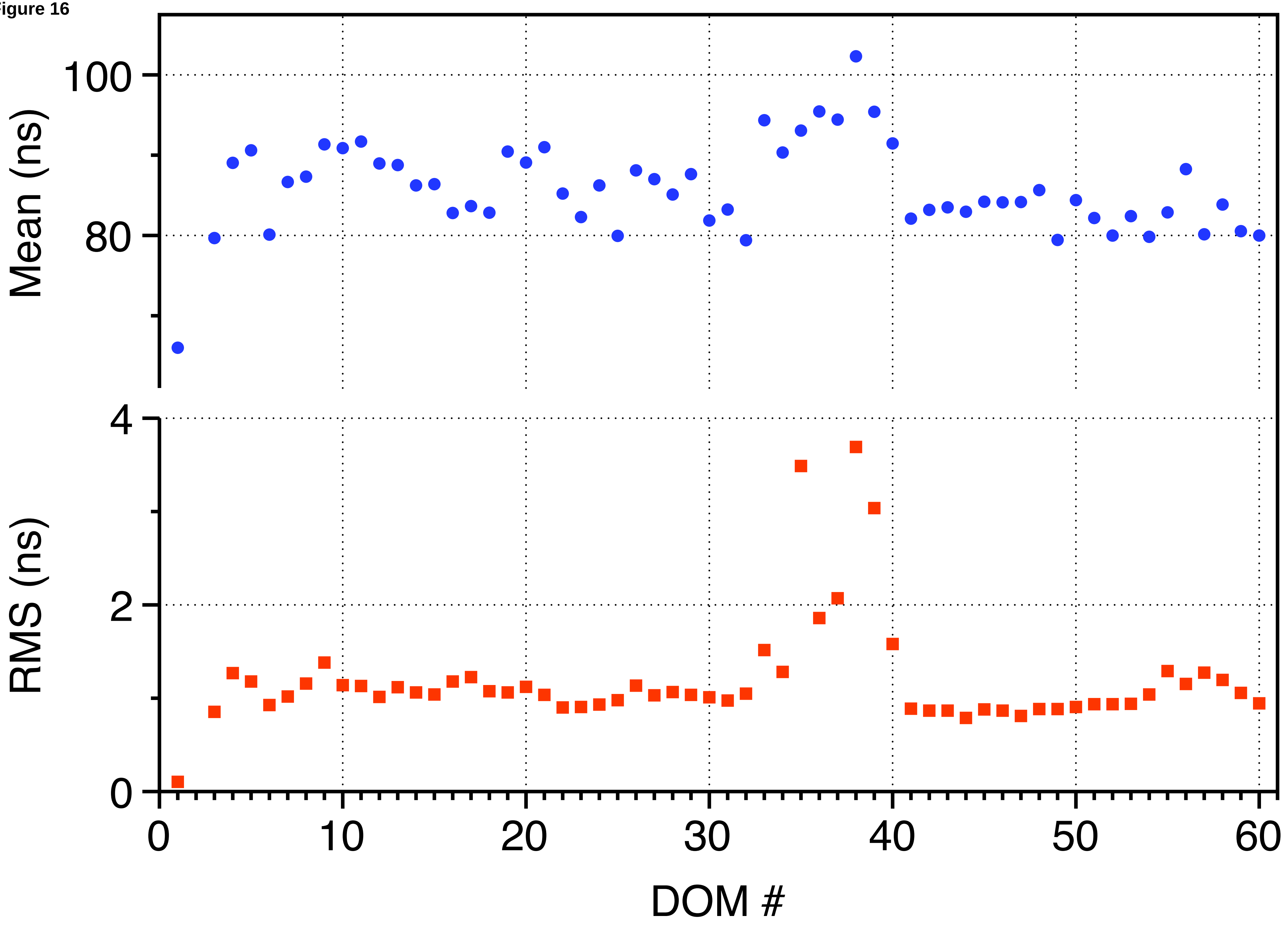

Figure 17

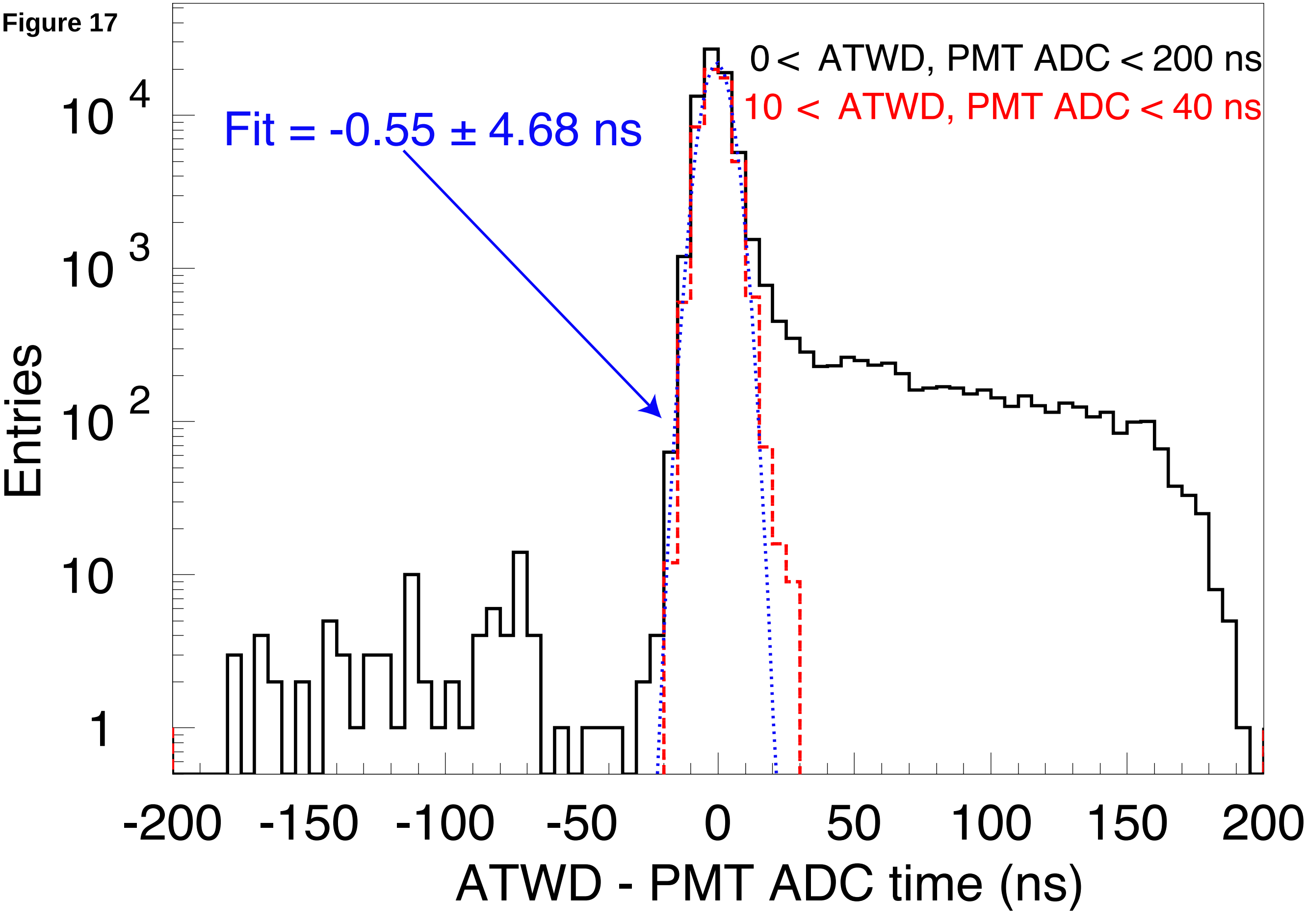

0 < ATWD, PMT ADC < 200 ns
10 < ATWD, PMT ADC < 40 ns
Fit = -0.55 ± 4.68 ns
Entries
1
10
10 2
10 3
10 4
-200
-150
-100
-50
0
50
100
150
200
ATWD - PMT ADC time (ns)

Figure 18

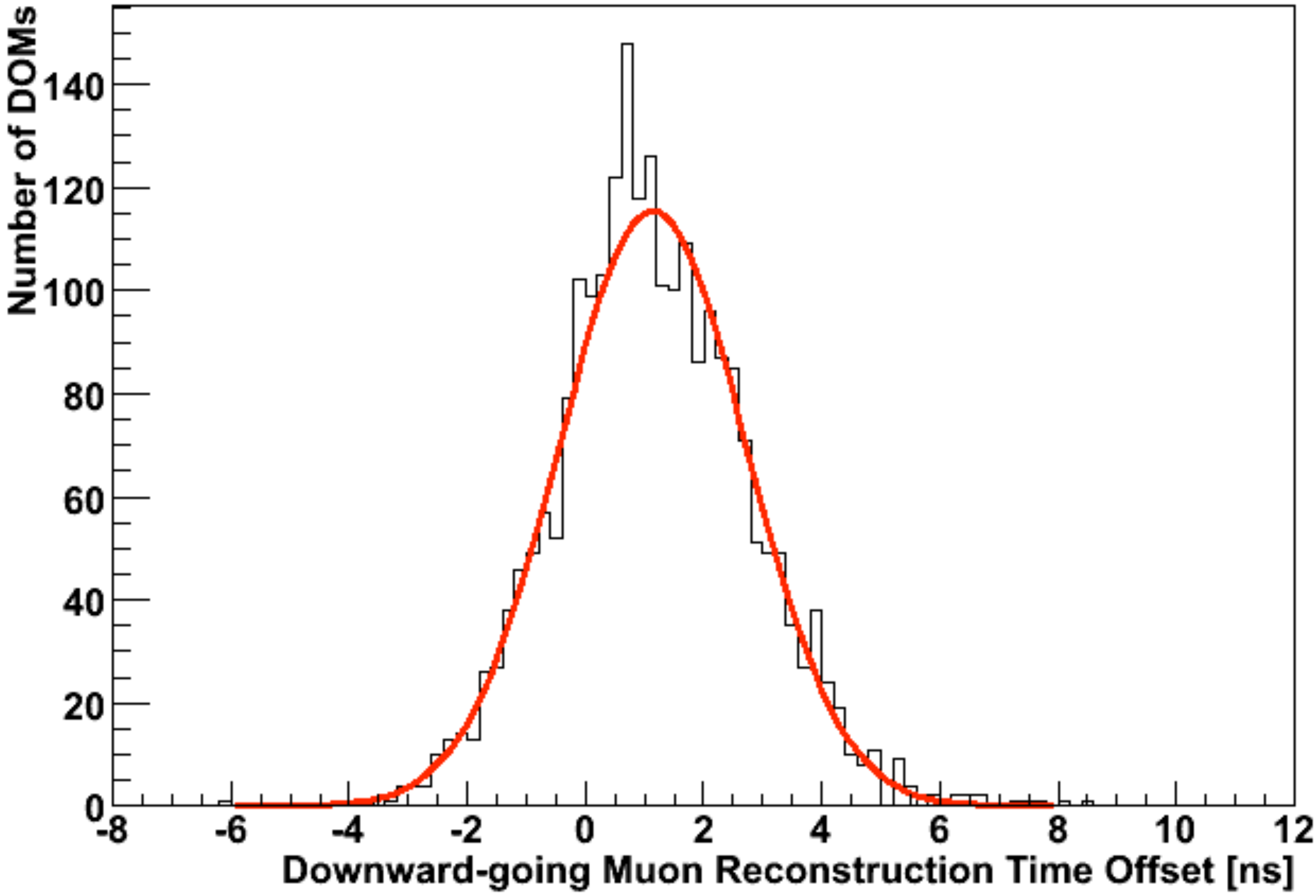

**Figure 19**

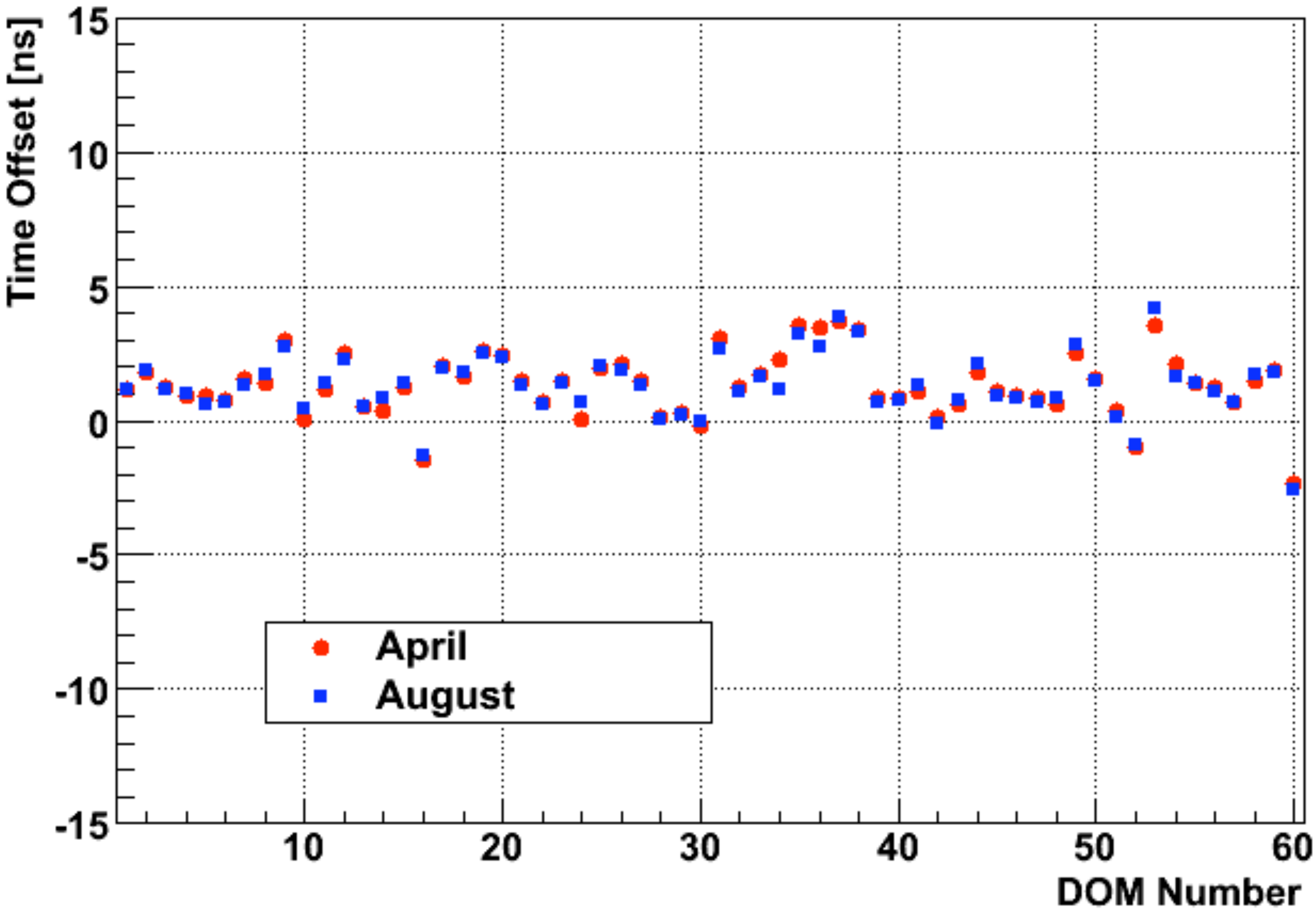

Figure 20

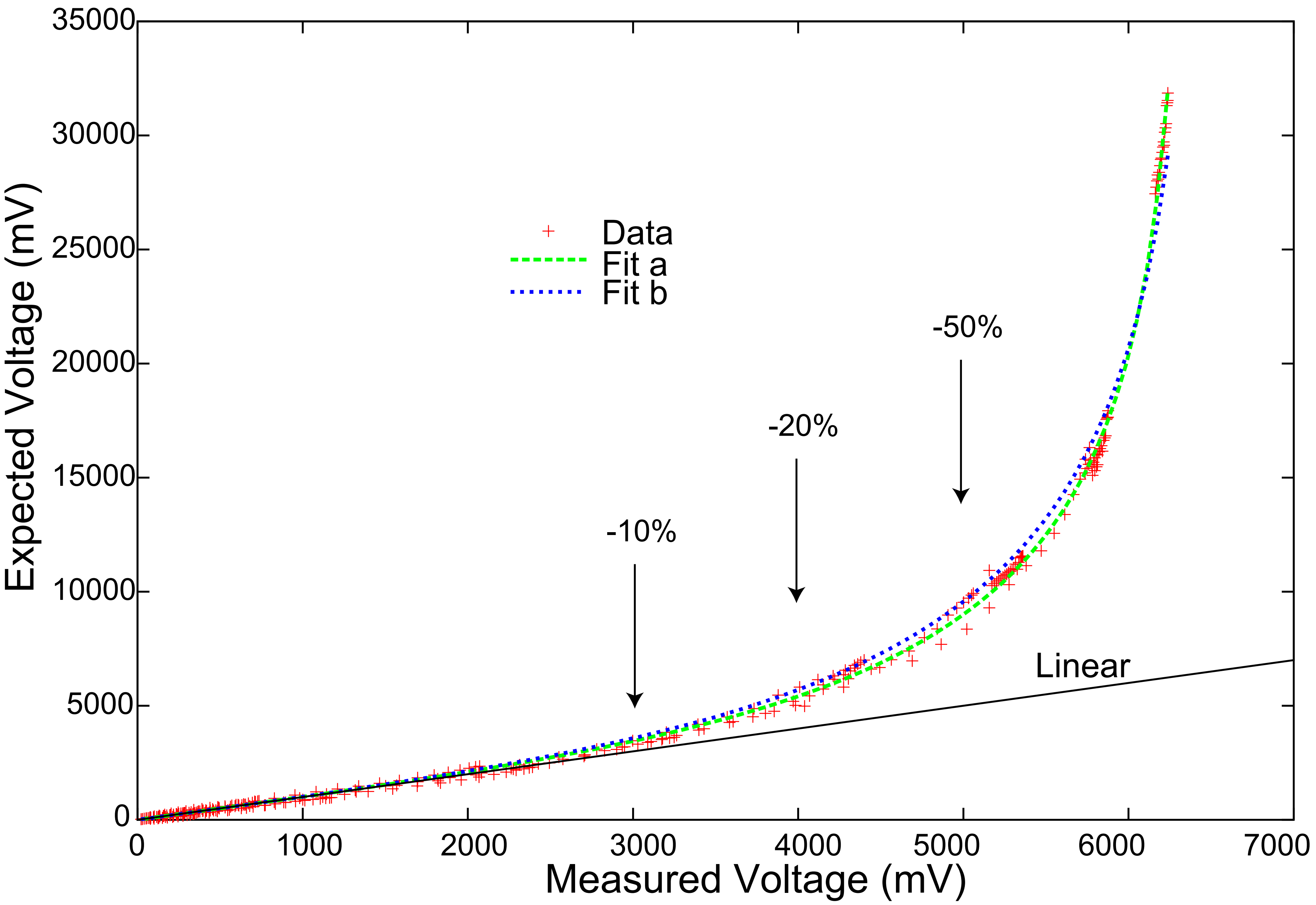